\documentclass[reprint,prl,aps,amsmath,amssymb,superscriptaddress]{revtex4-2}

\usepackage[T1]{fontenc}
\usepackage[utf8]{inputenc}
\usepackage[english]{babel}
\usepackage{amsmath,amssymb}   % ok with revtex4-2
\usepackage{bm}
\usepackage{graphicx}
\usepackage{units}
\usepackage{hyperref}
\hypersetup{
  colorlinks=true,
  citecolor=blue,
  urlcolor=blue,
  linkcolor=red,
  breaklinks=true
}

\begin{document}

\title{Twistraintronics in Square Moir\'{e} Superlattices of Stacked Graphene Layers}

\author{Roberto Carrasco}
\thanks{These authors contributed equally}
\affiliation{Departamento de F\'isica de la Materia Condensada, Universidad Aut\'{o}noma de Madrid, E-28049 Madrid, Spain}

\author{Federico Escudero}
\thanks{These authors contributed equally}
\affiliation{Imdea Nanoscience, Faraday 9, 28015 Madrid, Spain}

\author{Zhen Zhan}
\affiliation{Imdea Nanoscience, Faraday 9, 28015 Madrid, Spain}

\author{Eva Cort\'es-del R\'io}
\affiliation{Department of Physics, University of Hamburg, D-20355, Hamburg, Germany}

\author{Beatriz Viña-Baus\'a}
\affiliation{Departamento de F\'isica de la Materia Condensada, Universidad Aut\'{o}noma de Madrid, E-28049 Madrid, Spain}

\author{Yulia Maximenko}
\affiliation{Department of Physics, Colorado State University, Fort Collins, Colorado 80523, USA}

\author{Pierre A. Pantale\'on}
\email{pierre.pantaleon@imdea.org}
\affiliation{Imdea Nanoscience, Faraday 9, 28015 Madrid, Spain}
\affiliation{Department of Physics, Colorado State University, Fort Collins, Colorado 80523, USA}

\author{Francisco Guinea}
\affiliation{Imdea Nanoscience, Faraday 9, 28015 Madrid, Spain}
\affiliation{Donostia International Physics Center, Paseo Manuel de Lardiz\'{a}bal 4, 20018 San Sebastián, Spain}

\author{Iv\'an Brihuega}
\email{ivan.brihuega@uam.es}
\affiliation{Departamento de F\'isica de la Materia Condensada, Universidad Aut\'{o}noma de Madrid, E-28049 Madrid, Spain}
\affiliation{Condensed Matter Physics Center (IFIMAC), Universidad Aut\'onoma de Madrid, E-28049 Madrid, Spain.}
\affiliation{Instituto Nicol\'as Cabrera (INC), Universidad Aut\'onoma de Madrid, E-28049 Madrid, Spain.}

\begin{abstract}
We report the first observation of controlled, strain-induced square moiré patterns in stacked graphene. By selectively displacing native wrinkles, we drive a reversible transition from the usual trigonal to square moiré order. Scanning tunneling microscopy reveals elliptically shaped AA domains, while spectroscopy shows strong electronic correlation in the form of narrow bands with split Van Hove singularities near the Fermi level. A continuum model with electrostatic interactions reproduces these features under the specific twist–strain combination that minimizes elastic energy. This work demonstrates that the combination of twist and strain, or \textit{twistraintronics}, enables the realization of highly correlated electronic states in moiré heterostructures with geometries that were previously inaccessible.
\end{abstract}

\maketitle

\textit{Introduction}.--- 
The discovery of unconventional superconductivity and strongly correlated phases of matter in twisted bilayer graphene \cite{cao_unconventional_2018, cao_correlated_2018, kerelsky_maximized_2019, yankowitz_tuning_2019, oh_evidence_2021, xie_fractional_2021} has sparked great interest in moiré heterostructures \cite{andrei_marvels_2021,Andrei2020}. These behaviors depend critically on the electronic modulation induced by the moiré potential \cite{shallcross_quantum_2008,shallcross_electronic_2010, laissardiere_localization_2010,lopes_dos_santos_continuum_2012, Sboychakov2015, Dai2016}, particularly around the magic angle, where electronic correlations are greatly enhanced by the formation of very 
flat bands \cite{bistritzer_moire_2011,SuarezMorell2010,lopes_dos_santos_graphene_2007, Li2010,Brihuega_VHs_PRL2012,Yin2015}. In general, the moiré potential depends on the moiré interference created by the lattice mismatch in the system \cite{mele_commensuration_2010, Koshino2015}. While twist-only graphene configurations only yield trigonal moiré patterns  \cite{moon_optical_2014},
the addition of strain can lead to a plethora of different moiré geometries \cite{kazmierczak_strain_2021, sinner_strain_2022, kogl_moire_2023, Escudero2024}. Through the right combination of twist and strain one can actually have any moiré pattern \cite{Escudero2024}, each with unique properties \cite{lee_strain-shear_2017, huder_electronic_2018, bi_designing_2019, mesple_heterostrain_2021,wang_unusual_2023, Mesple2023}. 
The recent advancement in experimentally inducing and manipulating strain in two-dimensional materials created a path to what we call \textit{twistraintronics}---a method to tune electronic properties via the interplay of twist and strain 
\cite{jiang_visualizing_2017,Brzhezinskaya2021,kapfer_programming_2023,pena_moire_2023, Sequeira2024, Huang2025}. 

A particular example is the formation of square moiré patterns in stacked hexagonal lattices,
which have been theoretically predicted to arise under specific combinations of twist and strain \cite{kogl_moire_2023, Escudero2024}. Although some experimental works have reported the observation of rectangular
patterns in strained systems \cite{tilak_moire_2022-1,Boi2025,Ou2025}, so far there has been no report for the formation of square patterns induced by externally applied strains.

In this Letter, we report the first observation of controlled, strain-induced, square moiré patterns in stacked graphene layers. By straining the system through the displacement of wrinkles that arise during sample preparation, we drive a transition from the usual twist-only trigonal moiré pattern to square patterns.  Scanning tunneling microscopy (STM) reveals clear local regions approaching nearly perfect square order, with prominent, elliptically shaped AA-stacking domains induced by strain. We show that these square patterns can arise from a continuous set of twist and strain configurations. Further scanning tunneling spectroscopy (STS) along different paths of the square pattern reveals the presence of narrow bands with two prominent Van Hove singularities (VHs) around the Fermi energy, which are further split by the applied strain. The observed behavior is accurately captured by a continuum model that includes electrostatic interactions under the specific combination of twist and shear strain that minimizes the elastic energy.

\begin{figure*}[t!]
	\includegraphics[width=\linewidth]{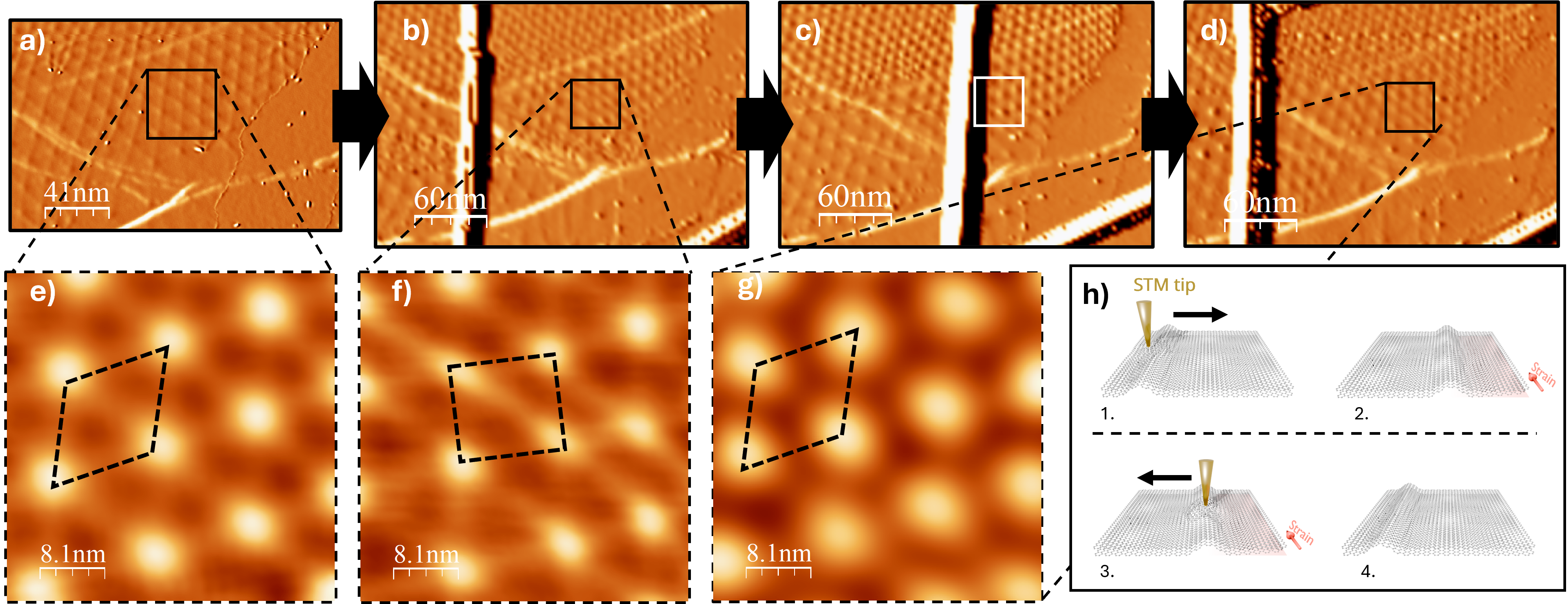}	
	\caption{Local strain control via STM-based manipulation of graphene wrinkles. \textbf{(a–d)} STM sequence showing reversible moiré switching (trigonal → square → trigonal) by laterally shifting a nearby wrinkle: \textbf{(a)} The region of interest (black square) exhibits a strain-free trigonal moiré geometry; \textbf{(b)} approaching the wrinkle yields a square geometry; \textbf{(c–d)} sweeping the wrinkle across the area and then retracting it releases strain and restores the trigonal geometry. \textbf{(e–g)} Zoomed-in views of the same area after each manipulation step; dashed parallelograms mark the moiré unit cells. \textbf{(h)} Schematic of the manipulation mechanism (see Supplemental Material, animation \texttt{wrinkle\_manipulation.mp4}). STM parameters: $I_T$ = $50\,\mathrm{pA}$, $V_{bias}$ = $600\,\mathrm{mV}$.
    }\label{fig:wrnkl}
\end{figure*}

\textit{Strain manipulation and STM/STS results}.--- 
The graphene samples were synthesized via thermal decomposition of 6H-SiC(000-1)--- a well-established method for obtaining  high-quality, large-area graphene domains~\cite{Varchon_GrapheneGrowth_PRB2008, Brihuega_VHs_PRL2012} (see Ref.~\cite{SM} for details). Within these domains, the rotations between the surface graphene layers give rise to moiré patterns (see Fig.~\ref{fig:wrnkl}a and Fig.~\textcolor{red}{S8} in Ref.~\cite{SM}). During growth, the graphene layers experience multiple sources of compression and dilation due to the difference in thermal expansion relative to the underlying SiC substrate. This induces significant mechanical stress, causing the graphene layers to buckle into wrinkles~\cite{Liu2011, Wang2013,Meng2013,Deng2016}. While the twist angle is generally fixed upon sample growth, the local strain, in contrast, offers a dynamic tuning parameter. 

Here, we introduce a method to modify local strain at the nanoscale based on the controlled displacement of ubiquitously present graphene wrinkles. By pushing these wrinkles laterally with the STM tip, we were able to manipulate them over distances greater than 100 nm, as shown in Fig.~\ref{fig:wrnkl}. As a consequence of such manipulations, the nearby moiré patterns exhibit evident changes in geometry~\cite{kogl_moire_2023, Escudero2024}. 

For instance, in Fig.~\ref{fig:wrnkl}(a-d) we can see how the moiré of a localized graphene region is reversibly modified from trigonal to square geometry, thereby modifying the electronic structure of the correlated system \textit{in situ}~\cite{bi_designing_2019, Escudero2024}. The initial state of the selected region displays a conventional trigonal moiré pattern (Fig.~\ref{fig:wrnkl}a). By approaching an adjacent wrinkle with the STM tip, we induce a strain field that transforms the superlattice into a well-defined square moiré (Fig.~\ref{fig:wrnkl}b). A subsequent manipulation first displaces the wrinkle over the region of study (Fig.~\ref{fig:wrnkl}c), then reverses the process, releasing the strain and restoring the trigonal pattern (Fig.~\ref{fig:wrnkl}d). This approach allows us to actively tune the properties of moiré superlattices within a single sample, providing a unique platform for investigating the effects of strain on correlated electronic states in twisted graphene.

\begin{figure*}[t]
	\includegraphics[width=\linewidth]{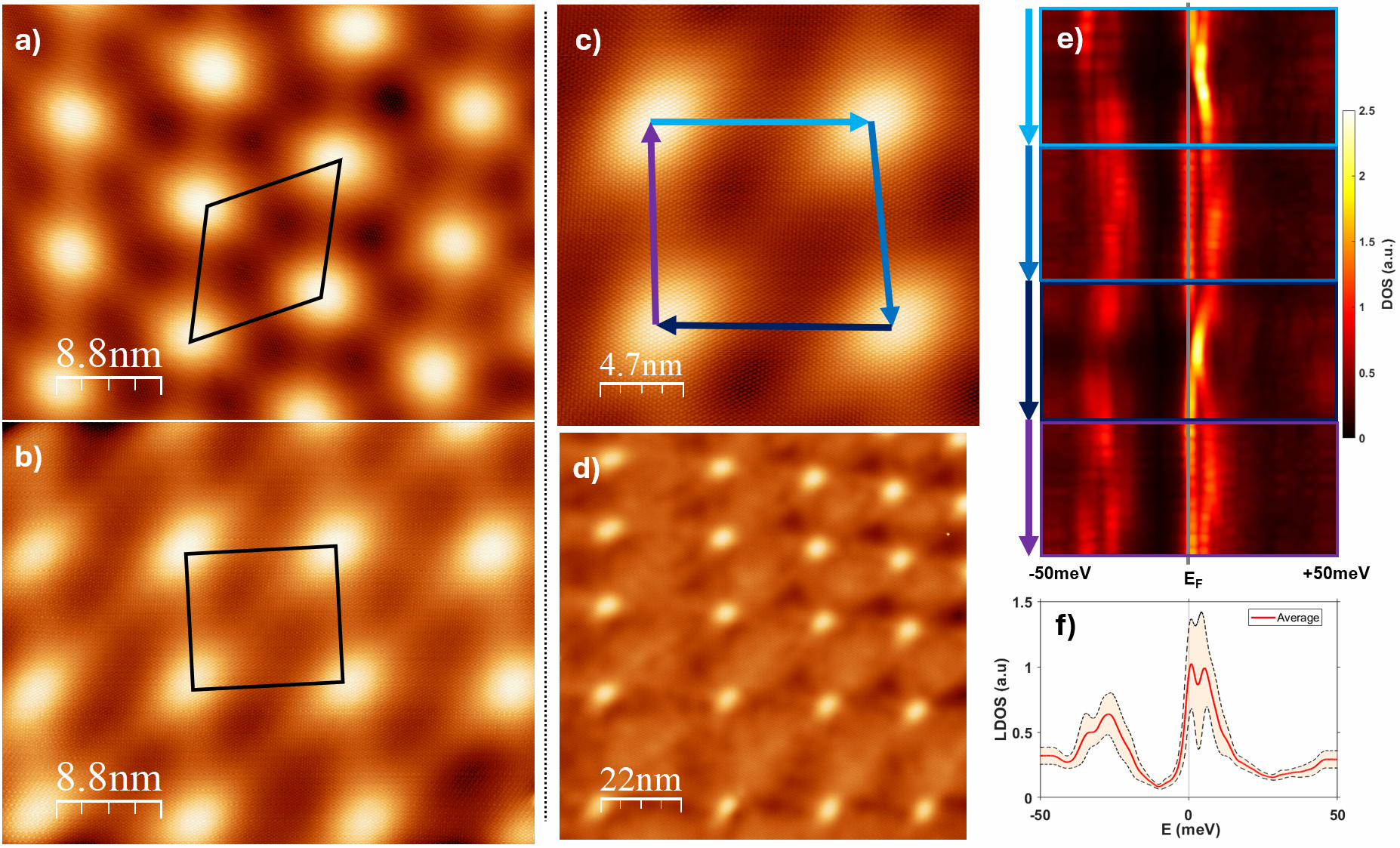}	
	\caption{STM topography and spatially-resolved spectroscopy of trigonal and square moiré superlattices in TBG. \textbf{(a-c)} Atomic-resolution STM images of trigonal \textbf{(a)} and square moiré patterns \textbf{(b,c)} , both with a periodicity of $\approx 12\,\mathrm{nm}$. In each panel, the black shapes outline the corresponding moiré unit cell. \textbf{(c)} Atomic-resolution STM image of a single square moiré unit cell. Colored arrows trace the closed-loop path along which differential conductance ($dI/dV$) spectra lines were acquired. \textbf{(d)} STM image of a large strained moiré. The non-uniform periodicity is caused by an inhomogeneous strain profile. (All STM data available with atomic resolution in the Supplementary Material). \textbf{(e)} Two-dimensional map of dI/dV intensity as a function of energy (horizontal axis)  and spatial position along the four segments of the path (vertical axis; color-coded to match the arrows in the left inset). Spectra show consistent features along vertical lines, indicating spatial homogeneity. Similar sharp features can be seen in the horizontal lines of spectra. \textbf{(f)} Average of all $dI/dV$ curves in (e), showing LDOS vs. Energy. The red curve shows the average; the area shaded in light orange between the black dashed curves is the one-standard deviation interval.  STM parameters: $I_T$ = $340\,\mathrm{pA}$, $V_{bias}$ = $50\,\mathrm{mV}$ \textbf{(a)}; $I_T$ = $50\,\mathrm{pA}$, $V_{bias}$ = $16\,\mathrm{mV}$ \textbf{(b)}; $I_T$ = $230\,\mathrm{pA}$, $V_{bias}$ = $10\,\mathrm{mV}$ \textbf{(c)}; $I_T$ = $50\,\mathrm{pA}$, $V_{bias}$ = $25\,\mathrm{mV}$ \textbf{(d)}.}\label{fig:exp}
\end{figure*}

In our samples, due to rotational disorder, we find a wide variety of moiré periodicities \cite{Brihuega_VHs_PRL2012, Beechem2014}. Most regions show minimal or negligible strain exhibiting typical trigonal moiré patterns, as shown in Fig.~\ref{fig:wrnkl}e, Fig.~\ref{fig:exp}a and Fig.~\textcolor{red}{S8} in Ref.~\cite{SM}, resulting purely from twist—i.e., \textit{twistronics}~\cite{Carr2017,Yang2020,Hennighausen2021}. When strain is applied on a trigonal moiré, three-fold symmetry is broken and the moiré geometry smoothly distorts towards a square unit cell, as shown in Fig.~\ref{fig:wrnkl}b, Fig.~\ref{fig:wrnkl}f and Fig.~\textcolor{red}{S9} in Ref.~\cite{SM}. If in \textit{twistronics} we use the twist angle as a lever to tune the electronic structure, we can now use strain as a second lever \cite{guinea_strain_2012, amorim_novel_2016, naumis_electronic_2017}, which is why we call it \textit{twistraintronics}. In as-grown samples, we also observe square-like, distorted moiré patterns, with uniform (Fig.~\ref{fig:exp}b) or non-uniform (Fig.~\ref{fig:exp}d) periodicities, in some cases spanning distances greater than 100 nm (Fig.~\ref{fig:exp}d and Fig.~\textcolor{red}{S10} in Ref.~\cite{SM}). We attribute this to a homogeneous or inhomogeneous strain profile, respectively, that modifies the moiré length (see the \textit{Theoretical Model} section) and locally alters the twist angle~\cite{Benschop2021, Nakatsuji2022, yu2024twist}, further modulating the periodicity. 

To probe the electronic properties of the square patterns, we carried out STS measurements, focusing on energies near the Fermi energy $E_F$. We obtained differential conductance $dI/dV$ spectra, proportional to the Local Density of States (LDOS) of the sample region directly below the tip at the atomic scale~\cite{Tersoff1985, Chen1988}. 
The conductance map $dI/dV(x,E)$ in Fig.~\ref{fig:exp}, plotted with respect to the position along the edge of the square moiré unit cell and the energy, reveals a peak at $E_{F}$ and another at around 20 mV below $E_F$. Each of these peaks is further split into two components, reflecting the strain-induced formation of multiple VHs
\cite{kerelsky_maximized_2019, choi_electronic_2019, bi_designing_2019}.

\begin{figure*}[t]
	\includegraphics[width=\linewidth]{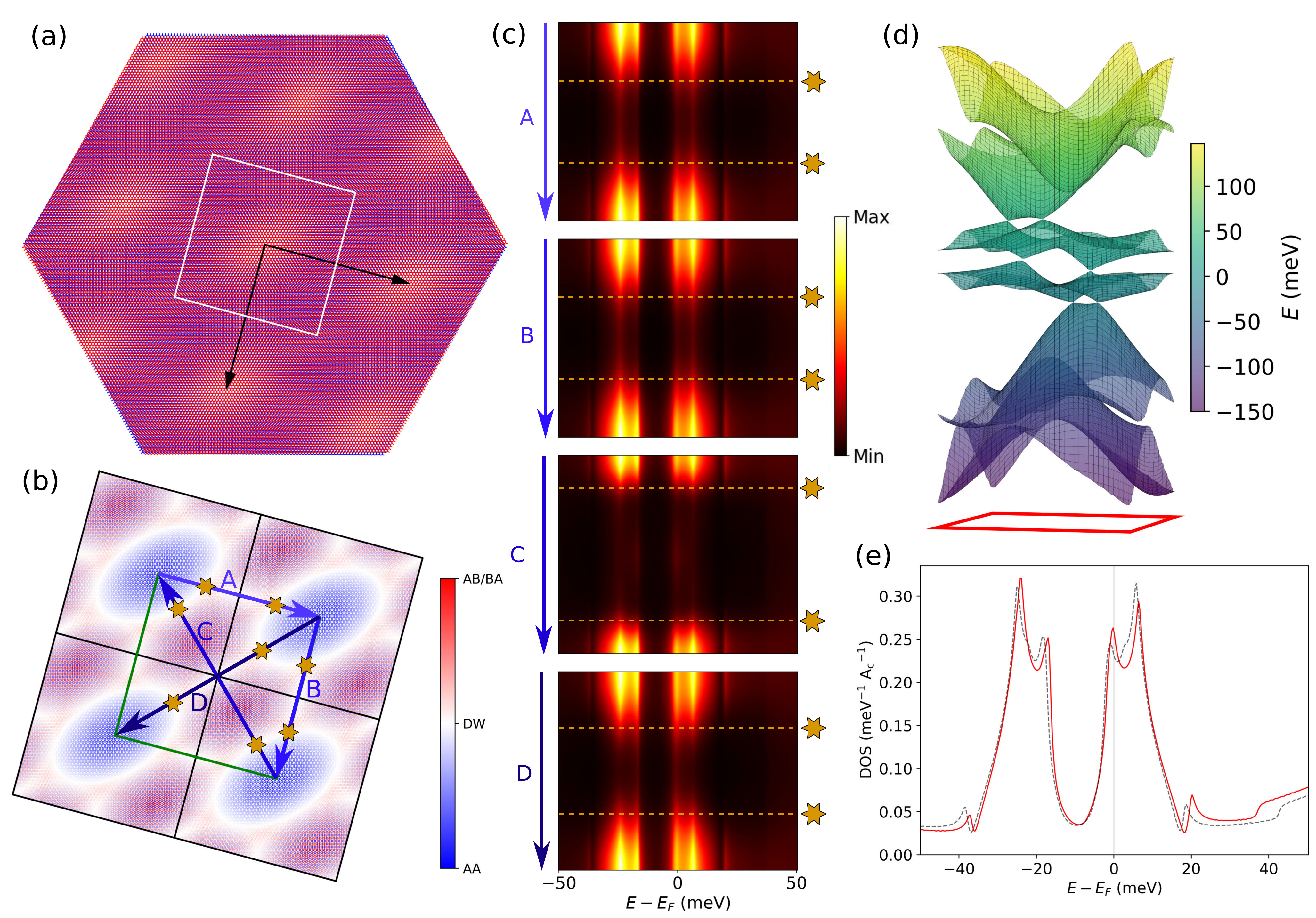}	
	\caption{\textbf{(a)} Square moiré pattern formed in a bilayer graphene configuration with a twist angle $\theta\approx1.125{}^{\circ}$ and shear strain with magnitude $\epsilon_s\approx-0.526\%$ and direction $\phi=30{}^{\circ}$ (see Ref.~\cite{SM}). The top and bottom layers are rotated by $\pm\theta/2$ and strained with equal magnitude but opposite direction. \textbf{(b)} Colormap of atomic positions in the square pattern, indicating the stacking regime of each atom relative to the closest atom in the other layer, raging from directly on top (AA), to in between (DW), to bernal stacking (AB/BA). The gold stars point the DW that indicates the transition from AA to AB/BA stacking. \textbf{(c)} LDOS along the four directions $A,B,C,D$ shown in \textbf{(b)}, for energies $E=E_{F}\pm50\,\mathrm{meV}$ around the Fermi energy $E_{F}$. \textbf{(d)} 3D plot of the band structure. \textbf{(e)} Total density of states with Hartree (red line) and without Hartree (gray dashed-line). The electronic properties are obtained from the continuum model with strain, including the electrostatic interactions (Hartree potential) with a filling of $\nu=+1$ electron per moiré unit cell (see Ref.~\cite{SM} for details). }\label{fig:theo}
\end{figure*}

\textit{Theoretical model}.---
The observation of square moiré patterns can be explained by a combined effect of twist and strain in the system \cite{bi_designing_2019, kogl_moire_2023, Escudero2024}.\ Previous theoretical studies have indeed highlighted the possibility of square moiré patterns in hexagonal moiré heterostructures, under particular combinations of twist and uniaxial heterostrain \cite{kogl_moire_2023, Escudero2024}. In general, there is a family of twist and strain configurations that lead to square moiré geometries. Their formation can be understood by identifying those combinations of twist and strain that result in perpendicular, equal-length moiré vectors (see Ref.~\cite{SM} for details). 

In the relevant regime of low twist and strain, the moiré length is much larger than the atomic length, and the family of strain configurations that yield square moiré patterns differ primarily in the orientation of the moiré vectors (see Fig.~\textcolor{red}{S2} in Ref.~\cite{SM}). Therefore, from STM images with a resolution at the moiré scale one can hardly discern, geometrically, the type of strain in the system. The moiré length of the square patterns generally reads
\begin{equation}
L\left(\theta,\epsilon\right)=\frac{a}{2\sin\left(\theta/2\right)}f\left(\epsilon\right),\label{eq:Lm}
\end{equation}
where $f\left(\epsilon\right)$ is a function that depends on a unique parameter $\epsilon$ accounting for the family of strain configurations that yield square patterns (see Ref.~\cite{SM}). Thus one cannot, unequivocally, determine the twist or strain in the system from solely the moiré length. 

To proceed, we note that for any given moiré length there is always a minimum strain configuration, for which the energy cost associated with the lattice deformation is minimum. In particular, we can ask what is the strain tensor $\mathcal{E}$ that minimizes the elastic energy
\begin{equation}
E_{\mathrm{elastic}}=\frac{\lambda}{2}\left(\mathrm{tr}\mathcal{E}\right)^{2}+\mu\left|\mathcal{E}\right|^{2},\label{eq:Elastic}
\end{equation}
where $\lambda$ and $\mu$ are the Lamé coefficients, while $\mathrm{tr}\mathcal{E}=\epsilon_{xx}+\epsilon_{yy}$ and $\left|\mathcal{E}\right|^{2}=\epsilon_{xx}^{2}+\epsilon_{yy}^{2}+2\epsilon_{xy}^{2}$ are the trace and modulus of the strain tensor. Taking into account \textit{all} the strain configurations that produce square patterns, we find that the minimum elastic energy corresponds to the particular shear strain solution ($\mathrm{tr}\mathcal{E}=0$), independently of of the Lamé coefficients. A shear strain is expected to minimize the elastic energy because it preserves the unit cell area and thus minimize any energy cost associated with expanding or contracting the lattices. In terms of the moiré length, the shear strain case corresponds to the particular parameter $\epsilon$ at which the function $f\left(\epsilon\right)$ in Eq.~\eqref{eq:Lm} takes its maximum value ($f\sim0.966$, see Fig.~\textcolor{red}{S1} in Ref.~\cite{SM}). We note that a recent experiment by Yu et al.~\cite{yu2024twist} has also found stretched moiré profiles consistent with a combination of twist and strain that minimize the elastic energy. In that work, local shear strain naturally emerged as part of the structural reorganization of the sample.

Considering the shear strain solution for the square pattern, we can then estimate the twist angle from the moiré length extracted from the experiment. Since the observed patterns are not perfect squares, the actual moiré length connecting the centers of the elliptical AA stacking regimes is not uniform along the four directions shown in Fig.~\ref{fig:exp}(c). As noted above, this is due to the rather inhomogeneous strain configuration in the system. To obtain a minimal, periodic model of the observed pattern, we consider the average of the moiré lengths in the four directions, which gives $\tilde{L}_{M}=12.1205\,\mathrm{nm}$. For the minimum shear strain configuration, this average moiré length corresponds to a twist angle $\theta\approx1.125^{\circ}$ and shear strain magnitude $\left|\epsilon_{s}\right|\sim0.526\%$ (see Fig.~\textcolor{red}{S1} in Ref.~\cite{SM}). 

Next, to model the electronic properties of the square pattern we made use of an extension of the continuum model of TBG under the presence of strain~\cite{bi_designing_2019, Escudero2024}. The strain modifies the electronic properties in two key ways: On one hand, it changes the moiré vectors which determine the coupling between the two layers~\cite{sinner_strain_2022, Escudero2024}; on the other hand, it introduces strain-induced scalar and gauge fields~\cite{suzuura_phonons_2002,choi_effects_2010, vozmediano_gauge_2010}, which shift the Dirac points both in momentum and energy~\cite{naumis_electronic_2017, bi_designing_2019, Escudero2024}. To better capture the properties of the STS measurements, we also included the Hartree potential that accounts for electrostatic interactions arising from charge inhomogeneities induced by the moiré potential~\cite{guinea_electrostatic_2018,cea_electrostatic_2022,Cea2019,Pantaleon2021}. We find that the best fit is for a filling of about $\nu=1$ electron per moiré unit cell.

Figure~\ref{fig:theo} shows the theoretical results for the square moiré pattern formed by twist and shear strain. The strain effectively leads to the elliptically shaped AA stacking regimes seen in the STM \cite{kerelsky_maximized_2019}, which always point in the direction of one corner of the square unit cell. This behavior is more clearly seen in Panel (b), which shows an atomic-scale colormap of the stacking regimes, calculated as the in-plane distance $d$ of each atom in one layer to the closest atom in the other layer; thus, AA stacking corresponds to one atom directly on top of the other $\left(d=0\right)$, while AB/BA stacking corresponds to an interatomic distance $\left(d=0.142\,\mathrm{nm}\right)$. The AA stackings transition to AB/BA stackings through the domain wall DW (indicated as gold stars), following the elliptical shape seen in the moiré pattern. 

Panels (c)-(e) in Fig.~\ref{fig:theo} show the corresponding continuum model results, with parameters $\hbar v/a=2.135\,\mathrm{eV}$, $u_{0}=0.0797\,\mathrm{eV}$, $u_{1}=0.0975\,\mathrm{eV}$ \cite{moon_optical_2014, Koshino2018}, and a filling $\nu=+1$ (one electron per moiré unit cell). The obtained LDOS, along the four directions $A,B,C,D$ shown in panel (b), qualitatively captures the two main peaks around the Fermi energy $E_{F}$ and $\sim E_{F}-30\,\mathrm{meV}$, as seen in the experiments. In all directions, one sees that the LDOS always peaks at the AA stacking regimes, up until one reaches the DW. Thus, the LDOS along the two diagonal directions $C$ and $D$ is different due to the elliptical shape of the AA stackings in the square pattern. The two main peaks at $E_{F}$ and at $\sim E_{F}-30\,\mathrm{meV}$ are further split into two, resembling the multiple peaks observed in Fig.~\ref{fig:exp}(e). We note that this particular splitting of the VHs %Y:abbreviation only
is only pronounced for the considered minimum strain configuration (shear strain); other strain configurations, which still yield square patterns, do not exhibit such VHs splitting (see Fig.~\textcolor{red}{S6} in Ref.~\cite{SM}). Therefore, the geometry of the moir\'e pattern can be uniquely determined by the results from both the STM image and STS spectra.  While the theory reproduces the most significant features of the experimental data, there are, however, some quantitative differences between the continuum model LDOS and the STS measurements of Figure~\ref{fig:exp}, see Ref.~\cite{SM} for a detailed discussion.

\textit{Conclusions.}---
We have reported the first clear experimental observation of strain-induced square moiré patterns in stacked graphene layers. The strain in the system was externally controlled by applying lateral forces to graphene wrinkles arising from the high-temperature graphitization process during sample preparation. Through STM measurements, we reveal local regions of almost perfect square patterns, with the charge density concentrated in elliptically shaped AA stacking regimes. STS measurements further reveal the presence of narrow bands with a small two-fold splitting of two prominent VHs around the Fermi energy. We show that these observations align with the square pattern arising from a particular combination of twist and shear strain that minimizes the elastic energy. 

In our previous theoretical work, we showed that the coexistence of a strong Hartree potential, charge localization, and a high density of states favors superconductivity~\cite{Long2024Evolution}. The present experimental results exhibit these same features, with the Fermi level near the van Hove singularities and a strong Hartree contribution, suggesting that this system is a promising platform for realizing anisotropic superconductivity in a square-symmetric moiré lattice.

Overall, our analysis reveals that a combination of both STM and STS measurements provides a clear fingerprint of the twist and strain configuration in the system. Our approach thus opens a path to \emph{twistraintronics} by actively tuning the properties of moiré heterostructures with the interplay of twist and strain.

\vspace{1ex}
\textit{Acknowledgments.}---
We thank Gerardo Naumis and Danna Liu for useful discussions. IMDEA Nanociencia acknowledges support from the \textquotedblleft Severo Ochoa\textquotedblright ~Programme for Centres of Excellence in R\&D (Grant No. SEV-2016-0686). R.C. acknowledges funding from MICIU/EU/AEI through the FPI grant
PRE2021-098139. F.E. acknowledges support funding from the European Union's Horizon 2020 research and innovation programme under the Marie Skłodowska-Curie grant agreement No 101210351. Z.Z. acknowledges support funding from the European Union's Horizon 2020 research and innovation programme under the Marie Skłodowska-Curie grant agreement No 101034431 and from the ``Severo Ochoa" Programme for Centres of Excellence in R\&D (CEX2020-001039-S / AEI / 10.13039/501100011033). E.C-R acknowledges funding from the Alexander von Humboldt Foundation via the Henriette Herz program. B.V-B acknowledges funding from the Spanish Ministerio de Universidades through the PhD scholarship No. FPU22/03675. P.A.P acknowledges funding by Grant No.\ JSF-24-05-0002 of the Julian Schwinger Foundation for Physics Research. P.A.P. acknowledges the hospitality of the Department of Physics at Colorado State University, where part of this work was developed. We acknowledge financial support from the Spanish Ministry of Science and Innovation, through project PID2023-149106NB-I00, the Mar\'ia de Maeztu Program for Units of Excellence in R\&D (grant no. CEX2023–001316-M), the Comunidad de Madrid and the Spanish State through the Recovery, Transformation and Resilience Plan [Materiales Disruptivos Bidimensionales (2D), (MAD2DCM)-UAM Materiales Avanzados], the European Union through the Next Generation EU funds and the BBVA Foundation under the Fundamentos Research Program 2024 ([Artificial Quantum Matter: From 2D Materials to Spin Lattice Systems]).

\let\oldaddcontentsline\addcontentsline
\renewcommand{\addcontentsline}[3]{}

%\bibliography{References.bib} 
%\bibliography{References.bib}


\let\addcontentsline\oldaddcontentsline

\clearpage
\onecolumngrid

\begin{center}
{\Large\emph{Supplemental Materials for}:\\
Twistraintronics in Square Moir\'{e} Superlattices of Stacked Graphene Layers}{\Large :}{\Large\par}
\par\end{center}

\begin{center}
Roberto Carrasco, Federico Escudero, Zhen Zhan, Eva Cort\'es-del Río, Beatriz Viña-Baus\'a, Yulia Maximenko, Pierre A. Pantale\'on, Francisco Guinea, and Iv\'an Brihuega
\par\end{center}

\setcounter{equation}{0}
\setcounter{figure}{0}
\setcounter{table}{0}
\setcounter{page}{1}
\makeatletter
\renewcommand{\theequation}{S\arabic{equation}}
\renewcommand{\thefigure}{S\arabic{figure}}
\setcounter{secnumdepth}{3}

\tableofcontents
\let\oldaddcontentsline\addcontentsline

\section{Experimental methods}

All sample preparation and experimental procedures were performed under ultra-high vacuum (UHV) conditions. The samples were grown by thermal decomposition of the carbon-face SiC at temperatures close to 1150 ºC in ultrahigh vacuum \cite{Varchon_GrapheneGrowth_PRB2008, Brihuega_VHs_PRL2012}. Unlike graphene grown on the Si-face of SiC, which typically forms monolayer or bilayer graphene with well-defined stacking \cite{Faugeras2008, Norimatsu2014, Yazdi2016, Mishra2016}, graphene on the C-face exhibits significant rotational disorder \cite{Brihuega_VHs_PRL2012, Beechem2014}. It is well known that large twist angles electronically decouple the $\pi$ bands of stacked graphene layers, resulting in weak interlayer coupling and effectively isolating the topmost layers from the SiC substrate \cite{Sprinkle_PRL2009}. As a result, the surface graphene layer remains essentially undoped, with the Dirac energy matching the Fermi energy. Rotational disorder in the sample also leads to the natural formation of TBG domains with a broad range of twist angles, including the magic angle.

% To study the effect of strain on moiré superlattices, targeted STM manipulation of wrinkles was performed. By applying lateral forces to graphene wrinkles adjacent to magic-angle moiré patterns, we induced controlled deformations in real space, which translated into modifications of the reciprocal lattice. This approach allowed us to actively tune the properties of the moiré superlattice, providing a unique platform to investigate strain effects on correlated electronic states in twisted bilayer graphene.

The STM and STS experiments were performed in a UHV system using a homemade low-temperature STM, at base temperatures of $T_{sample}=4$ K, and $T_{tip}=3$ K. In STM, areas corresponding to local AA stacking appear brighter due to topographic corrugation as well as a higher local density of states (LDOS).  The periodicity of these bright regions provides a method to quantify the twist angle and strain magnitude of the moiré superlattice \cite{Rosenberger2020,Halbertal2022,jong_imaging_2022}. Differential conductance $dI/dV$ curves were obtained by numerical differentiation of measured I-V curves. All STM/STS data were acquired and processed using the WSxM software \cite{Horcas2007}.

% STM images reveal single-crystal domains with lateral size of 100|500 nm.

%During the entire process|from the annealing of the sample, to the STM imaging of pristine graphene, and the manipulation of wrinkles|the sample never left the UHV environment.

\section{Strain-induced square moiré patterns}\label{sec:Theory_Square}

As noted above, in the experimental setup the topmost layers are practically isolated from the layers beneath. Therefore, to model the observed moiré patterns we consider two graphene monolayers with lattice vectors $\mathbf{a}_{1} =a\left(1,0\right)$ and $\mathbf{a}_{2}=a\left(1/2,\sqrt{3}/2\right)$ (where $a\simeq2.46\:\textrm{Å}$), stacked in a bilayer configuration. A relative twist between the layers leads to the emergence of trigonal moiré patterns (twisted bilayer graphene) \cite{Andrei2020}. The additional presence of strain in the system distorts such patterns and can, in particular, lead to the square moiré patterns \cite{kogl_moire_2023, Escudero2024}. In what follows we focus on this particular situation.

The formation of square moiré patterns can be accounted by identifying those combinations of twist and strain that result in perpendicular, equal-length moiré vectors. As noted in Ref. \cite{Escudero2024}, under strain the shortest (primitive) set of moiré vectors $\mathbf{g}_{i}$ are not necessarily given by the usual difference $\mathbf{g}_{i}=\mathbf{b}_{i}^{-}-\mathbf{b}_{i}^{+}$ between the twisted and strained reciprocal vectors $\mathbf{b}_{i}^{\pm}$ in each layer. A full account of all possible moiré geometries should take into consideration the correct moiré vectors construction that yields the primitive moiré vectors. We shall for the moment postpone this analysis, and continue with the particular construction $\mathbf{g}_{i}=\mathbf{b}_{i}^{-}-\mathbf{b}_{i}^{+}$. Later in Section \ref{sec:fullSQ} we generalize, through a simple symmetry argument, the obtained results to fully account all the possible solutions given by the different constructions of the moiré vectors.

Following Ref. \cite{Escudero2024}, the perpendicular, equal-length moiré vector conditions can be concisely stated in terms of a unique symmetric transformation
$\mathbf{F}=\mathbf{T}^{\mathrm{T}}\mathbf{T}$ acting on the reciprocal vector $\mathbf{b}_{i}$ of a honeycomb lattice. Here
\begin{equation}
\mathbf{T}=\left(\mathbb{I}-\mathcal{E}_{-}\right)\mathrm{R}\left(\theta_{-}\right)-\left(\mathbb{I}-\mathcal{E}_{+}\right)\mathrm{R}\left(\theta_{+}\right)
\end{equation}
is the transformation that determines the moiré vectors $\mathbf{g}_{i}$ from the reciprocal vectors, $\mathbf{g}_{i}=\mathbf{T}\mathbf{b}_{i}$, due to the combination of twist and strain (see Refs. \cite{sinner_strain_2022, Escudero2024}). In terms of the $\mathbf{F}$ matrix, the equal-length and perpendicular moiré vectors conditions can be generally stated as
\begin{align}
\left(\mathbf{F}\mathbf{b}_{1}\right)\cdot\mathbf{b}_{1} & =\left(\mathbf{F}\mathbf{b}_{2}\right)\cdot\mathbf{b}_{2},\\
\left(\mathbf{F}\mathbf{b}_{1}\right)\cdot\mathbf{b}_{2} & =0.
\end{align}
Given, for instance, honeycomb reciprocal vectors $\mathbf{b}_{1}=b\left(\frac{\sqrt{3}}{2},-\frac{1}{2}\right)$, $\mathbf{b}_{2}=b\left(0,1\right)$, the conditions above are satisfied if
\begin{equation}
\mathbf{F}=F\left(\begin{array}{cc}
\frac{5}{\sqrt{3}} & 1\\
1 & \sqrt{3}
\end{array}\right),\label{eq:Fsquare}
\end{equation}
for any scalar $F$. By relating this $\mathbf{F}$ to the transformation $\mathbf{T}$ one can then determine \emph{all} the twist and strain parameters that result in square moiré patterns.

Let's first consider the practical case of equal but opposite strain in each layer: $\mathcal{E}_{+}=-\mathcal{E}_{-}=\mathcal{E}/2$. If each layer is rotated by $\pm\theta/2$, the transformation $\mathbf{T}$ then reads
\begin{equation}
\mathbf{T}=\left(\mathbb{I}+\mathcal{E}/2\right)\mathrm{R}\left(-\theta/2\right)-\left(\mathbb{I}-\mathcal{E}/2\right)\mathrm{R}\left(\theta/2\right).\label{eq:T}
\end{equation}
For nonzero twist $\theta$ it is convenient to rewrite the strain magnitudes as
\begin{equation}
\epsilon_{ij}=E_{ij}\tan\left(\theta/2\right),
\end{equation}
 so that in general
\begin{equation}
\mathbf{F}=\sin^{2}\left(\theta/2\right)\left[\begin{array}{cc}
E_{xx}^{2}+\left(E_{xy}-2\right)^{2} & E_{xx}\left(E_{xy}+2\right)+E_{yy}\left(E_{xy}-2\right)\\
E_{xx}\left(E_{xy}+2\right)+E_{yy}\left(E_{xy}-2\right) & E_{yy}^{2}+\left(E_{xy}+2\right)^{2}\label{eq:Fsquare1}
\end{array}\right].
\end{equation}
The twist angle thus factor out of $\mathbf{F}$. Comparing then Eq. \eqref{eq:Fsquare} with Eq. \eqref{eq:Fsquare1} we find, in terms of $\theta$ and $\epsilon\equiv E_{xx}$, the lowest strain solutions
\begin{align}
\epsilon_{xx} & =\epsilon\tan\left(\theta/2\right),\label{eq:exx}\\
\epsilon_{yy} & =\frac{3\left[\left(2+\sqrt{3}\right)\epsilon-4\right]}{6+5\sqrt{3}}\tan\left(\theta/2\right),\label{eq:eyy}\\
\epsilon_{xy} & =\frac{12-10\sqrt{3}+3\epsilon}{6+5\sqrt{3}}\tan\left(\theta/2\right).\label{eq:exy}
\end{align}
The above solutions give\emph{ all }the smallest strain tensors that yield square moiré vectors. For a fixed twist angle, the strain magnitudes $\epsilon_{ij}$ are solely determined by the value of $\epsilon$.

The corresponding moiré vectors $\mathbf{g}_{i}=\mathbf{T}\mathbf{b}_{i}$ for the square solutions read
\begin{align}
\mathbf{g}_{1} & =-\frac{3b\sin\left(\theta/2\right)}{6+5\sqrt{3}}\left(4-2\epsilon+\sqrt{3}\epsilon,\epsilon+8\right),\label{eq:g1}\\
\mathbf{g}_{2} & =\frac{3b\sin\left(\theta/2\right)}{6+5\sqrt{3}}\left(\epsilon+8,2\epsilon+\sqrt{3}\epsilon-4\right),\label{eq:g2}
\end{align}
where $b=4\pi/\sqrt{3}a$ is the length of the honeycomb reciprocal vectors $\mathbf{b}_{i}$. The real space moiré vectors $\mathbf{g}_{i}^{R}$, which satisfy $\mathbf{g}_{i}^{R}\cdot\mathbf{g}_{i}=2\pi\delta_{ij}$, follow as
\begin{align}
\mathbf{g}_{1}^{R} & =-\frac{a}{2\sin\left(\theta/2\right)}f\left(\epsilon\right)\frac{\left(4-2\epsilon+\sqrt{3}\epsilon,\epsilon+8\right)}{2\sqrt{\left(\sqrt{3}+2\right)\epsilon^{2}-2\sqrt{3}\epsilon+20}},\\
\mathbf{g}_{2}^{R} & =\frac{a}{2\sin\left(\theta/2\right)}f\left(\epsilon\right)\frac{\left(\epsilon+8,2\epsilon+\sqrt{3}\epsilon-4\right)}{2\sqrt{\left(\sqrt{3}+2\right)\epsilon^{2}-2\sqrt{3}\epsilon+20}},
\end{align}
where
\begin{align}
f\left(\epsilon\right) & =\frac{\sqrt{3}+5/2}{\sqrt{\left(2+\sqrt{3}\right)\epsilon^{2}-2\sqrt{3}\epsilon+20}}.\label{eq:fe}
\end{align}
Since
\begin{equation}
\left|\frac{\left(\epsilon+8,2\epsilon+\sqrt{3}\epsilon-4\right)}{2\sqrt{\left(\sqrt{3}+2\right)\epsilon^{2}-2\sqrt{3}\epsilon+20}}\right|=1,
\end{equation}
it readily follows that the moiré length $L_{M}\left(\theta,\epsilon\right)=\left|\mathbf{g}_{i}^{R}\right|$ is given by Eq. \eqref{eq:Lm} in the main text:
\begin{align}
L_{M}\left(\theta,\epsilon\right) & =\frac{a}{2\sin\left(\theta/2\right)}f\left(\epsilon\right)\nonumber \\
 & =\frac{a}{2\sin\left(\theta/2\right)}\frac{\sqrt{3}+5/2}{\sqrt{\left(2+\sqrt{3}\right)\epsilon^{2}-2\sqrt{3}\epsilon+20}}.\label{eq:Lmoire}
\end{align}

\begin{figure}[t]
	\includegraphics[scale=0.2]{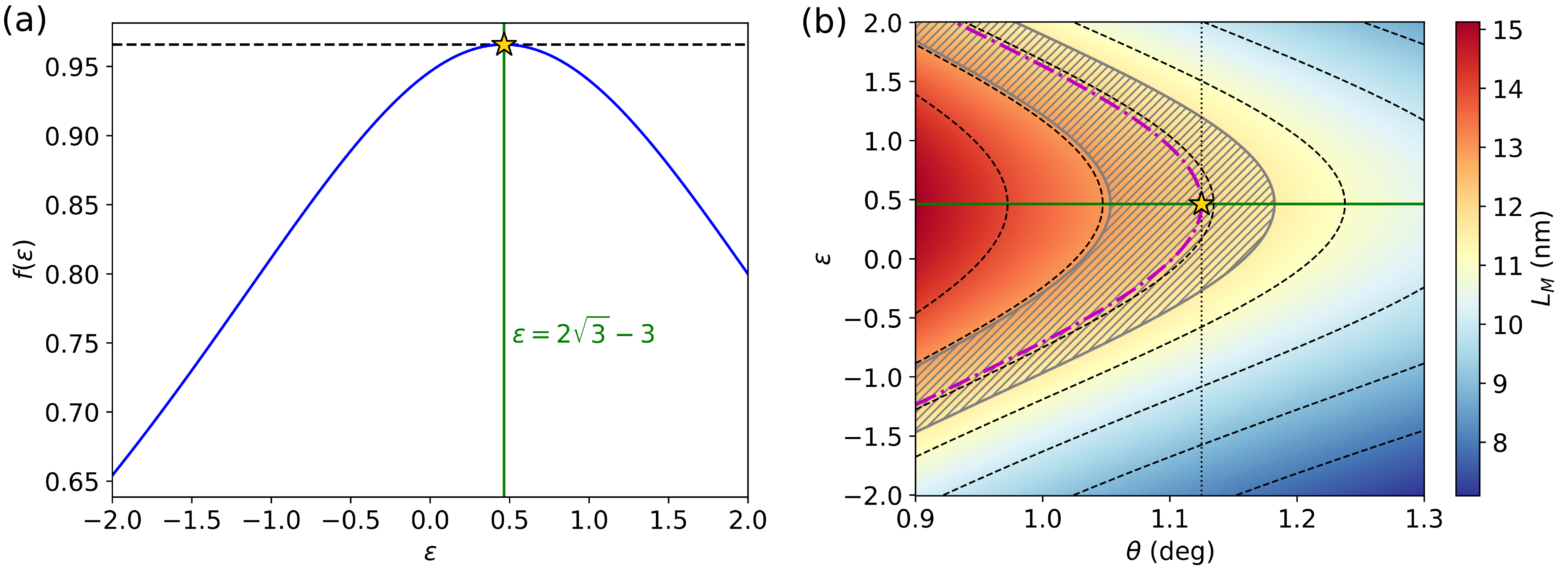}	
	\caption{(a) Plot of the function $f\left(\epsilon\right)$ given by Eq. \eqref{eq:fe}, which determines the moiré length $L_{M}$ through Eq. \eqref{eq:Lmoire}. The green vertical line at $\epsilon=2\sqrt{3}-3$ corresponds to the minimum (traceless) shear strain case, at which $f\simeq0.966$ takes its maximum value. (b) Evolution of the moiré length $L_{M}\left(\theta,\epsilon\right)$ as a function of the twist angle $\theta$ and $\epsilon$. The contour black dashed curves correspond to constant moiré lengths $L_{M}$ from 8 to 14 nm. The dashed gray region represent the range of moiré lengths extracted from the experiment results shown in Figure \ref{fig:exp}(c). The magenta dot-dashed curve corresponds to the average moiré length $\tilde{L}_{M}=12.1205\,\mathrm{nm}$. The horizontal green line corresponds to the value $\epsilon=2\sqrt{3}-3\simeq0.42641$ of minimum shear strain configuration, and the gold star indicates its intercept with the average moiré length $\tilde{L}_{M}$ at a twist angle $\theta\simeq1.125^{\circ}$.}\label{fig:moire_length}

\end{figure}

Figure \ref{fig:moire_length}(a) shows the variation of $f\left(\epsilon\right)$ as a function of $\epsilon$. Since $f\left(\epsilon\right)<1$ always, the moiré length of the square patterns generally correspond to larger twist angle than those without strain. Figure \ref{fig:moire_length}(b) shows the evolution of the moiré length $L_{M}\left(\theta,\epsilon\right)$ in the twist/strain plane of the square solutions. For a fixed twist angle there are two strain configurations that yield the same moiré length. Conversely, for a given strain parameter $\epsilon$ there is only one twist angle that yields a particular moiré length $L_{M}$. In general there is a continuous family of twist and strain configurations that yield the same moiré length (black dashed-lines in Figure \ref{fig:moire_length}), corresponding to all the possible square patterns that differ only by an overall rotation.

Eqs. \eqref{eq:exx}-\eqref{eq:exy} only hold under the assumption of equal but opposite strain in each layer. Another relevant case is when a net strain $\mathcal{E}'$ only acts in one layer, say the bottom layer. Assuming that each layer is still rotated by $\pm\theta/2$, the transformation in that situation becomes
\begin{equation}
\mathbf{T}'=\left(\mathbb{I}-\mathcal{E}'\right)\mathrm{R}\left(-\theta/2\right)-\mathrm{R}\left(\theta/2\right).
\end{equation}
The strain tensor $\mathcal{E}'$ that yields square moiré vector can be obtained by equating the above transformation with Eq. \eqref{eq:T}, as then both transformations would yield equal moiré vectors. The condition $\mathbf{T}'=\mathbf{T}$ implies
\begin{align}
\mathcal{E}' & =-\frac{\mathcal{E}}{2}\left[\mathrm{R}\left(-\theta/2\right)+\mathrm{R}\left(\theta/2\right)\right]\mathrm{R}\left(\theta/2\right)\nonumber \\
 & =-\cos\left(\theta/2\right)\mathcal{E}\mathrm{R}\left(\theta/2\right),
\end{align}
where in the last step we used that $\mathrm{R}\left(-\theta/2\right)+\mathrm{R}\left(\theta/2\right)=2\cos\left(\theta/2\right)\mathbb{I}$. By replacing the strain tensor components of $\mathcal{E}$ given by Eqs. \eqref{eq:exx}-\eqref{eq:exy}, from the above equation one can readily obtain the strain tensor $\mathcal{E}'$ that gives square moiré vectors when acting only on the bottom layer. At low twist angles it is easy to see that
\begin{equation}
\mathcal{E}'\sim-\mathcal{E}.
\end{equation}
This just means that a strain equal but opposite in each layer yields practically the same moiré geometry as the same net strain applied only on the bottom layer. Consequently, the cases of strain in both layers, or only in one layer, give almost identical geometrical and electronic properties (see Figure \ref{fig:Comparison}  below). Our results in the main text are thus readily generalized to configurations in which the considered strain rather acts only on one layer. 

\begin{figure}[t]
	\includegraphics[width=\linewidth]{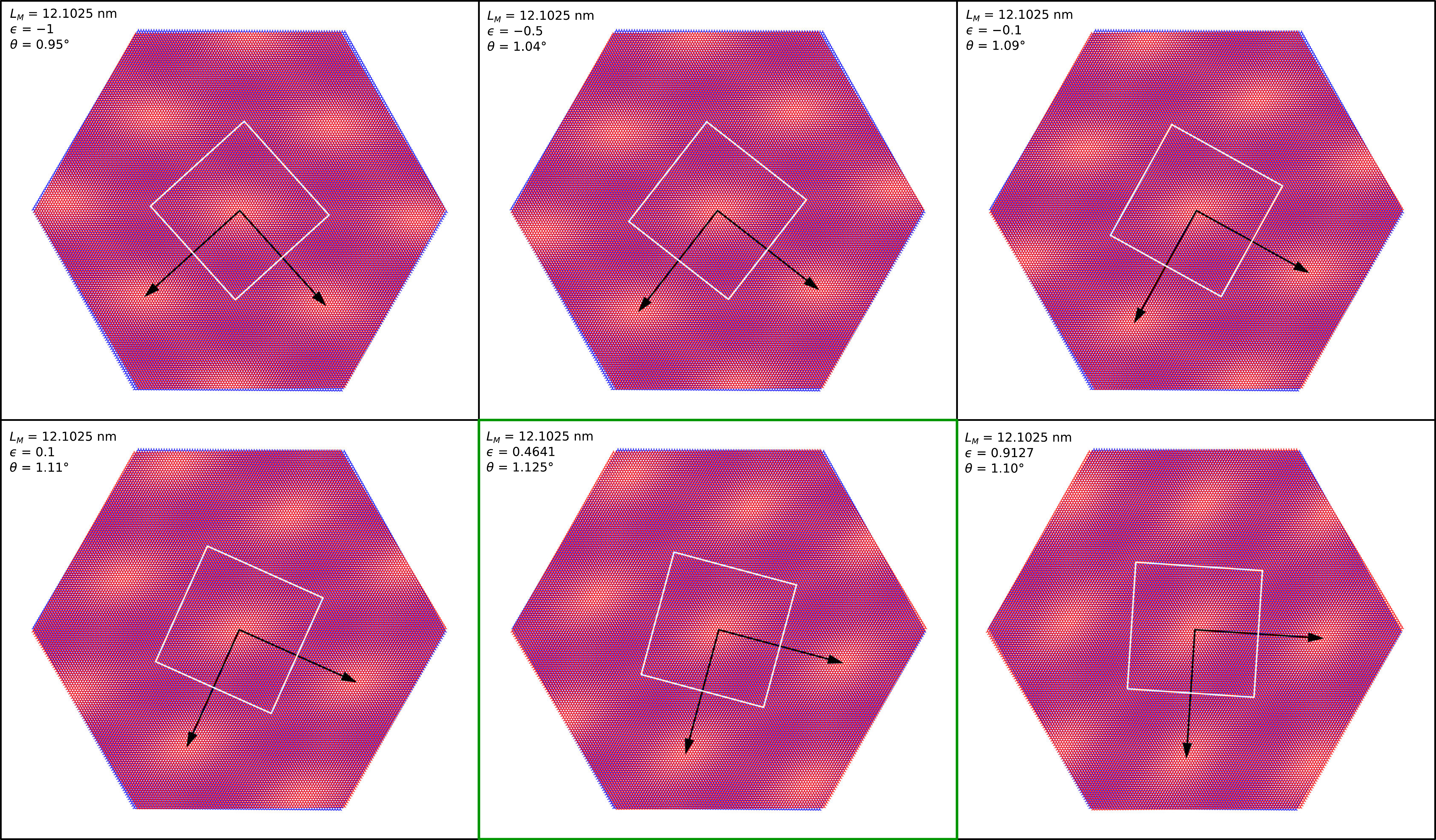}	
	\caption{Square patterns for the same moiré length $\tilde{L}_{M}=12.1025\,\mathrm{nm}$, and different parameters $\epsilon=-1,-0.5,-0.1,0.1,2\sqrt{3}-3,0.9127$. The case $\epsilon=2\sqrt{3}-3\approx0.42641$ of minimum strain (traceless, shear strain) is highlighted in a green box. The last case $\epsilon=0.9127$ corresponds to the uniaxial heterostrain. Note that in order to preserve the same moiré length, for each parameter $\epsilon$ there is a different twist angle $\theta$; cf. Eq. \eqref{eq:Lmoire}. As $L_{M}$ is the same in all cases, up to an overall rotation the moiré patterns look practically the same at the moiré scale. However, they have markedly different electronic properties, cf. Figure \ref{fig:DOS_all}.}\label{fig:MP_square}
\end{figure}

\subsection{Uniaxial heterostrain}\label{subsec:uniaxial}

In the particular case of uniaxial heterostrain, with magnitude $\epsilon_{u}$ along a direction $\phi$, one has
\begin{equation}
\mathcal{E}=\epsilon_{u}\left[\begin{array}{cc}
\cos^{2}\phi-\nu\sin^{2}\phi & \left(1+\nu\right)\sin\phi\cos\phi\\
\left(1+\nu\right)\sin\phi\cos\phi & \sin^{2}\phi-\nu\cos^{2}\phi
\end{array}\right],
\end{equation}
where $\nu=0.16$ is the Poisson's ratio in graphene. The corresponding square solution for equal length moiré vectors were obtained in Ref. \cite{Escudero2024}, and read 
\begin{align}
\epsilon_{u} & =\frac{4\left(2-\sqrt{3}\right)}{\left(1+\nu\right)\sqrt{\frac{\left(4\sqrt{3}-7\right)\left(1-\nu\right)^{2}}{\left(1+\nu\right)^{2}}+1}}\tan\left(\theta/2\right),\label{eq:eu_square}\\
\phi & =\frac{\pi}{6}-\frac{1}{2}\cos^{-1}\left[\frac{\left(2-\sqrt{3}\right)\left(1-\nu\right)}{1+\nu}\right].
\end{align}
Since $\epsilon_{xx}+\epsilon_{yy}=\epsilon_{u}\left(1-\nu\right)$,
from Eq. \eqref{eq:eu_square} and Eqs. \eqref{eq:exx}-\eqref{eq:eyy} we readily get
\begin{align}
\epsilon & =\frac{1}{4+\sqrt{3}}\left[\frac{13\left(2-\sqrt{3}\right)\left(1-\nu\right)}{\sqrt{\left(1+\nu\right)^{2}+\left(4\sqrt{3}-7\right)\left(1-\nu\right)^{2}}}+5\sqrt{3}-6\right]\approx0.9127.
\end{align}

\subsection{Shear heterostrain}\label{sec:shear}

The case of shear heterostrain, although similar to that of uniaxial strain, is not the same. The shear strain tensor, with magnitude $\epsilon_{s}$ perpendicular to a direction $\varphi$, is given by
\begin{equation}
\mathcal{E}=\epsilon_{s}\left(\begin{array}{cc}
-\sin2\varphi & \cos2\varphi\\
\cos2\varphi & \sin2\varphi
\end{array}\right).
\end{equation}
A particular set of square-patterns solutions can be found in the same exact same way as for uniaxial heterostrain \cite{Escudero2024}, with the simple result:
\begin{align}
\epsilon_{s} & =-2\left(2-\sqrt{3}\right)\tan\left(\theta/2\right),\label{eq:es_shear}\\
\varphi & =\frac{\pi}{6}.\label{eq:phi_shear}
\end{align}
Comparing with Eq. \eqref{eq:exx} we then have
\begin{equation}
\epsilon=\sqrt{3}\left(2-\sqrt{3}\right)\approx0.4641.
\end{equation}

Figure \ref{fig:MP_square} shows some examples of square moiré patterns for the same moiré length $\tilde{L}_{M}=12.1025\,\mathrm{nm}$,
and different parameters $\epsilon$. Although all cases seem to give practically the same square pattern (up to an overall rotation), their electronic properties are markedly different; cf. Figure \ref{fig:DOS_all}.

\subsection{Full family of square pattern solutions}\label{sec:fullSQ}

As discussed at the beginning of Section \ref{sec:Theory_Square}, our results so far only give the subset of square patterns associated with equal-length perpendicular moiré vectors $\mathbf{g}_{i}=\mathbf{b}_{i}^{-}-\mathbf{b}_{i}^{+}$. There are yet other strain configurations that also yield square patterns, but in which one of the shortest (primitive) moiré vectors is, e.g., $\mathbf{g}_{3}=\mathbf{g}_{1}-\mathbf{g}_{2}$, i.e., constructed in terms of the difference between the strained reciprocal vectors $\mathbf{b}_{3}^{\pm}=\mathbf{b}_{1}^{\pm}-\mathbf{b}_{2}^{\pm}$. The complete account of these different constructions can be simplified by noting that they are related by the underlying $C_{6}$ symmetry of the honeycomb lattices, so that any rotation of the system by $60^{\circ}$ leads to the same moiré geometry (up to an overall rotation) \cite{Escudero2024}. Consequently, we can readily generalize the strain tensor that yields all possible square pattern by simply considering all $60^{\circ}$ rotations of the strain tensor with components given by Eqs. \eqref{eq:exx}-\eqref{eq:exy}. Thus, for any nonzero twist angle $\theta$, in general
the square patterns result from a strain tensor
\begin{equation}
\mathcal{E}\left(\theta,\epsilon,n\right)=\mathrm{R}\left(n\pi/3\right)\mathcal{E}\left(\theta,\epsilon\right)\mathrm{R}\left(-n\pi/3\right),\label{eq:Esq_n}
\end{equation}
where $n$ is an integer and
\begin{equation}
\mathcal{E}\left(\theta,\epsilon\right)=\left(\begin{array}{cc}
\epsilon\left(6+5\sqrt{3}\right) & 12-10\sqrt{3}+3\epsilon\\
12-10\sqrt{3}+3\epsilon & 3\left[\left(2+\sqrt{3}\right)\epsilon-4\right]
\end{array}\right)\frac{\tan\left(\theta/2\right)}{6+5\sqrt{3}}.\label{eq:Esq_general}
\end{equation}
It is important to note that for $n\neq0$ the two moiré vectors would not longer be given by Eqs. \eqref{eq:g1} and \eqref{eq:g2}, (i.e., equal to the construction $\mathbf{g}_{i}=\mathbf{b}_{i}^{-}-\mathbf{b}_{i}^{+}$). Rather, at least one moiré vector would be given by a combination of $\mathbf{g}_{1}$ and $\mathbf{g}_{2}$. As these different constructions are related by simple rotations of the local atomic positions, the electronic properties of the square patterns for different $n$ in Eq. \eqref{eq:Esq_n} is the same.

As an example, let's consider the shear strain case (Section \ref{sec:shear}). The solution given by Eqs. \eqref{eq:es_shear} and \eqref{eq:phi_shear} then generalizes to
\begin{equation}
\mathcal{E}=-2\left(2-\sqrt{3}\right)\left(\begin{array}{cc}
-\sin2\varphi_{n} & \cos2\varphi_{n}\\
\cos2\varphi_{n} & \sin2\varphi_{n}
\end{array}\right)\tan\left(\theta/2\right),
\end{equation}
where now 
\begin{equation}
\varphi_{n}=\frac{\pi}{6}+\frac{n\pi}{3}.
\end{equation}
Thus, there is a family of shear strain, with equal magnitude and different direction, that give square patterns. 

\begin{figure}[t]
	\includegraphics[width=\linewidth]{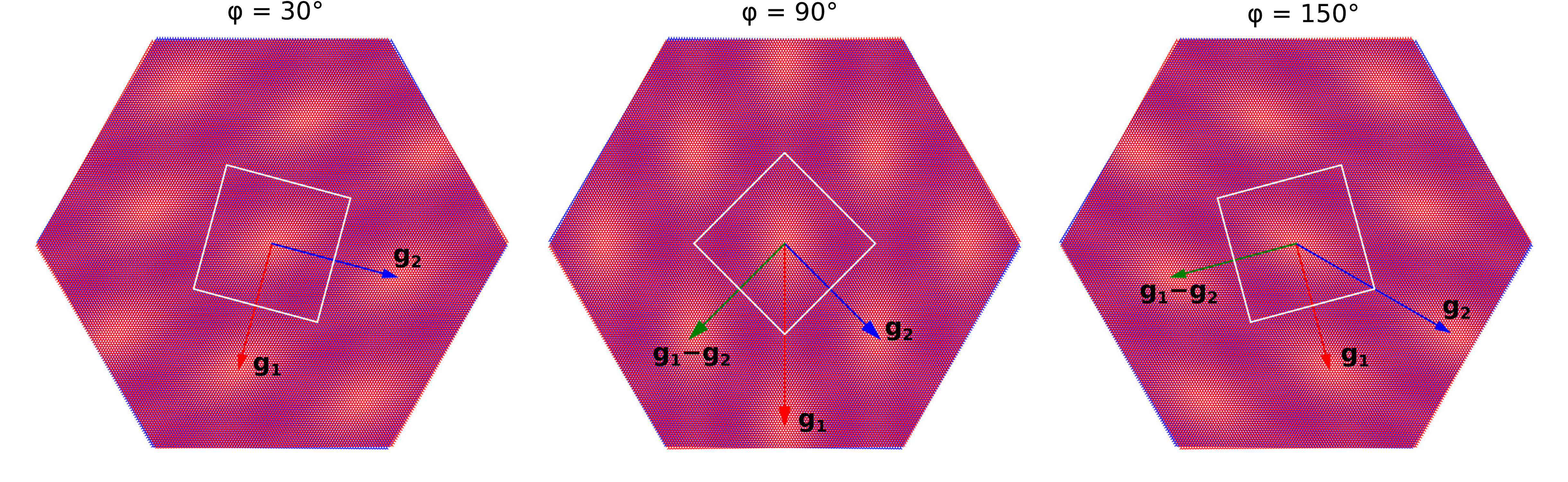}	
	\caption{Square patterns of moiré length $\tilde{L}_{M}=12.1205\,\mathrm{nm}$, arising from twist $\theta\approx1.125^{\circ}$ and shear strain with magnitude $\left|\epsilon_{s}\right|\sim0.526\%$ and directions $\varphi=30^{\circ},90^{\circ},150^{\circ}$. In the first case $\varphi=30^{\circ}$, the primitive moiré vectors are $\mathbf{g}_{1}$ and $\mathbf{g}_{2}$ (red and blue arrows), obtained from the usual difference $\mathbf{g}_{i}=\mathbf{b}_{i}^{-}-\mathbf{b}_{i}^{+}$ between the twisted and strained reciprocal vectors $\mathbf{b}_{i}^{\pm}$ in each layer (see Section \ref{sec:Theory_Square}). However, in the other cases $\varphi=90^{\circ},150^{\circ}$, one of the primitive moiré vectors is $\mathbf{g}_{3}=\mathbf{g}_{1}-\mathbf{g}_{2}$ (green arrow), which is obtained by taking the difference between the reciprocal vectors $\left(\mathbf{b}_{1}^{\pm}-\mathbf{b}_{2}^{\pm}\right)$. The other equal-length, perpendicular moiré vector is $\mathbf{g}_{2}$ for $\varphi=90^{\circ}$, and $\mathbf{g}_{1}$ for $\varphi=150^{\circ}$.}\label{fig:shear_rotated}
\end{figure}

Figure \ref{fig:shear_rotated} shows three examples of square moiré patterns arising from shear strain with magnitude $\epsilon_{s}=-2\left(2-\sqrt{3}\right)\tan\left(\theta/2\right)$, and three different orientations $\varphi=30^{\circ},90^{\circ},150^{\circ}$. Note that in the last two cases one of the primitive moiré vectors is indeed no longer obtained from the difference $\mathbf{b}_{i}^{-}-\mathbf{b}_{i}^{+}$, but rather a combination of those.

\subsection{Minimum elastic energy}\label{sec:traceless}

As noted in the main text, despite there being a continuous family of strains that produce square patterns, there is only a particular configuration that minimizes the elastic energy [cf. Eq. \eqref{eq:Elastic}]
\begin{equation}
E_{\mathrm{elastic}}=\frac{\lambda}{2}\left(\epsilon_{xx}+\epsilon_{yy}\right)^{2}+\mu\left(\epsilon_{xx}^{2}+\epsilon_{yy}^{2}+2\epsilon_{xy}^{2}\right).
\end{equation}
Note that both the trace $\mathrm{tr}\mathcal{E}=\epsilon_{xx}+\epsilon_{yy}$ and the modulus $\left|\mathcal{E}\right|^{2}=\epsilon_{xx}^{2}+\epsilon_{yy}^{2}+2\epsilon_{xy}^{2}$ do not depend on the symmetry angle $n\pi/3$ of the general solutions given by Eq. \eqref{eq:Esq_general}. Thus,
replacing the strain tensors given by Eqs. \eqref{eq:exx}-\eqref{eq:exy} gives, for the square
patterns, 
\begin{align}
E_{\mathrm{elastic}} & =\frac{24\tan^{2}\left(\theta/2\right)}{\left(6+5\sqrt{3}\right)^{2}}\left\{ \epsilon^{2}\left[2\left(4\sqrt{3}+7\right)\lambda+4\left(\sqrt{3}+2\right)\mu\right]\right.\nonumber \\
 & \qquad\left.-\epsilon\left[8\sqrt{3}\left(\mu+\lambda\right)+12\lambda\right]+\left(43-20\sqrt{3}\right)\mu+6\lambda\right\} .
\end{align}
The minimum of the elastic energy depends solely on the parameter $\epsilon$. We have
\begin{align}
\frac{dE_{\mathrm{elastic}}}{d\epsilon} & =\frac{24\tan^{2}\left(\theta/2\right)}{\left(6+5\sqrt{3}\right)^{2}}\left\{ 2\epsilon\left[2\left(4\sqrt{3}+7\right)\lambda+4\left(\sqrt{3}+2\right)\mu\right]-\left[8\sqrt{3}\left(\mu+\lambda\right)+12\lambda\right]\right\} ,\\
\frac{d^{2}E_{\mathrm{elastic}}}{d\epsilon^{2}} & =\frac{24\tan^{2}\left(\theta/2\right)}{\left(6+5\sqrt{3}\right)^{2}}\left\{ 2\left[2\left(4\sqrt{3}+7\right)\lambda+4\left(\sqrt{3}+2\right)\mu\right]\right\} .
\end{align}
Since $d^{2}E_{\mathrm{elastic}}/d\epsilon^{2}>0$ always, the minimum condition occurs when $dE_{\mathrm{elastic}}/d\epsilon=0$, which implies the shear strain solution [cf. Section \ref{sec:shear}]
\begin{equation}
\epsilon=\sqrt{3}\left(2-\sqrt{3}\right)\approx0.4641,
\end{equation}
independently of the Lamé coefficients. This result is expected because shear strains minimize the elastic energy by deforming the lattices without changing their unit area. Note that since a shear strain corresponds to a traceless tensor, the minimum elastic energy coincides with the minimum of the strain tensor modulus $\left|\mathcal{E}\right|^{2}=\epsilon_{xx}^{2}+\epsilon_{yy}^{2}+2\epsilon_{xy}^{2}$.

As seen in Figure \ref{fig:moire_length}(a), at $\epsilon=2\sqrt{3}-3$ the function $f\left(\epsilon\right)$ given by Eq. \eqref{eq:fe} takes its maximum value 
\begin{equation}
f\left(2\sqrt{3}-3\right)=\frac{\sqrt{2-\sqrt{3}}}{2}\approx0.966,
\end{equation}
for which the moiré length $L_{M}$ is then maximum at a given twist angle, cf. Eq. \eqref{eq:Lmoire}. In other words, shear strains minimize the variation of moiré length with respect to the only twist configuration. 

% \begin{figure}[t]
% 	\includegraphics[scale=0.6]{FigureSM_Emod.png}	
% 	\caption{Modulus strain tensor. Modulus of the strain tensor, $\left|\mathcal{E}\right|^{2}=\epsilon_{xx}^{2}+\epsilon_{yy}^{2}+2\epsilon_{xy}^{2}$, as a function of the twist $\theta$ and the parameter $\epsilon$. The minimum strain takes place at the shear strain case $\epsilon=2\sqrt{3}-3\simeq0.42641$ (horizontal green line). Other lines and curves as in Figure 3 of the main text.}
% \end{figure}

\section{Electronic properties}

\subsection{Effective continuum model}

To model the electronic properties of the square patterns we use the continuum model of TBG \cite{lopes_dos_santos_graphene_2007,  Andrei2020,bistritzer_moire_2011, moon_optical_2014, Koshino2015}, extended to account for the strain in the system \cite{bi_designing_2019, Escudero2024}. We neglect couplings between different valleys in each layer (negligible at low energies), and consider the continuum model Hamiltonian for the $K$ valley (the one for the $K'$ valley being related by time-reversal symmetry). The Hamiltonian takes the form
\begin{equation}
H=\left(\begin{array}{cc}
h_{b}\left(\mathbf{k}\right)+\mathcal{S}_{b} & U^{\dagger}\left(\mathbf{r}\right)\\
U\left(\mathbf{r}\right) & h_{t}\left(\mathbf{k}\right)+\mathcal{S}_{t}
\end{array}\right).
\end{equation}
Here $h_{\ell}\left(\mathbf{k}\right)$ is the Dirac Hamiltonian relative to the twisted and strained Dirac points in each $\ell=b,t$ layer,
\begin{equation}
h_{\ell}\left(\mathbf{k}\right)=-\hbar v\boldsymbol{\sigma}\cdot R_{\ell}\left(-\theta_{\ell}\right)\left(1+\mathcal{E}_{\ell}\right)\left(\mathbf{k}-\mathbf{K}_{\ell}\right),
\end{equation}
where $v$ is the Fermi velocity in monolayer graphene, $\boldsymbol{\sigma}=\left(\sigma_{x},\sigma_{y}\right)$ are the Dirac matrices and $\mathbf{K}_{\ell}=\left(1-\mathcal{E}_{\ell}\right)R_{\ell}\left(\theta_{\ell}\right)\mathbf{K}^{0}$, where $\mathbf{K}_{\zeta}^{0}=-\left(2\mathbf{b}_{1}+\mathbf{b}_{2}\right)/3$ is the $K$-valley Dirac point of a honeycomb layer. The additional term $\mathcal{S}_{\ell}$ takes into account the strain-induced deformation and gauge potentials in each layer \cite{suzuura_phonons_2002, vozmediano_gauge_2010} 
\begin{equation}
\mathcal{S}_{\ell}=\mathbb{I}V_{\ell}-\hbar v\boldsymbol{\sigma}\cdot R_{\ell}\left(-\theta_{\ell}\right)\left(1+\mathcal{E}_{\ell}\right)\mathbf{A}_{\ell},
\end{equation}
where
\begin{align}
V_{\ell} & =g\left(\epsilon_{xx}^{\ell}+\epsilon_{yy}^{\ell}\right),\\
\mathbf{A}_{\ell} & =\frac{\sqrt{3}}{2a}\beta\left(\epsilon_{xx}^{\ell}-\epsilon_{yy}^{\ell},-2\epsilon_{xy}^{\ell}\right),
\end{align}
with $g=4$ eV and $\beta=3.14$ for graphene \cite{choi_effects_2010}. The scalar potential $V_{\ell}$ shifts the Dirac points in energy, resembling the effect of a perpendicular electric field \cite{long_atomistic_2022}. The vector potential $\mathbf{A}_{\ell}$ accounts for the strain-induced change in the hopping energies within the Dirac approximation \cite{naumis_electronic_2017}. 

The moiré-induced coupling potential $U\left(\mathbf{r}\right)$ depends on the interplay between twist and strain through its Fourier expansion in terms of the moiré vectors \cite{Escudero2024}. At small deformations the Fourier expansion can be truncated to the first three leading order terms,
\begin{equation}
U\left(\mathbf{r}\right)=U_{1}+U_{2}e^{i\mathbf{g}_{1}\cdot\mathbf{r}}+U_{3}e^{i\left(\mathbf{g}_{1}+\mathbf{g}_{2}\right)\cdot\mathbf{r}},
\end{equation}
where
\begin{equation}
U_{j}=\left(\begin{array}{cc}
u_{0} & u_{1}e^{-\omega_{j}}\\
u_{1}e^{\omega_{j}} & u_{0}
\end{array}\right),
\end{equation}
with $\omega_{j}=\left(j-1\right)2\pi/3$. Here $u_{0}$ and $u_{1}$ are the effective AA and AB/BA hopping amplitudes. In the main text we use the TBG parameters $\hbar v/a=2.1354\,\mathrm{eV}$, $u_{0}=0.0797\,\mathrm{eV}$ and $u_{1}=0.0975\,\mathrm{eV}$ \cite{moon_optical_2014, Koshino2018}, which already give results in relatively good agreement with the experiments. We note, however, that the electronic properties of the moiré heterostructures are highly sensitivity to small variations in the effective continuum model parameters, especially at the relevant regime of low deformations (small twist and strain) \cite{bi_designing_2019, Escudero2024}. Therefore, we do not conclusively rule out that a better agreement could be achieved by more realistic fits of the hopping energies, e.g., taking into account their possible local variations due to twist angle disorders or strain inhomogeneities. 

\subsection{Hartree potential}

\begin{figure}[t]
	\includegraphics[scale=0.5]{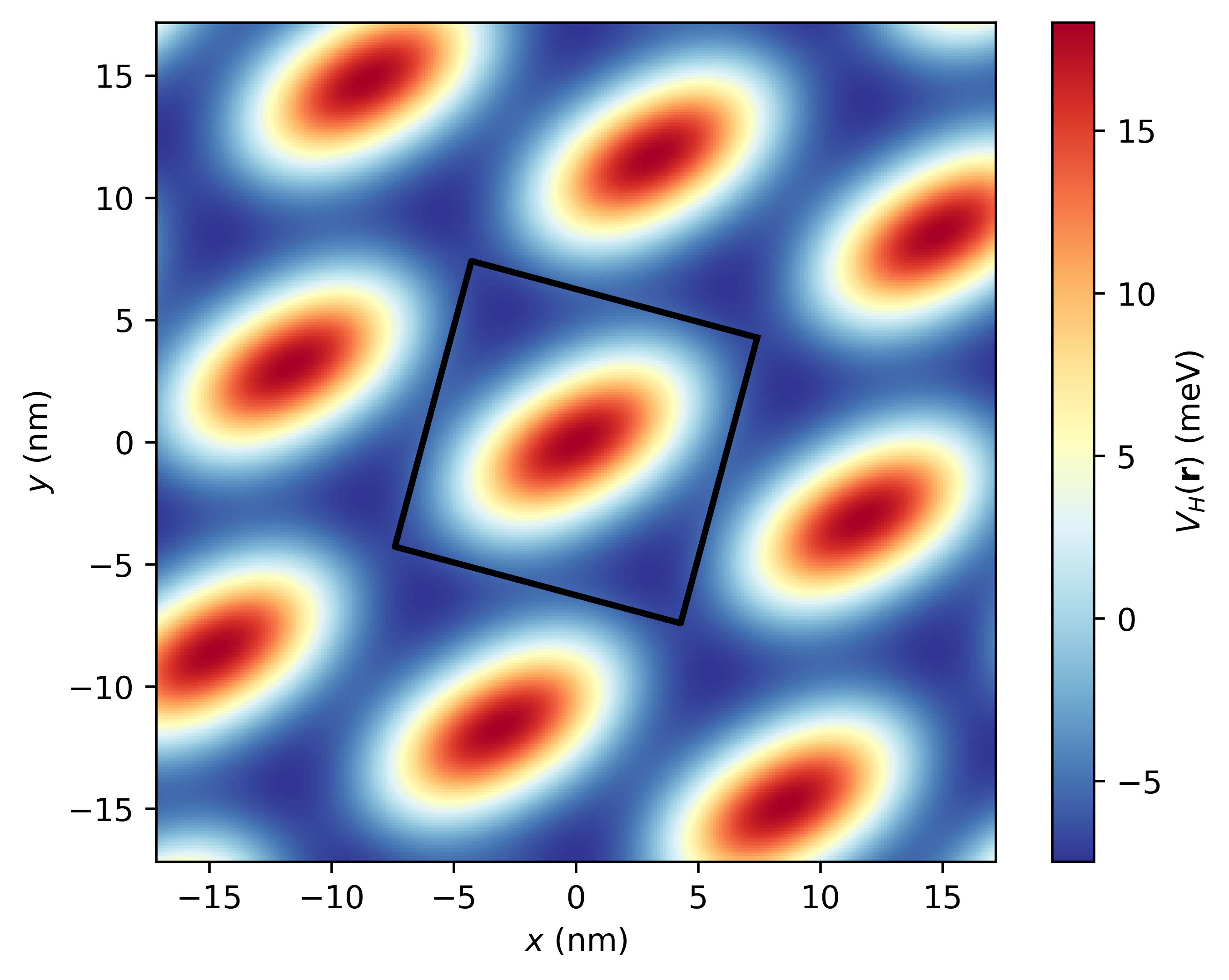}	
	\caption{Real space Hartree potential in the square pattern arising from twist and shear strain (blue green case in Figure \ref{fig:MP_square}), for a filling $\nu=1$ (one electron per unit cell).}\label{fig:Hartree}
\end{figure}

We further take into account the effect of electrostatic interactions through the Hartree potential. In moiré system, the Hartree potential accounts for the effect of charge inhomogeneites induced by the moiré potential \cite{guinea_electrostatic_2018}. The Hartree effect becomes particularly prominent at low twist angles, where the bandwidth of the central bands is minimized, enhancing the electronic interactions. Within a jellium model for a periodic system, the Hartree potential is simply the direct interaction term
\begin{equation}
V_{H}\left(\mathbf{r}\right)=\int d\mathbf{r}'v_{C}\left(\mathbf{r}-\mathbf{r}'\right)\delta\rho\left(\mathbf{r}'\right),
\end{equation}
where $v_{C}\left(\mathbf{r}-\mathbf{r}'\right)$ is the bare Coulomb potential and $\delta\rho\left(\mathbf{r}'\right)$ is the electronic charge density with respect to the charge neutrality (CN) point. As usual, we expand the Bloch states of the system as
\begin{align}
\psi_{n,\mathbf{k},\eta,i}\left(\mathbf{r}\right) & =\frac{1}{\sqrt{A_{c}}}\sum_{\mathbf{g}}u_{n,\mathbf{k},\eta,i}\left(\mathbf{g}\right)e^{i\left(\mathbf{k}+\mathbf{g}\right)\cdot\mathbf{r}},
\end{align}
where $A_{c}$ is the moiré unit cell area, and $n,\eta,i$ are the band, valley/spin and layer/sublattice indices, respectively. $\mathbf{k}$ is a momentum in the moiré Brillouin zone (mBZ), and $\mathbf{g}$ are the reciprocal moiré vectors of the twisted and strained moiré pattern. The Fourier coefficients are normalized as \cite{cea_electrostatic_2022}
\begin{equation}
\sum_{\mathbf{g},i}u_{n,\mathbf{k},\eta,i}^{*}\left(\mathbf{g}\right)u_{m,\mathbf{k},\eta,i}\left(\mathbf{g}\right)=\delta_{n,m},
\end{equation}
so that $\sum_{i}\int_{\mathrm{unit\,cell}}d\mathbf{r}\left|\psi_{n,\mathbf{k},\eta,i}\left(\mathbf{r}\right)\right|^{2}=1$.
The charge density is given by 
\begin{equation}
\delta\rho\left(\mathbf{r}\right)=\sum_{\mathbf{k}}\sum_{n,\eta,i}^{\prime}\left|\psi_{n,\mathbf{k},\eta,i}\left(\mathbf{r}\right)\right|^{2},
\end{equation}
where the primed summation implies taking only occupied (or unoccupied) states from CN. Replacing the Bloch states yields
\begin{align}
\delta\rho\left(\mathbf{r}\right) & =\sum_{\mathbf{g}}\delta\rho\left(\mathbf{g}\right)e^{-i\mathbf{g}\cdot\mathbf{r}},\\
\delta\rho\left(\mathbf{g}\right) & =A_{c}^{-1}\sum_{\mathbf{k},\mathbf{g}'}\sum_{n,\eta,i}^{\prime}u_{n,\mathbf{k},\eta,i}^{*}\left(\mathbf{g}'+\mathbf{g}\right)u_{n,\mathbf{k},\eta,i}\left(\mathbf{g}'\right),
\end{align}
The Hartree potential reads
\begin{align}
V_{H}\left(\mathbf{r}\right) & =\sum_{\mathbf{g}\neq0}V_{H}\left(\mathbf{g}\right)e^{-i\mathbf{g}\cdot\mathbf{r}},\\
V_{H}\left(\mathbf{g}\right) & =\frac{v_{C}\left(\mathbf{g}\right)}{A_{c}}\sum_{\mathbf{k},\mathbf{g}'}\sum_{n,\eta,i}^{\prime}u_{n,\mathbf{k},\eta,i}^{*}\left(\mathbf{g}'+\mathbf{g}\right)u_{n,\mathbf{k},\eta,i}\left(\mathbf{g}'\right),
\end{align}
where $v_{C}\left(\mathbf{g}\right)$ is the Fourier transform of the bare Coulomb potential
\begin{equation}
v_{C}\left(\mathbf{g}\right)=\frac{e^{2}}{2\varepsilon_{0}\varepsilon_{r}}\frac{1}{\left|\mathbf{g}\right|},
\end{equation}
where $\varepsilon_{r}$ is the relative permitivity of the system. For the numerical calculations we consider $\varepsilon_{r}=10$. Note that the $\mathbf{g}=0$ term in $V_{H}\left(\mathbf{r}\right)$ is neglected as it is canceled by the background positive charge (jellium model).

The Hartree potential is diagonal in the valley/spin and sublattice/layer flavors, with matrix elements \cite{cea_electrostatic_2022,Cea2019,Pantaleon2021}
\begin{equation}
\left\langle \mathbf{k}+\mathbf{g}'-\mathbf{g},\eta',i'\left|\hat{V}_{H}\right|\mathbf{k}+\mathbf{g}',\eta,i\right\rangle =\delta_{\eta\eta'}\delta_{ii'}V_{H}\left(\mathbf{g}\right).
\end{equation}
The total continuum model Hamiltonian plus the Hartree interaction is then solved self-consistently. Due to the strain effect, the Fourier coefficients $V_{H}\left(\mathbf{g}\right)$ are generally not equal for the first six reciprocal vectors, as in unstrained TBG \cite{cea_electrostatic_2022}. Therefore, in the numerical calculations we included all the potentials coefficients $V_{H}\left(\mathbf{g}\right)$ within the considered reciprocal vectors $\mathbf{g}$ in the continuum model. To ensure convergence, we consider a cutoff up to four closest set of reciprocal vectors.

Figure \ref{fig:Hartree} shows a density plot of the Hartree potential in the square pattern arising from twist and shear strain (blue green case in Figure \ref{fig:MP_square}), for a filling $\nu=1$ (one electron per unit cell). As seen, the Hartree potential follows the shape of the moiré pattern. In particular, it always peak around the elliptical-shape AA stacking regimes.

\begin{figure}[t]
	\includegraphics[width=\linewidth]{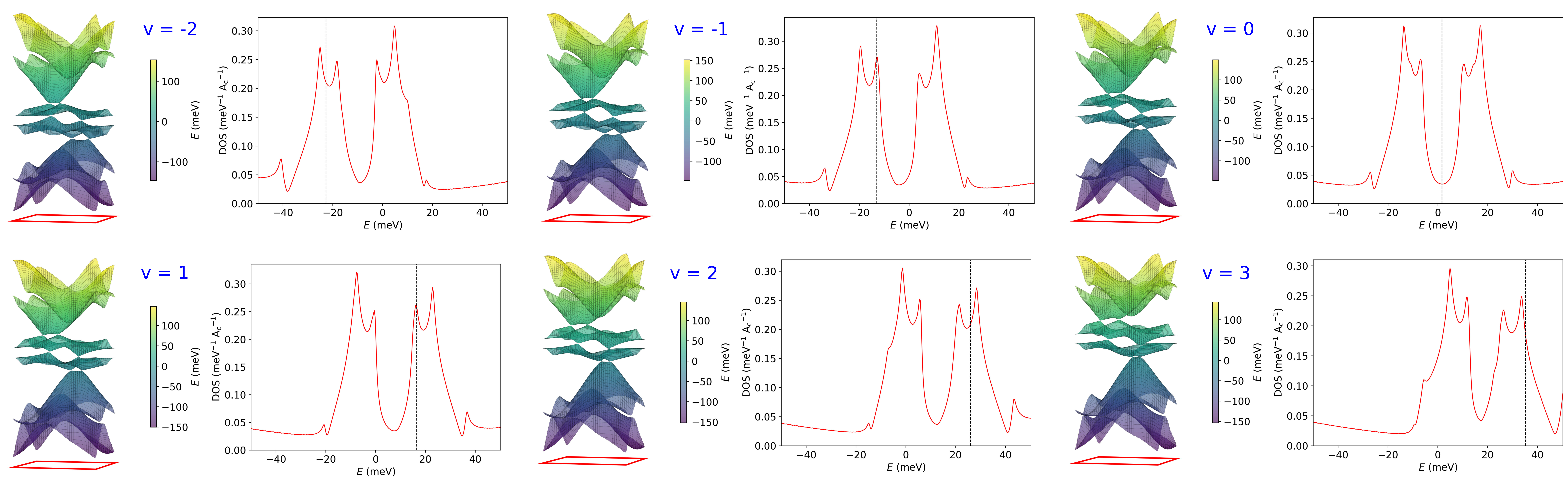}	
	\caption{Band structure and DOS for filling factors $\nu=-2,-1,0,1,2,3$. All cases correspond to the square pattern with moiré length $\tilde{L}_{M}=12.1205\,\mathrm{nm}$, arising from twist and shear strain (see Section \ref{sec:Theory_Square}). The vertical black dashed-line in the DOS indicates the Fermi energy. }\label{fig:Hartree_nu}
\end{figure}

Figure \ref{fig:Hartree_nu} shows a 3D plot of the band structure and total density of states, for the square pattern arising from twist and shear strain with moiré length $\tilde{L}_{M}=12.1205\,\mathrm{nm}$, and different filling factors $\nu$ (electrons per moiré unit cell, with respect to CN). The CN case $\nu=0$ corresponds to no Hartree potential. As seen, the Hartree effect is not as pronounced as in magic angle TBG, mainly because of the increase in the bandwidth due to the strain. The Hartree shifts the bands and the Fermi energy, and modifies the splitting profile of the VHs. Interestingly, there is still a pinning of the Fermi energy to the VHs at $\nu\neq0$.

\subsection{Local density of states}

\begin{figure}[t]
	\includegraphics[width=\linewidth]{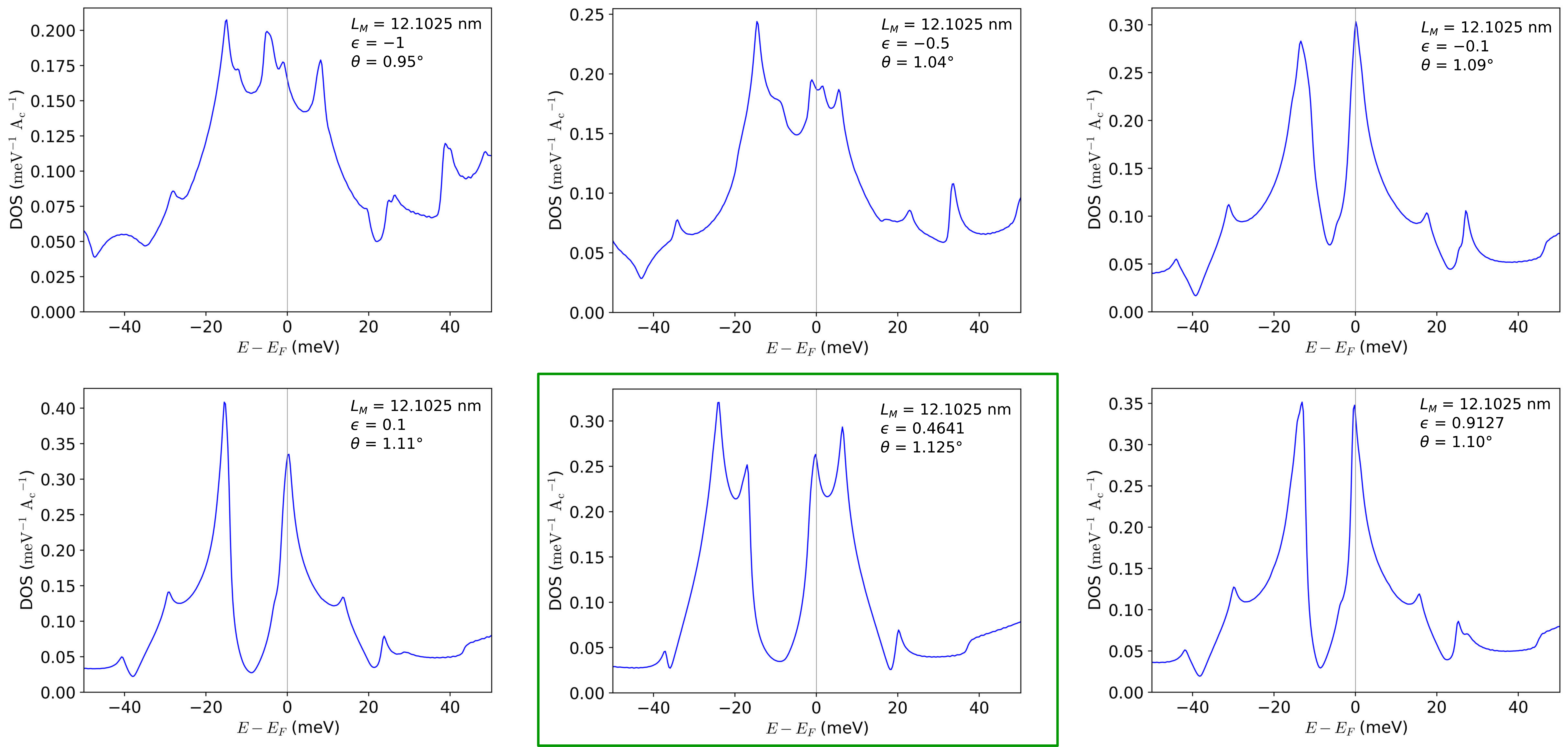}	
	\caption{Total density of states (DOS) for the square patterns shown in Figure \ref{fig:MP_square}. The minimum strain case $\epsilon=2\sqrt{3}-3\approx0.42641$ (shear strain) is highlighted in a green box. Despite the similar square pattern in all cases, the DOS is markedly different. In particular, only the shear strain (green box) captures the splitting of the two main peaks at $E_{F}$ and $E_{F}-30\,\mathrm{meV}$, seen in the experiments; see Figure \ref{fig:exp}(f).}\label{fig:DOS_all}
\end{figure}

The local density of states at energy $E$ is given by
\begin{equation}
\rho\left(\mathbf{r},E\right)=\sum_{n,\mathbf{k},\eta,i}\left|\psi_{n,\mathbf{k},\eta,i}\left(\mathbf{r}\right)\right|^{2}\delta\left(E-E_{n,\mathbf{k},\eta}\right).
\end{equation}
Replacing the Bloch states gives
\begin{align}
\rho\left(\mathbf{r},E\right) & =\sum_{\mathbf{g}}\rho\left(\mathbf{g},E\right)e^{-i\mathbf{g}\cdot\mathbf{r}},\\
\rho\left(\mathbf{g},E\right) & =A_{c}^{-1}\sum_{\mathbf{k},\mathbf{g}'}\sum_{n,\eta,i}u_{n,\mathbf{k},\eta,i}^{*}\left(\mathbf{g}'+\mathbf{g}\right)u_{n,\mathbf{k},\eta,i}\left(\mathbf{g}'\right)\delta\left(E-E_{n,\mathbf{k},\eta}\right).
\end{align}
The total density of states is
\begin{equation}
\rho\left(E\right)=\int_{\mathrm{unit\,cell}}d\mathbf{r}\rho\left(\mathbf{r},E\right)=\sum_{n,\mathbf{k},\eta}\delta\left(E-E_{n,\mathbf{k},\eta}\right).
\end{equation}
For the numerical calculations we model the Dirac delta as a Lorentzian $\delta\left(x\right)\rightarrow\eta\left(x^{2}+\eta^{2}\right)^{-1}/\pi$, with a sufficiently small width $\eta$. Since the strain preserves the $\mathcal{C}_{2}\mathcal{T}$ symmetry \cite{bi_designing_2019}, it holds that both valleys are related by 
\begin{align}
E_{n,\mathbf{k},K} & =E_{n,-\mathbf{k},K'},\\
\psi_{n,\mathbf{k},K,i}\left(\mathbf{r}\right) & =\psi_{n,-\mathbf{k},K',i}^{*}\left(\mathbf{r}\right).
\end{align}
Therefore all computations can be done by working entirely with one valley and including a fourfold valley and spin degeneracy.

Figure \ref{fig:DOS_all} shows the total DOS for the same square patterns shown in Figure \ref{fig:MP_square}, corresponding to the average moiré length $\tilde{L}_{M}=12.1025\,\mathrm{nm}$ in Figure \ref{fig:exp}(c). The results highlight that although all these cases correspond to practically the same square moiré pattern (up to an overall rotation), their electronic properties are noticeable different. Thus the electronic properties provide a more clear fingerprint of the strain and twist in the system. Note that only the minimum shear strain case (green box) captures the splitting of the two main DOS peaks seen in Figure \ref{fig:exp}.

\begin{figure}[t]
	\includegraphics[width=\linewidth]{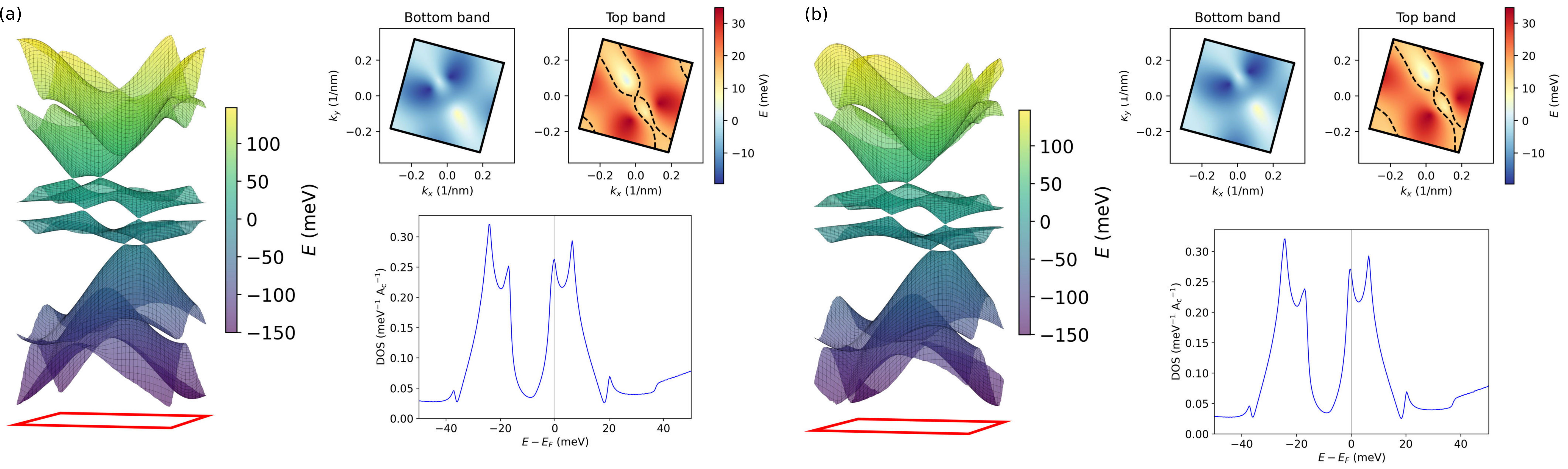}	
	\caption{Plots of the 3D band structure, the density plot of the top and bottom middle moiré bands, and the total DOS, for the twist and shear strain square pattern arising from: (a) strain applied in the two layers (with equal magnitude and opposite direction), and (b) strain applied in only one layer (see Section \ref{sec:Theory_Square}). Other parameters as in Fig. \ref{fig:theo} of the main text.}\label{fig:Comparison}
\end{figure}

Figure \ref{fig:Comparison} shows a comparison of the electronic properties between: (a) strain applied in both layers (with equal magnitude and opposite direction), and (b) strain applied in only one layer, for the square pattern arising from twist and shear strain. As anticipated in Section \ref{sec:Theory_Square}, both cases give practically the same results. Their main difference lies in the shifting of the Dirac points within the unit cell (mainly due to the asymmetrical gauge potential $\mathbf{A}$ when the strain is applied in only one layer), and the consequently shifting of the
Fermi energy. 

\subsection{Quantitative differences between the continuum model LDOS and the STS measurements}

Here we address the quantitative differences between the continuum model LDOS and the STS measurements of Figure \ref{fig:exp}. Before we address these differences, let us point out that the system behavior is highly sensitive to many parameters, many of which cannot even be estimated in the experimental setup (e.g., local variations in the effective hopping energies between different sublattices, or residual charges that influence the doping filling and the electrostatic interactions). For instance, the small variations observed in the measurements along the boundaries of the square unit cell can be attributed to several factors, such as the existence of a slight inhomogeneity in the strain. Even if the strain variation across a unit cell is small enough to preserve the square pattern, it may still induce significant differences in the LDOS due to the magnifying effect of the moiré (see e.g. Figure \ref{fig:DOS_all} in SM).  Thus our discussion will only focus on the overall trends seen in the experiments. Quite generally, we identify three main differences between the measured STS and the LDOS of the continuum model. 

First, we see that the STS measurements (see Figure \ref{fig:exp}) always reflect a larger absolute magnitude around the Fermi energy (zero bias voltage); the second peak (around $V_{b}\sim-30\,\mathrm{meV}$) has a lower magnitude. In contrast, the LDOS in Figure \ref{fig:theo} exhibits those two peaks with practically equal magnitudes. There could be several explanations for this difference. A smaller DOS at energies lower than the Fermi energy could signal that for electronic doping the valence bands are more dispersive than the conduction bands. Such a behavior can partially result from relaxation effects, which are typically stronger for the valence bands and may lead to an increase in its bandwidth \cite{Nam2017, guinea_continuum_2019}. Another contributing factor can be that the STS signal is sensitive to the bias voltage $V_{b}$ due to the dependence on the tunneling between the tip and the sample, so that the signal strength decreases as $V_{b}$ increases \cite{Tersoff1985,Chen1988}. 

Second, the continuum model naturally yields equal LDOS along equivalent real space paths. This contrasts with the STS measurements, which show a quantitative difference even in what would be equivalent paths in a periodic system (e.g., the two horizontal or the vertical paths along the square unit cell). We can safely attribute these variations to the strain configuration, upon which the LDOS can be highly sensitive (cf. Figure \ref{fig:DOS_all}). Small local changes in the strain magnitude or direction along different paths of the unit cell can preserve the overall square pattern, and yet yield noticeable discrepancies in the electronic properties in each case. One would actually expect this from STM measurements in Figure \ref{fig:exp} d), which show that there are small variations in the lattice vectors (connecting AA stackings) at different points (essentially reflecting a strain inhomogeneity).

Lastly, in Figure \ref{fig:exp} we see that along the two horizontal directions (top and bottom of the square unit cell) there is a prominent secondary peak close to the Fermi energy (at about $V_{b}\sim+6\,\mathrm{meV}$), which importantly, peaks between the AA staking regimes (i.e., around the DW; cf. Figure \ref{fig:theo}). As discussed above, the continuum model captures the appearance of secondary peaks for shear strain, but only around the AA stacking regime rather than the observed DW regime. In general one would expect the LDOS to not peak at the AA stacking only for remote bands at which the charge density is not concentrated around the AA centers. In our case, the remote bands are typically more than $\sim20$ meV away from the Fermi energy, which is far apart from the observed secondary peak at $V_{b}=+6\,\mathrm{meV}$.

\section{Supplemental STM data}

In the following, we present atomically resolved, magnified views of the data shown in the manuscript. 

\begin{figure*}
	\includegraphics[width=0.75\linewidth]{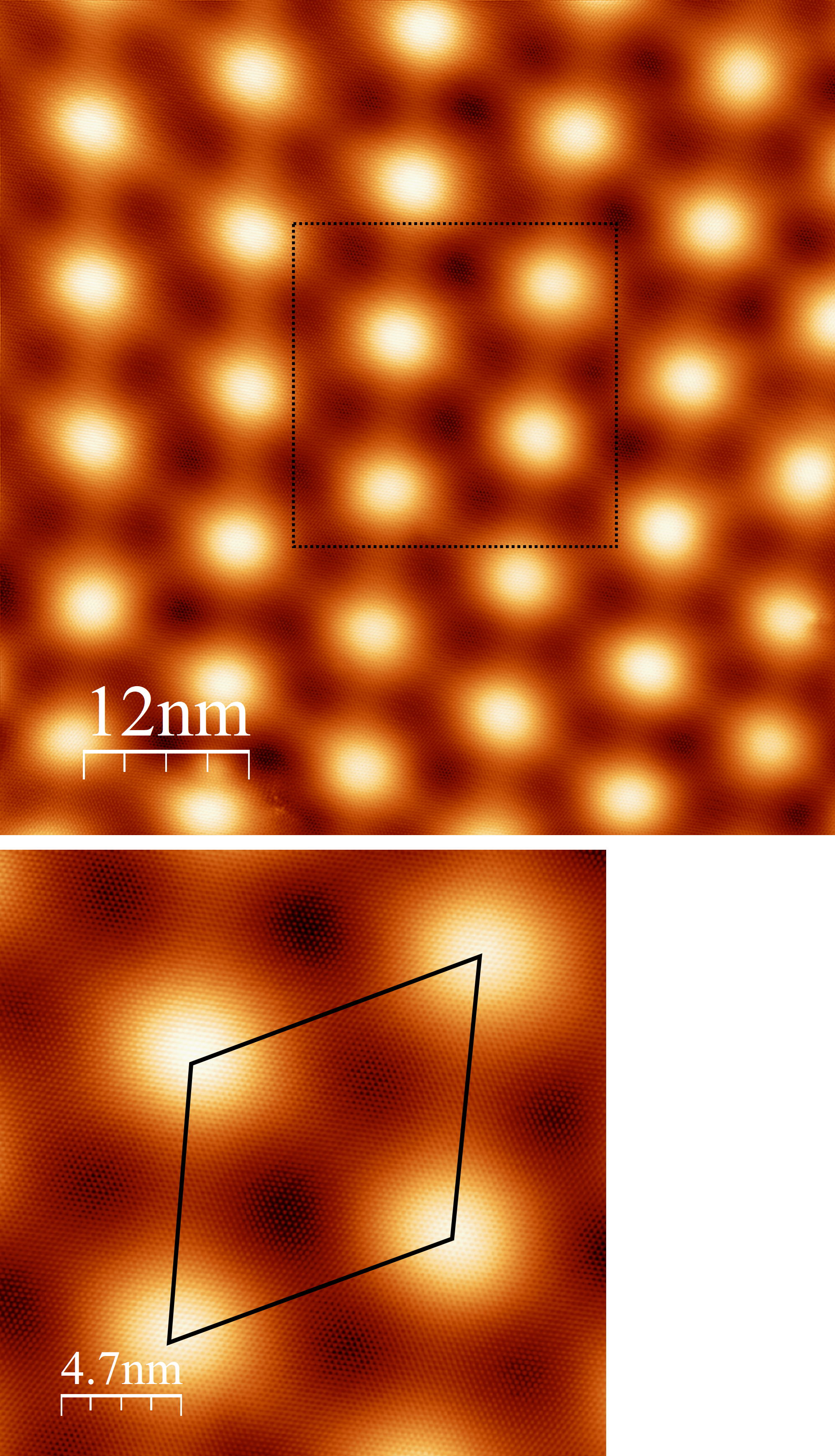}	
	\caption{Atomic-resolution STM image of a region with negligible strain exhibiting magic-angle  trigonal moiré superlattice.  STM parameters: $I_T$ = $340\,\mathrm{pA}$, $V_{bias}$ = $50\,\mathrm{mV}$. }\label{fig:S1}
\end{figure*}

\begin{figure*}
	\includegraphics[width=0.85\linewidth]{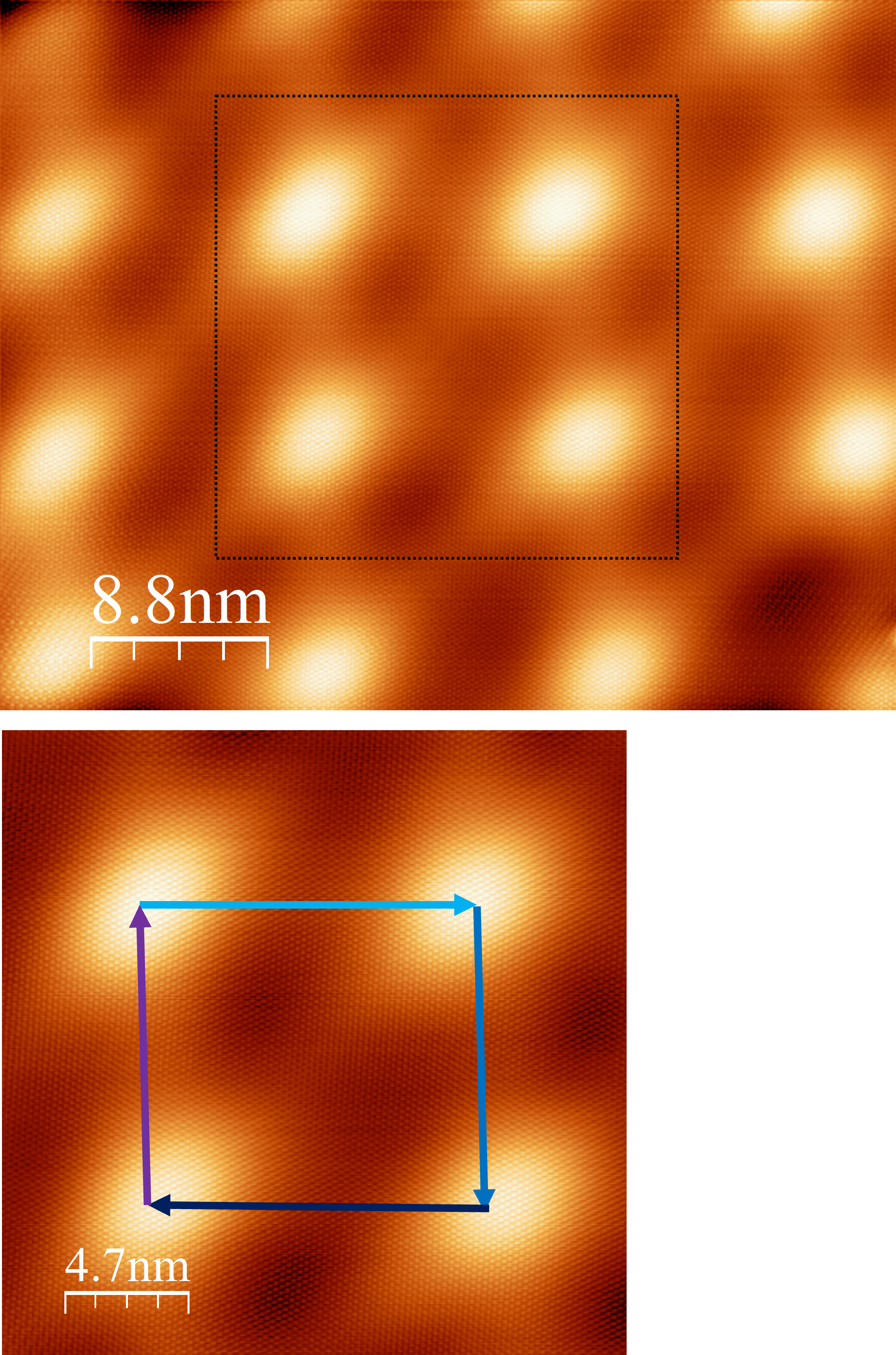}	
	\caption{Atomic-resolution STM image of a magic-angle square moiré superlattice (periodicity of $\approx12\;\mathrm{nm}$). STM parameters: $I_T$ = $50\,\mathrm{pA}$, $V_{bias}$ = $16\,\mathrm{mV}$ , top; $I_T$ = $230\,\mathrm{pA}$, $V_{bias}$ = $10\,\mathrm{mV}$, bottom.}\label{fig:S2}
\end{figure*}

\begin{figure*}
	\includegraphics[width=0.7\linewidth]{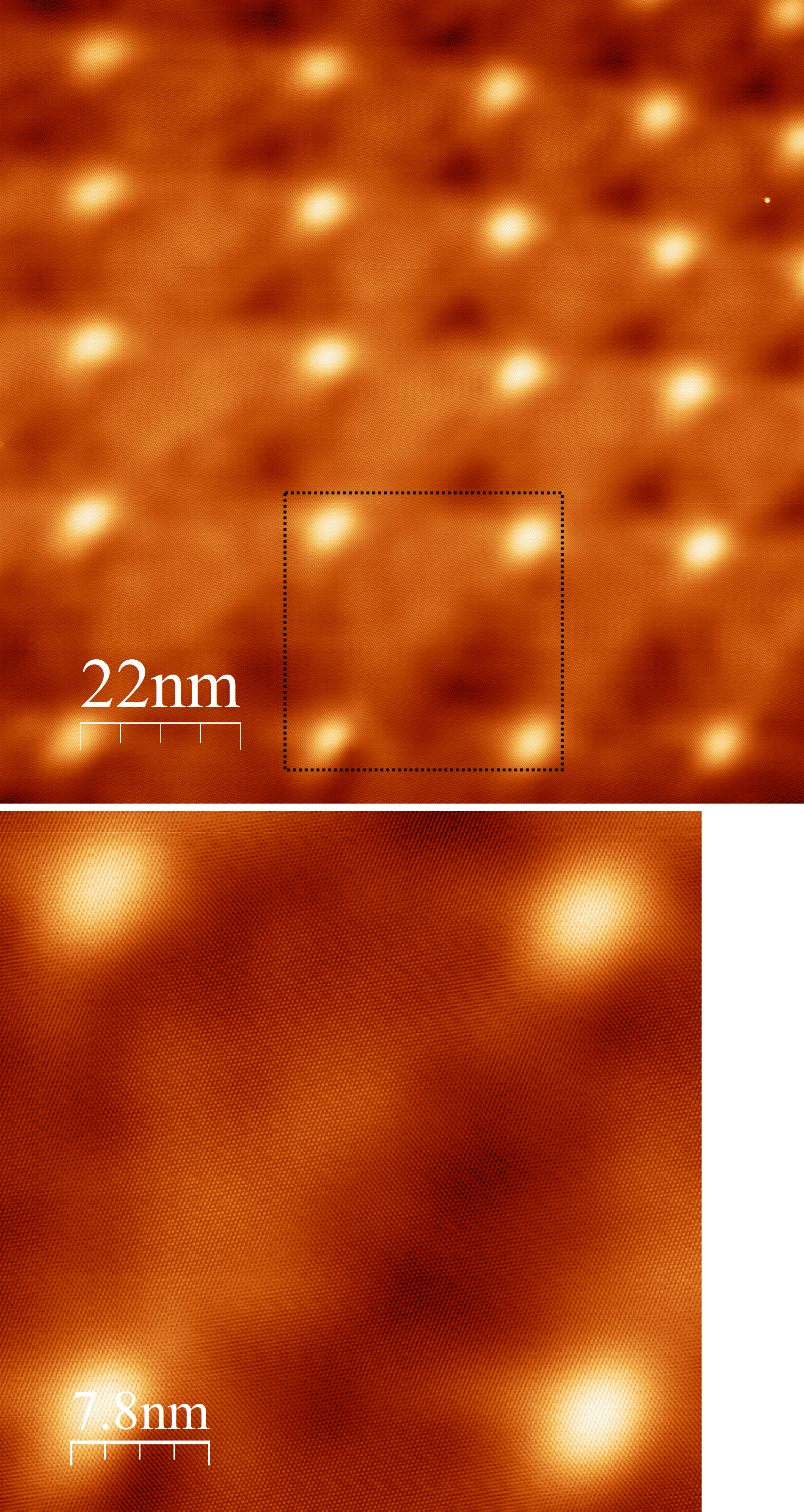}	
	\caption{Atomic-resolution STM image of a large strained moiré. The nonuniform periodicity is caused by an inhomogeneous strain profile. STM parameters: $I_T$ = $50\,\mathrm{pA}$, $V_{bias}$ = $25\,\mathrm{mV}$}.\label{fig:S3}
\end{figure*}


\begin{thebibliography}{82}%
\makeatletter
\providecommand \@ifxundefined [1]{%
 \@ifx{#1\undefined}
}%
\providecommand \@ifnum [1]{%
 \ifnum #1\expandafter \@firstoftwo
 \else \expandafter \@secondoftwo
 \fi
}%
\providecommand \@ifx [1]{%
 \ifx #1\expandafter \@firstoftwo
 \else \expandafter \@secondoftwo
 \fi
}%
\providecommand \natexlab [1]{#1}%
\providecommand \enquote  [1]{``#1''}%
\providecommand \bibnamefont  [1]{#1}%
\providecommand \bibfnamefont [1]{#1}%
\providecommand \citenamefont [1]{#1}%
\providecommand \href@noop [0]{\@secondoftwo}%
\providecommand \href [0]{\begingroup \@sanitize@url \@href}%
\providecommand \@href[1]{\@@startlink{#1}\@@href}%
\providecommand \@@href[1]{\endgroup#1\@@endlink}%
\providecommand \@sanitize@url [0]{\catcode `\\12\catcode `\$12\catcode `\&12\catcode `\#12\catcode `\^12\catcode `\_12\catcode `\%12\relax}%
\providecommand \@@startlink[1]{}%
\providecommand \@@endlink[0]{}%
\providecommand \url  [0]{\begingroup\@sanitize@url \@url }%
\providecommand \@url [1]{\endgroup\@href {#1}{\urlprefix }}%
\providecommand \urlprefix  [0]{URL }%
\providecommand \Eprint [0]{\href }%
\providecommand \doibase [0]{https://doi.org/}%
\providecommand \selectlanguage [0]{\@gobble}%
\providecommand \bibinfo  [0]{\@secondoftwo}%
\providecommand \bibfield  [0]{\@secondoftwo}%
\providecommand \translation [1]{[#1]}%
\providecommand \BibitemOpen [0]{}%
\providecommand \bibitemStop [0]{}%
\providecommand \bibitemNoStop [0]{.\EOS\space}%
\providecommand \EOS [0]{\spacefactor3000\relax}%
\providecommand \BibitemShut  [1]{\csname bibitem#1\endcsname}%
\let\auto@bib@innerbib\@empty
%</preamble>
\bibitem [{\citenamefont {Cao}\ \emph {et~al.}(2018{\natexlab{a}})\citenamefont {Cao}, \citenamefont {Fatemi}, \citenamefont {Fang}, \citenamefont {Watanabe}, \citenamefont {Taniguchi}, \citenamefont {Kaxiras},\ and\ \citenamefont {Jarillo-Herrero}}]{cao_unconventional_2018}%
  \BibitemOpen
  \bibfield  {author} {\bibinfo {author} {\bibfnamefont {Y.}~\bibnamefont {Cao}}, \bibinfo {author} {\bibfnamefont {V.}~\bibnamefont {Fatemi}}, \bibinfo {author} {\bibfnamefont {S.}~\bibnamefont {Fang}}, \bibinfo {author} {\bibfnamefont {K.}~\bibnamefont {Watanabe}}, \bibinfo {author} {\bibfnamefont {T.}~\bibnamefont {Taniguchi}}, \bibinfo {author} {\bibfnamefont {E.}~\bibnamefont {Kaxiras}},\ and\ \bibinfo {author} {\bibfnamefont {P.}~\bibnamefont {Jarillo-Herrero}},\ }\bibfield  {title} {\bibinfo {title} {Unconventional superconductivity in magic-angle graphene superlattices},\ }\href {https://doi.org/10.1038/nature26160} {\bibfield  {journal} {\bibinfo  {journal} {Nature}\ }\textbf {\bibinfo {volume} {556}},\ \bibinfo {pages} {43} (\bibinfo {year} {2018}{\natexlab{a}})}\BibitemShut {NoStop}%
\bibitem [{\citenamefont {Cao}\ \emph {et~al.}(2018{\natexlab{b}})\citenamefont {Cao}, \citenamefont {Fatemi}, \citenamefont {Demir}, \citenamefont {Fang}, \citenamefont {Tomarken}, \citenamefont {Luo}, \citenamefont {Sanchez-Yamagishi}, \citenamefont {Watanabe}, \citenamefont {Taniguchi}, \citenamefont {Kaxiras}, \citenamefont {Ashoori},\ and\ \citenamefont {Jarillo-Herrero}}]{cao_correlated_2018}%
  \BibitemOpen
  \bibfield  {author} {\bibinfo {author} {\bibfnamefont {Y.}~\bibnamefont {Cao}}, \bibinfo {author} {\bibfnamefont {V.}~\bibnamefont {Fatemi}}, \bibinfo {author} {\bibfnamefont {A.}~\bibnamefont {Demir}}, \bibinfo {author} {\bibfnamefont {S.}~\bibnamefont {Fang}}, \bibinfo {author} {\bibfnamefont {S.~L.}\ \bibnamefont {Tomarken}}, \bibinfo {author} {\bibfnamefont {J.~Y.}\ \bibnamefont {Luo}}, \bibinfo {author} {\bibfnamefont {J.~D.}\ \bibnamefont {Sanchez-Yamagishi}}, \bibinfo {author} {\bibfnamefont {K.}~\bibnamefont {Watanabe}}, \bibinfo {author} {\bibfnamefont {T.}~\bibnamefont {Taniguchi}}, \bibinfo {author} {\bibfnamefont {E.}~\bibnamefont {Kaxiras}}, \bibinfo {author} {\bibfnamefont {R.~C.}\ \bibnamefont {Ashoori}},\ and\ \bibinfo {author} {\bibfnamefont {P.}~\bibnamefont {Jarillo-Herrero}},\ }\bibfield  {title} {\bibinfo {title} {Correlated insulator behaviour at half-filling in magic-angle graphene superlattices},\ }\href {https://doi.org/10.1038/nature26154} {\bibfield  {journal} {\bibinfo  {journal}
  {Nature}\ }\textbf {\bibinfo {volume} {556}},\ \bibinfo {pages} {80} (\bibinfo {year} {2018}{\natexlab{b}})}\BibitemShut {NoStop}%
\bibitem [{\citenamefont {Kerelsky}\ \emph {et~al.}(2019)\citenamefont {Kerelsky}, \citenamefont {McGilly}, \citenamefont {Kennes}, \citenamefont {Xian}, \citenamefont {Yankowitz}, \citenamefont {Chen}, \citenamefont {Watanabe}, \citenamefont {Taniguchi}, \citenamefont {Hone}, \citenamefont {Dean}, \citenamefont {Rubio},\ and\ \citenamefont {Pasupathy}}]{kerelsky_maximized_2019}%
  \BibitemOpen
  \bibfield  {author} {\bibinfo {author} {\bibfnamefont {A.}~\bibnamefont {Kerelsky}}, \bibinfo {author} {\bibfnamefont {L.~J.}\ \bibnamefont {McGilly}}, \bibinfo {author} {\bibfnamefont {D.~M.}\ \bibnamefont {Kennes}}, \bibinfo {author} {\bibfnamefont {L.}~\bibnamefont {Xian}}, \bibinfo {author} {\bibfnamefont {M.}~\bibnamefont {Yankowitz}}, \bibinfo {author} {\bibfnamefont {S.}~\bibnamefont {Chen}}, \bibinfo {author} {\bibfnamefont {K.}~\bibnamefont {Watanabe}}, \bibinfo {author} {\bibfnamefont {T.}~\bibnamefont {Taniguchi}}, \bibinfo {author} {\bibfnamefont {J.}~\bibnamefont {Hone}}, \bibinfo {author} {\bibfnamefont {C.}~\bibnamefont {Dean}}, \bibinfo {author} {\bibfnamefont {A.}~\bibnamefont {Rubio}},\ and\ \bibinfo {author} {\bibfnamefont {A.~N.}\ \bibnamefont {Pasupathy}},\ }\bibfield  {title} {\bibinfo {title} {Maximized electron interactions at the magic angle in twisted bilayer graphene},\ }\href {https://doi.org/10.1038/s41586-019-1431-9} {\bibfield  {journal} {\bibinfo  {journal} {Nature}\ }\textbf
  {\bibinfo {volume} {572}},\ \bibinfo {pages} {95} (\bibinfo {year} {2019})}\BibitemShut {NoStop}%
\bibitem [{\citenamefont {Yankowitz}\ \emph {et~al.}(2019)\citenamefont {Yankowitz}, \citenamefont {Chen}, \citenamefont {Polshyn}, \citenamefont {Zhang}, \citenamefont {Watanabe}, \citenamefont {Taniguchi}, \citenamefont {Graf}, \citenamefont {Young},\ and\ \citenamefont {Dean}}]{yankowitz_tuning_2019}%
  \BibitemOpen
  \bibfield  {author} {\bibinfo {author} {\bibfnamefont {M.}~\bibnamefont {Yankowitz}}, \bibinfo {author} {\bibfnamefont {S.}~\bibnamefont {Chen}}, \bibinfo {author} {\bibfnamefont {H.}~\bibnamefont {Polshyn}}, \bibinfo {author} {\bibfnamefont {Y.}~\bibnamefont {Zhang}}, \bibinfo {author} {\bibfnamefont {K.}~\bibnamefont {Watanabe}}, \bibinfo {author} {\bibfnamefont {T.}~\bibnamefont {Taniguchi}}, \bibinfo {author} {\bibfnamefont {D.}~\bibnamefont {Graf}}, \bibinfo {author} {\bibfnamefont {A.~F.}\ \bibnamefont {Young}},\ and\ \bibinfo {author} {\bibfnamefont {C.~R.}\ \bibnamefont {Dean}},\ }\bibfield  {title} {\bibinfo {title} {Tuning superconductivity in twisted bilayer graphene},\ }\href {https://doi.org/10.1126/science.aav1910} {\bibfield  {journal} {\bibinfo  {journal} {Science}\ }\textbf {\bibinfo {volume} {363}},\ \bibinfo {pages} {1059} (\bibinfo {year} {2019})}\BibitemShut {NoStop}%
\bibitem [{\citenamefont {Oh}\ \emph {et~al.}(2021)\citenamefont {Oh}, \citenamefont {Nuckolls}, \citenamefont {Wong}, \citenamefont {Lee}, \citenamefont {Liu}, \citenamefont {Watanabe}, \citenamefont {Taniguchi},\ and\ \citenamefont {Yazdani}}]{oh_evidence_2021}%
  \BibitemOpen
  \bibfield  {author} {\bibinfo {author} {\bibfnamefont {M.}~\bibnamefont {Oh}}, \bibinfo {author} {\bibfnamefont {K.~P.}\ \bibnamefont {Nuckolls}}, \bibinfo {author} {\bibfnamefont {D.}~\bibnamefont {Wong}}, \bibinfo {author} {\bibfnamefont {R.~L.}\ \bibnamefont {Lee}}, \bibinfo {author} {\bibfnamefont {X.}~\bibnamefont {Liu}}, \bibinfo {author} {\bibfnamefont {K.}~\bibnamefont {Watanabe}}, \bibinfo {author} {\bibfnamefont {T.}~\bibnamefont {Taniguchi}},\ and\ \bibinfo {author} {\bibfnamefont {A.}~\bibnamefont {Yazdani}},\ }\bibfield  {title} {\bibinfo {title} {Evidence for unconventional superconductivity in twisted bilayer graphene},\ }\href {https://doi.org/10.1038/s41586-021-04121-x} {\bibfield  {journal} {\bibinfo  {journal} {Nature}\ }\textbf {\bibinfo {volume} {600}},\ \bibinfo {pages} {240} (\bibinfo {year} {2021})}\BibitemShut {NoStop}%
\bibitem [{\citenamefont {Xie}\ \emph {et~al.}(2021)\citenamefont {Xie}, \citenamefont {Pierce}, \citenamefont {Park}, \citenamefont {Parker}, \citenamefont {Khalaf}, \citenamefont {Ledwith}, \citenamefont {Cao}, \citenamefont {Lee}, \citenamefont {Chen}, \citenamefont {Forrester}, \citenamefont {Watanabe}, \citenamefont {Taniguchi}, \citenamefont {Vishwanath}, \citenamefont {Jarillo-Herrero},\ and\ \citenamefont {Yacoby}}]{xie_fractional_2021}%
  \BibitemOpen
  \bibfield  {author} {\bibinfo {author} {\bibfnamefont {Y.}~\bibnamefont {Xie}}, \bibinfo {author} {\bibfnamefont {A.~T.}\ \bibnamefont {Pierce}}, \bibinfo {author} {\bibfnamefont {J.~M.}\ \bibnamefont {Park}}, \bibinfo {author} {\bibfnamefont {D.~E.}\ \bibnamefont {Parker}}, \bibinfo {author} {\bibfnamefont {E.}~\bibnamefont {Khalaf}}, \bibinfo {author} {\bibfnamefont {P.}~\bibnamefont {Ledwith}}, \bibinfo {author} {\bibfnamefont {Y.}~\bibnamefont {Cao}}, \bibinfo {author} {\bibfnamefont {S.~H.}\ \bibnamefont {Lee}}, \bibinfo {author} {\bibfnamefont {S.}~\bibnamefont {Chen}}, \bibinfo {author} {\bibfnamefont {P.~R.}\ \bibnamefont {Forrester}}, \bibinfo {author} {\bibfnamefont {K.}~\bibnamefont {Watanabe}}, \bibinfo {author} {\bibfnamefont {T.}~\bibnamefont {Taniguchi}}, \bibinfo {author} {\bibfnamefont {A.}~\bibnamefont {Vishwanath}}, \bibinfo {author} {\bibfnamefont {P.}~\bibnamefont {Jarillo-Herrero}},\ and\ \bibinfo {author} {\bibfnamefont {A.}~\bibnamefont {Yacoby}},\ }\bibfield  {title} {\bibinfo
  {title} {Fractional {Chern} insulators in magic-angle twisted bilayer graphene},\ }\href {https://doi.org/10.1038/s41586-021-04002-3} {\bibfield  {journal} {\bibinfo  {journal} {Nature}\ }\textbf {\bibinfo {volume} {600}},\ \bibinfo {pages} {439} (\bibinfo {year} {2021})}\BibitemShut {NoStop}%
\bibitem [{\citenamefont {Andrei}\ \emph {et~al.}(2021)\citenamefont {Andrei}, \citenamefont {Efetov}, \citenamefont {Jarillo-Herrero}, \citenamefont {MacDonald}, \citenamefont {Mak}, \citenamefont {Senthil}, \citenamefont {Tutuc}, \citenamefont {Yazdani},\ and\ \citenamefont {Young}}]{andrei_marvels_2021}%
  \BibitemOpen
  \bibfield  {author} {\bibinfo {author} {\bibfnamefont {E.~Y.}\ \bibnamefont {Andrei}}, \bibinfo {author} {\bibfnamefont {D.~K.}\ \bibnamefont {Efetov}}, \bibinfo {author} {\bibfnamefont {P.}~\bibnamefont {Jarillo-Herrero}}, \bibinfo {author} {\bibfnamefont {A.~H.}\ \bibnamefont {MacDonald}}, \bibinfo {author} {\bibfnamefont {K.~F.}\ \bibnamefont {Mak}}, \bibinfo {author} {\bibfnamefont {T.}~\bibnamefont {Senthil}}, \bibinfo {author} {\bibfnamefont {E.}~\bibnamefont {Tutuc}}, \bibinfo {author} {\bibfnamefont {A.}~\bibnamefont {Yazdani}},\ and\ \bibinfo {author} {\bibfnamefont {A.~F.}\ \bibnamefont {Young}},\ }\bibfield  {title} {\bibinfo {title} {The marvels of moiré materials},\ }\href {https://doi.org/10.1038/s41578-021-00284-1} {\bibfield  {journal} {\bibinfo  {journal} {Nature Reviews Materials}\ }\textbf {\bibinfo {volume} {6}},\ \bibinfo {pages} {201} (\bibinfo {year} {2021})}\BibitemShut {NoStop}%
\bibitem [{\citenamefont {Andrei}\ and\ \citenamefont {MacDonald}(2020)}]{Andrei2020}%
  \BibitemOpen
  \bibfield  {author} {\bibinfo {author} {\bibfnamefont {E.~Y.}\ \bibnamefont {Andrei}}\ and\ \bibinfo {author} {\bibfnamefont {A.~H.}\ \bibnamefont {MacDonald}},\ }\bibfield  {title} {\bibinfo {title} {Graphene bilayers with a twist},\ }\href {https://doi.org/10.1038/s41563-020-00840-0} {\bibfield  {journal} {\bibinfo  {journal} {Nature Materials}\ }\textbf {\bibinfo {volume} {19}},\ \bibinfo {pages} {1265} (\bibinfo {year} {2020})}\BibitemShut {NoStop}%
\bibitem [{\citenamefont {Shallcross}\ \emph {et~al.}(2008)\citenamefont {Shallcross}, \citenamefont {Sharma},\ and\ \citenamefont {Pankratov}}]{shallcross_quantum_2008}%
  \BibitemOpen
  \bibfield  {author} {\bibinfo {author} {\bibfnamefont {S.}~\bibnamefont {Shallcross}}, \bibinfo {author} {\bibfnamefont {S.}~\bibnamefont {Sharma}},\ and\ \bibinfo {author} {\bibfnamefont {O.~A.}\ \bibnamefont {Pankratov}},\ }\bibfield  {title} {\bibinfo {title} {Quantum {Interference} at the {Twist} {Boundary} in {Graphene}},\ }\href {https://doi.org/10.1103/PhysRevLett.101.056803} {\bibfield  {journal} {\bibinfo  {journal} {Phys. Rev. Lett.}\ }\textbf {\bibinfo {volume} {101}},\ \bibinfo {pages} {056803} (\bibinfo {year} {2008})}\BibitemShut {NoStop}%
\bibitem [{\citenamefont {Shallcross}\ \emph {et~al.}(2010)\citenamefont {Shallcross}, \citenamefont {Sharma}, \citenamefont {Kandelaki},\ and\ \citenamefont {Pankratov}}]{shallcross_electronic_2010}%
  \BibitemOpen
  \bibfield  {author} {\bibinfo {author} {\bibfnamefont {S.}~\bibnamefont {Shallcross}}, \bibinfo {author} {\bibfnamefont {S.}~\bibnamefont {Sharma}}, \bibinfo {author} {\bibfnamefont {E.}~\bibnamefont {Kandelaki}},\ and\ \bibinfo {author} {\bibfnamefont {O.~A.}\ \bibnamefont {Pankratov}},\ }\bibfield  {title} {\bibinfo {title} {Electronic structure of turbostratic graphene},\ }\href {https://doi.org/10.1103/PhysRevB.81.165105} {\bibfield  {journal} {\bibinfo  {journal} {Phys. Rev. B}\ }\textbf {\bibinfo {volume} {81}},\ \bibinfo {pages} {165105} (\bibinfo {year} {2010})}\BibitemShut {NoStop}%
\bibitem [{\citenamefont {Laissardière}\ \emph {et~al.}(2010)\citenamefont {Laissardière}, \citenamefont {Mayou},\ and\ \citenamefont {Magaud}}]{laissardiere_localization_2010}%
  \BibitemOpen
  \bibfield  {author} {\bibinfo {author} {\bibfnamefont {G.~T.~d.}\ \bibnamefont {Laissardière}}, \bibinfo {author} {\bibfnamefont {D.}~\bibnamefont {Mayou}},\ and\ \bibinfo {author} {\bibfnamefont {L.}~\bibnamefont {Magaud}},\ }\bibfield  {title} {\bibinfo {title} {Localization of {Dirac} {Electrons} in {Rotated} {Graphene} {Bilayers}},\ }\href {https://doi.org/10.1021/nl902948m} {\bibfield  {journal} {\bibinfo  {journal} {Nano Letters}\ }\textbf {\bibinfo {volume} {10}},\ \bibinfo {pages} {804} (\bibinfo {year} {2010})}\BibitemShut {NoStop}%
\bibitem [{\citenamefont {Lopes~dos Santos}\ \emph {et~al.}(2012)\citenamefont {Lopes~dos Santos}, \citenamefont {Peres},\ and\ \citenamefont {Castro~Neto}}]{lopes_dos_santos_continuum_2012}%
  \BibitemOpen
  \bibfield  {author} {\bibinfo {author} {\bibfnamefont {J.~M.~B.}\ \bibnamefont {Lopes~dos Santos}}, \bibinfo {author} {\bibfnamefont {N.~M.~R.}\ \bibnamefont {Peres}},\ and\ \bibinfo {author} {\bibfnamefont {A.~H.}\ \bibnamefont {Castro~Neto}},\ }\bibfield  {title} {\bibinfo {title} {Continuum model of the twisted graphene bilayer},\ }\href {https://doi.org/10.1103/PhysRevB.86.155449} {\bibfield  {journal} {\bibinfo  {journal} {Phys. Rev. B}\ }\textbf {\bibinfo {volume} {86}},\ \bibinfo {pages} {155449} (\bibinfo {year} {2012})}\BibitemShut {NoStop}%
\bibitem [{\citenamefont {Sboychakov}\ \emph {et~al.}(2015)\citenamefont {Sboychakov}, \citenamefont {Rakhmanov}, \citenamefont {Rozhkov},\ and\ \citenamefont {Nori}}]{Sboychakov2015}%
  \BibitemOpen
  \bibfield  {author} {\bibinfo {author} {\bibfnamefont {A.~O.}\ \bibnamefont {Sboychakov}}, \bibinfo {author} {\bibfnamefont {A.~L.}\ \bibnamefont {Rakhmanov}}, \bibinfo {author} {\bibfnamefont {A.~V.}\ \bibnamefont {Rozhkov}},\ and\ \bibinfo {author} {\bibfnamefont {F.}~\bibnamefont {Nori}},\ }\bibfield  {title} {\bibinfo {title} {Electronic spectrum of twisted bilayer graphene},\ }\href {https://doi.org/10.1103/PhysRevB.92.075402} {\bibfield  {journal} {\bibinfo  {journal} {Physical Review B}\ }\textbf {\bibinfo {volume} {92}},\ \bibinfo {pages} {075402} (\bibinfo {year} {2015})}\BibitemShut {NoStop}%
\bibitem [{\citenamefont {Dai}\ \emph {et~al.}(2016)\citenamefont {Dai}, \citenamefont {Xiang},\ and\ \citenamefont {Srolovitz}}]{Dai2016}%
  \BibitemOpen
  \bibfield  {author} {\bibinfo {author} {\bibfnamefont {S.}~\bibnamefont {Dai}}, \bibinfo {author} {\bibfnamefont {Y.}~\bibnamefont {Xiang}},\ and\ \bibinfo {author} {\bibfnamefont {D.~J.}\ \bibnamefont {Srolovitz}},\ }\bibfield  {title} {\bibinfo {title} {Twisted {Bilayer} {Graphene}: {Moiré} with a {Twist}},\ }\href {https://doi.org/10.1021/acs.nanolett.6b02870} {\bibfield  {journal} {\bibinfo  {journal} {Nano Letters}\ }\textbf {\bibinfo {volume} {16}},\ \bibinfo {pages} {5923} (\bibinfo {year} {2016})}\BibitemShut {NoStop}%
\bibitem [{\citenamefont {Bistritzer}\ and\ \citenamefont {MacDonald}(2011)}]{bistritzer_moire_2011}%
  \BibitemOpen
  \bibfield  {author} {\bibinfo {author} {\bibfnamefont {R.}~\bibnamefont {Bistritzer}}\ and\ \bibinfo {author} {\bibfnamefont {A.~H.}\ \bibnamefont {MacDonald}},\ }\bibfield  {title} {\bibinfo {title} {Moiré bands in twisted double-layer graphene},\ }\href {https://doi.org/10.1073/pnas.1108174108} {\bibfield  {journal} {\bibinfo  {journal} {Proceedings of the National Academy of Sciences}\ }\textbf {\bibinfo {volume} {108}},\ \bibinfo {pages} {12233} (\bibinfo {year} {2011})}\BibitemShut {NoStop}%
\bibitem [{\citenamefont {Suárez~Morell}\ \emph {et~al.}(2010)\citenamefont {Suárez~Morell}, \citenamefont {Correa}, \citenamefont {Vargas}, \citenamefont {Pacheco},\ and\ \citenamefont {Barticevic}}]{SuarezMorell2010}%
  \BibitemOpen
  \bibfield  {author} {\bibinfo {author} {\bibfnamefont {E.}~\bibnamefont {Suárez~Morell}}, \bibinfo {author} {\bibfnamefont {J.~D.}\ \bibnamefont {Correa}}, \bibinfo {author} {\bibfnamefont {P.}~\bibnamefont {Vargas}}, \bibinfo {author} {\bibfnamefont {M.}~\bibnamefont {Pacheco}},\ and\ \bibinfo {author} {\bibfnamefont {Z.}~\bibnamefont {Barticevic}},\ }\bibfield  {title} {\bibinfo {title} {Flat bands in slightly twisted bilayer graphene: {Tight}-binding calculations},\ }\href {https://doi.org/10.1103/PhysRevB.82.121407} {\bibfield  {journal} {\bibinfo  {journal} {Physical Review B}\ }\textbf {\bibinfo {volume} {82}},\ \bibinfo {pages} {121407} (\bibinfo {year} {2010})}\BibitemShut {NoStop}%
\bibitem [{\citenamefont {Lopes~dos Santos}\ \emph {et~al.}(2007)\citenamefont {Lopes~dos Santos}, \citenamefont {Peres},\ and\ \citenamefont {Castro~Neto}}]{lopes_dos_santos_graphene_2007}%
  \BibitemOpen
  \bibfield  {author} {\bibinfo {author} {\bibfnamefont {J.~M.~B.}\ \bibnamefont {Lopes~dos Santos}}, \bibinfo {author} {\bibfnamefont {N.~M.~R.}\ \bibnamefont {Peres}},\ and\ \bibinfo {author} {\bibfnamefont {A.~H.}\ \bibnamefont {Castro~Neto}},\ }\bibfield  {title} {\bibinfo {title} {Graphene {Bilayer} with a {Twist}: {Electronic} {Structure}},\ }\href {https://doi.org/10.1103/PhysRevLett.99.256802} {\bibfield  {journal} {\bibinfo  {journal} {Phys. Rev. Lett.}\ }\textbf {\bibinfo {volume} {99}},\ \bibinfo {pages} {256802} (\bibinfo {year} {2007})}\BibitemShut {NoStop}%
\bibitem [{\citenamefont {Li}\ \emph {et~al.}(2010)\citenamefont {Li}, \citenamefont {Luican}, \citenamefont {{Lopes dos Santos}}, \citenamefont {{Castro Neto}}, \citenamefont {Reina}, \citenamefont {Kong},\ and\ \citenamefont {Andrei}}]{Li2010}%
  \BibitemOpen
  \bibfield  {author} {\bibinfo {author} {\bibfnamefont {G.}~\bibnamefont {Li}}, \bibinfo {author} {\bibfnamefont {A.}~\bibnamefont {Luican}}, \bibinfo {author} {\bibfnamefont {J.~M.~B.}\ \bibnamefont {{Lopes dos Santos}}}, \bibinfo {author} {\bibfnamefont {A.~H.}\ \bibnamefont {{Castro Neto}}}, \bibinfo {author} {\bibfnamefont {A.}~\bibnamefont {Reina}}, \bibinfo {author} {\bibfnamefont {J.}~\bibnamefont {Kong}},\ and\ \bibinfo {author} {\bibfnamefont {E.~Y.}\ \bibnamefont {Andrei}},\ }\bibfield  {title} {\bibinfo {title} {Observation of van hove singularities in twisted graphene layers},\ }\href {https://doi.org/10.1038/nphys1463} {\bibfield  {journal} {\bibinfo  {journal} {Nature Physics}\ }\textbf {\bibinfo {volume} {6}},\ \bibinfo {pages} {109} (\bibinfo {year} {2010})}\BibitemShut {NoStop}%
\bibitem [{\citenamefont {Brihuega}\ \emph {et~al.}(2012)\citenamefont {Brihuega}, \citenamefont {Mallet}, \citenamefont {Gonz\'alez-Herrero}, \citenamefont {Trambly~de Laissardi\`ere}, \citenamefont {Ugeda}, \citenamefont {Magaud}, \citenamefont {G\'omez-Rodr\'{\i}guez}, \citenamefont {Yndur\'ain},\ and\ \citenamefont {Veuillen}}]{Brihuega_VHs_PRL2012}%
  \BibitemOpen
  \bibfield  {author} {\bibinfo {author} {\bibfnamefont {I.}~\bibnamefont {Brihuega}}, \bibinfo {author} {\bibfnamefont {P.}~\bibnamefont {Mallet}}, \bibinfo {author} {\bibfnamefont {H.}~\bibnamefont {Gonz\'alez-Herrero}}, \bibinfo {author} {\bibfnamefont {G.}~\bibnamefont {Trambly~de Laissardi\`ere}}, \bibinfo {author} {\bibfnamefont {M.~M.}\ \bibnamefont {Ugeda}}, \bibinfo {author} {\bibfnamefont {L.}~\bibnamefont {Magaud}}, \bibinfo {author} {\bibfnamefont {J.~M.}\ \bibnamefont {G\'omez-Rodr\'{\i}guez}}, \bibinfo {author} {\bibfnamefont {F.}~\bibnamefont {Yndur\'ain}},\ and\ \bibinfo {author} {\bibfnamefont {J.-Y.}\ \bibnamefont {Veuillen}},\ }\bibfield  {title} {\bibinfo {title} {Unraveling the intrinsic and robust nature of van hove singularities in twisted bilayer graphene by scanning tunneling microscopy and theoretical analysis},\ }\href {https://doi.org/10.1103/PhysRevLett.109.196802} {\bibfield  {journal} {\bibinfo  {journal} {Phys. Rev. Lett.}\ }\textbf {\bibinfo {volume} {109}},\ \bibinfo {pages}
  {196802} (\bibinfo {year} {2012})}\BibitemShut {NoStop}%
\bibitem [{\citenamefont {Yin}\ \emph {et~al.}(2015)\citenamefont {Yin}, \citenamefont {Qiao}, \citenamefont {Wang}, \citenamefont {Zuo}, \citenamefont {Yan}, \citenamefont {Xu}, \citenamefont {Dou}, \citenamefont {Nie},\ and\ \citenamefont {He}}]{Yin2015}%
  \BibitemOpen
  \bibfield  {author} {\bibinfo {author} {\bibfnamefont {L.-J.}\ \bibnamefont {Yin}}, \bibinfo {author} {\bibfnamefont {J.-B.}\ \bibnamefont {Qiao}}, \bibinfo {author} {\bibfnamefont {W.-X.}\ \bibnamefont {Wang}}, \bibinfo {author} {\bibfnamefont {W.-J.}\ \bibnamefont {Zuo}}, \bibinfo {author} {\bibfnamefont {W.}~\bibnamefont {Yan}}, \bibinfo {author} {\bibfnamefont {R.}~\bibnamefont {Xu}}, \bibinfo {author} {\bibfnamefont {R.-F.}\ \bibnamefont {Dou}}, \bibinfo {author} {\bibfnamefont {J.-C.}\ \bibnamefont {Nie}},\ and\ \bibinfo {author} {\bibfnamefont {L.}~\bibnamefont {He}},\ }\bibfield  {title} {\bibinfo {title} {Landau quantization and {Fermi} velocity renormalization in twisted graphene bilayers},\ }\href {https://doi.org/10.1103/PhysRevB.92.201408} {\bibfield  {journal} {\bibinfo  {journal} {Physical Review B}\ }\textbf {\bibinfo {volume} {92}},\ \bibinfo {pages} {201408} (\bibinfo {year} {2015})}\BibitemShut {NoStop}%
\bibitem [{\citenamefont {Mele}(2010)}]{mele_commensuration_2010}%
  \BibitemOpen
  \bibfield  {author} {\bibinfo {author} {\bibfnamefont {E.~J.}\ \bibnamefont {Mele}},\ }\bibfield  {title} {\bibinfo {title} {Commensuration and interlayer coherence in twisted bilayer graphene},\ }\href {https://doi.org/10.1103/PhysRevB.81.161405} {\bibfield  {journal} {\bibinfo  {journal} {Phys. Rev. B}\ }\textbf {\bibinfo {volume} {81}},\ \bibinfo {pages} {161405(R)} (\bibinfo {year} {2010})}\BibitemShut {NoStop}%
\bibitem [{\citenamefont {Koshino}(2015)}]{Koshino2015}%
  \BibitemOpen
  \bibfield  {author} {\bibinfo {author} {\bibfnamefont {M.}~\bibnamefont {Koshino}},\ }\bibfield  {title} {\bibinfo {title} {Interlayer interaction in general incommensurate atomic layers},\ }\href {https://doi.org/10.1088/1367-2630/17/1/015014} {\bibfield  {journal} {\bibinfo  {journal} {New Journal of Physics}\ }\textbf {\bibinfo {volume} {17}},\ \bibinfo {pages} {015014} (\bibinfo {year} {2015})}\BibitemShut {NoStop}%
\bibitem [{\citenamefont {Moon}\ \emph {et~al.}(2014)\citenamefont {Moon}, \citenamefont {Son},\ and\ \citenamefont {Koshino}}]{moon_optical_2014}%
  \BibitemOpen
  \bibfield  {author} {\bibinfo {author} {\bibfnamefont {P.}~\bibnamefont {Moon}}, \bibinfo {author} {\bibfnamefont {Y.-W.}\ \bibnamefont {Son}},\ and\ \bibinfo {author} {\bibfnamefont {M.}~\bibnamefont {Koshino}},\ }\bibfield  {title} {\bibinfo {title} {Optical absorption of twisted bilayer graphene with interlayer potential asymmetry},\ }\href {https://doi.org/10.1103/PhysRevB.90.155427} {\bibfield  {journal} {\bibinfo  {journal} {Phys. Rev. B}\ }\textbf {\bibinfo {volume} {90}},\ \bibinfo {pages} {155427} (\bibinfo {year} {2014})}\BibitemShut {NoStop}%
\bibitem [{\citenamefont {Kazmierczak}\ \emph {et~al.}(2021)\citenamefont {Kazmierczak}, \citenamefont {Winkle}, \citenamefont {Ophus}, \citenamefont {Bustillo}, \citenamefont {Carr}, \citenamefont {Brown}, \citenamefont {Ciston}, \citenamefont {Taniguchi}, \citenamefont {Watanabe},\ and\ \citenamefont {Bediako}}]{kazmierczak_strain_2021}%
  \BibitemOpen
  \bibfield  {author} {\bibinfo {author} {\bibfnamefont {N.~P.}\ \bibnamefont {Kazmierczak}}, \bibinfo {author} {\bibfnamefont {M.~V.}\ \bibnamefont {Winkle}}, \bibinfo {author} {\bibfnamefont {C.}~\bibnamefont {Ophus}}, \bibinfo {author} {\bibfnamefont {K.~C.}\ \bibnamefont {Bustillo}}, \bibinfo {author} {\bibfnamefont {S.}~\bibnamefont {Carr}}, \bibinfo {author} {\bibfnamefont {H.~G.}\ \bibnamefont {Brown}}, \bibinfo {author} {\bibfnamefont {J.}~\bibnamefont {Ciston}}, \bibinfo {author} {\bibfnamefont {T.}~\bibnamefont {Taniguchi}}, \bibinfo {author} {\bibfnamefont {K.}~\bibnamefont {Watanabe}},\ and\ \bibinfo {author} {\bibfnamefont {D.~K.}\ \bibnamefont {Bediako}},\ }\bibfield  {title} {\bibinfo {title} {Strain fields in twisted bilayer graphene},\ }\href {https://doi.org/10.1038/s41563-021-00973-w} {\bibfield  {journal} {\bibinfo  {journal} {Nature Materials}\ }\textbf {\bibinfo {volume} {20}},\ \bibinfo {pages} {956} (\bibinfo {year} {2021})}\BibitemShut {NoStop}%
\bibitem [{\citenamefont {Sinner}\ \emph {et~al.}(2022)\citenamefont {Sinner}, \citenamefont {Pantaleón},\ and\ \citenamefont {Guinea}}]{sinner_strain_2022}%
  \BibitemOpen
  \bibfield  {author} {\bibinfo {author} {\bibfnamefont {A.}~\bibnamefont {Sinner}}, \bibinfo {author} {\bibfnamefont {P.~A.}\ \bibnamefont {Pantaleón}},\ and\ \bibinfo {author} {\bibfnamefont {F.}~\bibnamefont {Guinea}},\ }\bibfield  {title} {\bibinfo {title} {Strain induced quasi-unidimensional channels in twisted moiré lattices},\ }\href {https://doi.org/10.48550/arXiv.2210.07262} {\bibfield  {journal} {\bibinfo  {journal} {arXiv}\ } (\bibinfo {year} {2022})}\BibitemShut {NoStop}%
\bibitem [{\citenamefont {Kögl}\ \emph {et~al.}(2023)\citenamefont {Kögl}, \citenamefont {Soubelet}, \citenamefont {Brotons-Gisbert}, \citenamefont {Stier}, \citenamefont {Gerardot},\ and\ \citenamefont {Finley}}]{kogl_moire_2023}%
  \BibitemOpen
  \bibfield  {author} {\bibinfo {author} {\bibfnamefont {M.}~\bibnamefont {Kögl}}, \bibinfo {author} {\bibfnamefont {P.}~\bibnamefont {Soubelet}}, \bibinfo {author} {\bibfnamefont {M.}~\bibnamefont {Brotons-Gisbert}}, \bibinfo {author} {\bibfnamefont {A.~V.}\ \bibnamefont {Stier}}, \bibinfo {author} {\bibfnamefont {B.~D.}\ \bibnamefont {Gerardot}},\ and\ \bibinfo {author} {\bibfnamefont {J.~J.}\ \bibnamefont {Finley}},\ }\bibfield  {title} {\bibinfo {title} {Moiré straintronics: a universal platform for reconfigurable quantum materials},\ }\bibfield  {journal} {\bibinfo  {journal} {npj 2D Materials and Applications}\ }\textbf {\bibinfo {volume} {7}},\ \href {https://doi.org/10.1038/s41699-023-00382-4} {10.1038/s41699-023-00382-4} (\bibinfo {year} {2023})\BibitemShut {NoStop}%
\bibitem [{\citenamefont {Escudero}\ \emph {et~al.}(2024)\citenamefont {Escudero}, \citenamefont {Sinner}, \citenamefont {Zhan}, \citenamefont {Pantaleón},\ and\ \citenamefont {Guinea}}]{Escudero2024}%
  \BibitemOpen
  \bibfield  {author} {\bibinfo {author} {\bibfnamefont {F.}~\bibnamefont {Escudero}}, \bibinfo {author} {\bibfnamefont {A.}~\bibnamefont {Sinner}}, \bibinfo {author} {\bibfnamefont {Z.}~\bibnamefont {Zhan}}, \bibinfo {author} {\bibfnamefont {P.~A.}\ \bibnamefont {Pantaleón}},\ and\ \bibinfo {author} {\bibfnamefont {F.}~\bibnamefont {Guinea}},\ }\bibfield  {title} {\bibinfo {title} {Designing moiré patterns by strain},\ }\href {https://doi.org/10.1103/physrevresearch.6.023203} {\bibfield  {journal} {\bibinfo  {journal} {Physical Review Research}\ }\textbf {\bibinfo {volume} {6}},\ \bibinfo {pages} {023203} (\bibinfo {year} {2024})}\BibitemShut {NoStop}%
\bibitem [{\citenamefont {Lee}\ \emph {et~al.}(2017)\citenamefont {Lee}, \citenamefont {Woo}, \citenamefont {Park}, \citenamefont {Park}, \citenamefont {Son},\ and\ \citenamefont {Cheong}}]{lee_strain-shear_2017}%
  \BibitemOpen
  \bibfield  {author} {\bibinfo {author} {\bibfnamefont {J.-U.}\ \bibnamefont {Lee}}, \bibinfo {author} {\bibfnamefont {S.}~\bibnamefont {Woo}}, \bibinfo {author} {\bibfnamefont {J.}~\bibnamefont {Park}}, \bibinfo {author} {\bibfnamefont {H.~C.}\ \bibnamefont {Park}}, \bibinfo {author} {\bibfnamefont {Y.-W.}\ \bibnamefont {Son}},\ and\ \bibinfo {author} {\bibfnamefont {H.}~\bibnamefont {Cheong}},\ }\bibfield  {title} {\bibinfo {title} {Strain-shear coupling in bilayer {MoS2}},\ }\bibfield  {journal} {\bibinfo  {journal} {Nature Communications}\ }\textbf {\bibinfo {volume} {8}},\ \href {https://doi.org/10.1038/s41467-017-01487-3} {10.1038/s41467-017-01487-3} (\bibinfo {year} {2017})\BibitemShut {NoStop}%
\bibitem [{\citenamefont {Huder}\ \emph {et~al.}(2018)\citenamefont {Huder}, \citenamefont {Artaud}, \citenamefont {Le~Quang}, \citenamefont {de~Laissardière}, \citenamefont {Jansen}, \citenamefont {Lapertot}, \citenamefont {Chapelier},\ and\ \citenamefont {Renard}}]{huder_electronic_2018}%
  \BibitemOpen
  \bibfield  {author} {\bibinfo {author} {\bibfnamefont {L.}~\bibnamefont {Huder}}, \bibinfo {author} {\bibfnamefont {A.}~\bibnamefont {Artaud}}, \bibinfo {author} {\bibfnamefont {T.}~\bibnamefont {Le~Quang}}, \bibinfo {author} {\bibfnamefont {G.~T.}\ \bibnamefont {de~Laissardière}}, \bibinfo {author} {\bibfnamefont {A.~G.~M.}\ \bibnamefont {Jansen}}, \bibinfo {author} {\bibfnamefont {G.}~\bibnamefont {Lapertot}}, \bibinfo {author} {\bibfnamefont {C.}~\bibnamefont {Chapelier}},\ and\ \bibinfo {author} {\bibfnamefont {V.~T.}\ \bibnamefont {Renard}},\ }\bibfield  {title} {\bibinfo {title} {Electronic {Spectrum} of {Twisted} {Graphene} {Layers} under {Heterostrain}},\ }\href {https://doi.org/10.1103/PhysRevLett.120.156405} {\bibfield  {journal} {\bibinfo  {journal} {Phys. Rev. Lett.}\ }\textbf {\bibinfo {volume} {120}},\ \bibinfo {pages} {156405} (\bibinfo {year} {2018})}\BibitemShut {NoStop}%
\bibitem [{\citenamefont {Bi}\ \emph {et~al.}(2019)\citenamefont {Bi}, \citenamefont {Yuan},\ and\ \citenamefont {Fu}}]{bi_designing_2019}%
  \BibitemOpen
  \bibfield  {author} {\bibinfo {author} {\bibfnamefont {Z.}~\bibnamefont {Bi}}, \bibinfo {author} {\bibfnamefont {N.~F.~Q.}\ \bibnamefont {Yuan}},\ and\ \bibinfo {author} {\bibfnamefont {L.}~\bibnamefont {Fu}},\ }\bibfield  {title} {\bibinfo {title} {Designing flat bands by strain},\ }\bibfield  {journal} {\bibinfo  {journal} {Physical Review B}\ }\textbf {\bibinfo {volume} {100}},\ \href {https://doi.org/10.1103/physrevb.100.035448} {10.1103/physrevb.100.035448} (\bibinfo {year} {2019})\BibitemShut {NoStop}%
\bibitem [{\citenamefont {Mesple}\ \emph {et~al.}(2021)\citenamefont {Mesple}, \citenamefont {Missaoui}, \citenamefont {Cea}, \citenamefont {Huder}, \citenamefont {Guinea}, \citenamefont {Trambly~de Laissardière}, \citenamefont {Chapelier},\ and\ \citenamefont {Renard}}]{mesple_heterostrain_2021}%
  \BibitemOpen
  \bibfield  {author} {\bibinfo {author} {\bibfnamefont {F.}~\bibnamefont {Mesple}}, \bibinfo {author} {\bibfnamefont {A.}~\bibnamefont {Missaoui}}, \bibinfo {author} {\bibfnamefont {T.}~\bibnamefont {Cea}}, \bibinfo {author} {\bibfnamefont {L.}~\bibnamefont {Huder}}, \bibinfo {author} {\bibfnamefont {F.}~\bibnamefont {Guinea}}, \bibinfo {author} {\bibfnamefont {G.}~\bibnamefont {Trambly~de Laissardière}}, \bibinfo {author} {\bibfnamefont {C.}~\bibnamefont {Chapelier}},\ and\ \bibinfo {author} {\bibfnamefont {V.~T.}\ \bibnamefont {Renard}},\ }\bibfield  {title} {\bibinfo {title} {Heterostrain {Determines} {Flat} {Bands} in {Magic}-{Angle} {Twisted} {Graphene} {Layers}},\ }\href {https://doi.org/10.1103/PhysRevLett.127.126405} {\bibfield  {journal} {\bibinfo  {journal} {Phys. Rev. Lett.}\ }\textbf {\bibinfo {volume} {127}},\ \bibinfo {pages} {126405} (\bibinfo {year} {2021})}\BibitemShut {NoStop}%
\bibitem [{\citenamefont {Wang}\ \emph {et~al.}(2023)\citenamefont {Wang}, \citenamefont {Finney}, \citenamefont {Sharpe}, \citenamefont {Rodenbach}, \citenamefont {Hsueh}, \citenamefont {Watanabe}, \citenamefont {Taniguchi}, \citenamefont {Kastner}, \citenamefont {Vafek},\ and\ \citenamefont {Goldhaber-Gordon}}]{wang_unusual_2023}%
  \BibitemOpen
  \bibfield  {author} {\bibinfo {author} {\bibfnamefont {X.}~\bibnamefont {Wang}}, \bibinfo {author} {\bibfnamefont {J.}~\bibnamefont {Finney}}, \bibinfo {author} {\bibfnamefont {A.~L.}\ \bibnamefont {Sharpe}}, \bibinfo {author} {\bibfnamefont {L.~K.}\ \bibnamefont {Rodenbach}}, \bibinfo {author} {\bibfnamefont {C.~L.}\ \bibnamefont {Hsueh}}, \bibinfo {author} {\bibfnamefont {K.}~\bibnamefont {Watanabe}}, \bibinfo {author} {\bibfnamefont {T.}~\bibnamefont {Taniguchi}}, \bibinfo {author} {\bibfnamefont {M.~A.}\ \bibnamefont {Kastner}}, \bibinfo {author} {\bibfnamefont {O.}~\bibnamefont {Vafek}},\ and\ \bibinfo {author} {\bibfnamefont {D.}~\bibnamefont {Goldhaber-Gordon}},\ }\bibfield  {title} {\bibinfo {title} {Unusual magnetotransport in twisted bilayer graphene from strain-induced open {Fermi} surfaces},\ }\bibfield  {journal} {\bibinfo  {journal} {Proceedings of the National Academy of Sciences}\ }\textbf {\bibinfo {volume} {120}},\ \href {https://doi.org/10.1073/pnas.2307151120} {10.1073/pnas.2307151120}
  (\bibinfo {year} {2023})\BibitemShut {NoStop}%
\bibitem [{\citenamefont {Mesple}\ \emph {et~al.}(2023)\citenamefont {Mesple}, \citenamefont {Walet}, \citenamefont {Trambly~de Laissardière}, \citenamefont {Guinea}, \citenamefont {Došenović}, \citenamefont {Okuno}, \citenamefont {Paillet}, \citenamefont {Michon}, \citenamefont {Chapelier},\ and\ \citenamefont {Renard}}]{Mesple2023}%
  \BibitemOpen
  \bibfield  {author} {\bibinfo {author} {\bibfnamefont {F.}~\bibnamefont {Mesple}}, \bibinfo {author} {\bibfnamefont {N.~R.}\ \bibnamefont {Walet}}, \bibinfo {author} {\bibfnamefont {G.}~\bibnamefont {Trambly~de Laissardière}}, \bibinfo {author} {\bibfnamefont {F.}~\bibnamefont {Guinea}}, \bibinfo {author} {\bibfnamefont {D.}~\bibnamefont {Došenović}}, \bibinfo {author} {\bibfnamefont {H.}~\bibnamefont {Okuno}}, \bibinfo {author} {\bibfnamefont {C.}~\bibnamefont {Paillet}}, \bibinfo {author} {\bibfnamefont {A.}~\bibnamefont {Michon}}, \bibinfo {author} {\bibfnamefont {C.}~\bibnamefont {Chapelier}},\ and\ \bibinfo {author} {\bibfnamefont {V.~T.}\ \bibnamefont {Renard}},\ }\bibfield  {title} {\bibinfo {title} {Giant {Atomic} {Swirl} in {Graphene} {Bilayers} with {Biaxial} {Heterostrain}},\ }\href {https://doi.org/10.1002/adma.202306312} {\bibfield  {journal} {\bibinfo  {journal} {Advanced Materials}\ }\textbf {\bibinfo {volume} {35}},\ \bibinfo {pages} {2306312} (\bibinfo {year} {2023})}\BibitemShut
  {NoStop}%
\bibitem [{\citenamefont {Jiang}\ \emph {et~al.}(2017)\citenamefont {Jiang}, \citenamefont {Mao}, \citenamefont {Duan}, \citenamefont {Lai}, \citenamefont {Watanabe}, \citenamefont {Taniguchi},\ and\ \citenamefont {Andrei}}]{jiang_visualizing_2017}%
  \BibitemOpen
  \bibfield  {author} {\bibinfo {author} {\bibfnamefont {Y.}~\bibnamefont {Jiang}}, \bibinfo {author} {\bibfnamefont {J.}~\bibnamefont {Mao}}, \bibinfo {author} {\bibfnamefont {J.}~\bibnamefont {Duan}}, \bibinfo {author} {\bibfnamefont {X.}~\bibnamefont {Lai}}, \bibinfo {author} {\bibfnamefont {K.}~\bibnamefont {Watanabe}}, \bibinfo {author} {\bibfnamefont {T.}~\bibnamefont {Taniguchi}},\ and\ \bibinfo {author} {\bibfnamefont {E.~Y.}\ \bibnamefont {Andrei}},\ }\bibfield  {title} {\bibinfo {title} {Visualizing {Strain}-{Induced} {Pseudomagnetic} {Fields} in {Graphene} through an {hBN} {Magnifying} {Glass}},\ }\href {https://doi.org/10.1021/acs.nanolett.6b05228} {\bibfield  {journal} {\bibinfo  {journal} {Nano Letters}\ }\textbf {\bibinfo {volume} {17}},\ \bibinfo {pages} {2839} (\bibinfo {year} {2017})}\BibitemShut {NoStop}%
\bibitem [{\citenamefont {Brzhezinskaya}\ \emph {et~al.}(2021)\citenamefont {Brzhezinskaya}, \citenamefont {Kononenko}, \citenamefont {Matveev}, \citenamefont {Zotov}, \citenamefont {Khodos}, \citenamefont {Levashov}, \citenamefont {Volkov}, \citenamefont {Bozhko}, \citenamefont {Chekmazov},\ and\ \citenamefont {Roshchupkin}}]{Brzhezinskaya2021}%
  \BibitemOpen
  \bibfield  {author} {\bibinfo {author} {\bibfnamefont {M.}~\bibnamefont {Brzhezinskaya}}, \bibinfo {author} {\bibfnamefont {O.}~\bibnamefont {Kononenko}}, \bibinfo {author} {\bibfnamefont {V.}~\bibnamefont {Matveev}}, \bibinfo {author} {\bibfnamefont {A.}~\bibnamefont {Zotov}}, \bibinfo {author} {\bibfnamefont {I.~I.}\ \bibnamefont {Khodos}}, \bibinfo {author} {\bibfnamefont {V.}~\bibnamefont {Levashov}}, \bibinfo {author} {\bibfnamefont {V.}~\bibnamefont {Volkov}}, \bibinfo {author} {\bibfnamefont {S.~I.}\ \bibnamefont {Bozhko}}, \bibinfo {author} {\bibfnamefont {S.~V.}\ \bibnamefont {Chekmazov}},\ and\ \bibinfo {author} {\bibfnamefont {D.}~\bibnamefont {Roshchupkin}},\ }\bibfield  {title} {\bibinfo {title} {Engineering of {Numerous} {Moiré} {Superlattices} in {Twisted} {Multilayer} {Graphene} for {Twistronics} and {Straintronics} {Applications}},\ }\href {https://doi.org/10.1021/acsnano.1c04286} {\bibfield  {journal} {\bibinfo  {journal} {ACS Nano}\ }\textbf {\bibinfo {volume} {15}},\ \bibinfo {pages}
  {12358} (\bibinfo {year} {2021})}\BibitemShut {NoStop}%
\bibitem [{\citenamefont {Kapfer}\ \emph {et~al.}(2023)\citenamefont {Kapfer}, \citenamefont {Jessen}, \citenamefont {Eisele}, \citenamefont {Fu}, \citenamefont {Danielsen}, \citenamefont {Darlington}, \citenamefont {Moore}, \citenamefont {Finney}, \citenamefont {Marchese}, \citenamefont {Hsieh}, \citenamefont {Majchrzak}, \citenamefont {Jiang}, \citenamefont {Biswas}, \citenamefont {Dudin}, \citenamefont {Avila}, \citenamefont {Watanabe}, \citenamefont {Taniguchi}, \citenamefont {Ulstrup}, \citenamefont {Bøggild}, \citenamefont {Schuck}, \citenamefont {Basov}, \citenamefont {Hone},\ and\ \citenamefont {Dean}}]{kapfer_programming_2023}%
  \BibitemOpen
  \bibfield  {author} {\bibinfo {author} {\bibfnamefont {M.}~\bibnamefont {Kapfer}}, \bibinfo {author} {\bibfnamefont {B.~S.}\ \bibnamefont {Jessen}}, \bibinfo {author} {\bibfnamefont {M.~E.}\ \bibnamefont {Eisele}}, \bibinfo {author} {\bibfnamefont {M.}~\bibnamefont {Fu}}, \bibinfo {author} {\bibfnamefont {D.~R.}\ \bibnamefont {Danielsen}}, \bibinfo {author} {\bibfnamefont {T.~P.}\ \bibnamefont {Darlington}}, \bibinfo {author} {\bibfnamefont {S.~L.}\ \bibnamefont {Moore}}, \bibinfo {author} {\bibfnamefont {N.~R.}\ \bibnamefont {Finney}}, \bibinfo {author} {\bibfnamefont {A.}~\bibnamefont {Marchese}}, \bibinfo {author} {\bibfnamefont {V.}~\bibnamefont {Hsieh}}, \bibinfo {author} {\bibfnamefont {P.}~\bibnamefont {Majchrzak}}, \bibinfo {author} {\bibfnamefont {Z.}~\bibnamefont {Jiang}}, \bibinfo {author} {\bibfnamefont {D.}~\bibnamefont {Biswas}}, \bibinfo {author} {\bibfnamefont {P.}~\bibnamefont {Dudin}}, \bibinfo {author} {\bibfnamefont {J.}~\bibnamefont {Avila}}, \bibinfo {author} {\bibfnamefont
  {K.}~\bibnamefont {Watanabe}}, \bibinfo {author} {\bibfnamefont {T.}~\bibnamefont {Taniguchi}}, \bibinfo {author} {\bibfnamefont {S.}~\bibnamefont {Ulstrup}}, \bibinfo {author} {\bibfnamefont {P.}~\bibnamefont {Bøggild}}, \bibinfo {author} {\bibfnamefont {P.~J.}\ \bibnamefont {Schuck}}, \bibinfo {author} {\bibfnamefont {D.~N.}\ \bibnamefont {Basov}}, \bibinfo {author} {\bibfnamefont {J.}~\bibnamefont {Hone}},\ and\ \bibinfo {author} {\bibfnamefont {C.~R.}\ \bibnamefont {Dean}},\ }\bibfield  {title} {\bibinfo {title} {Programming twist angle and strain profiles in {2D} materials},\ }\href {https://doi.org/10.1126/science.ade9995} {\bibfield  {journal} {\bibinfo  {journal} {Science}\ }\textbf {\bibinfo {volume} {381}},\ \bibinfo {pages} {677} (\bibinfo {year} {2023})}\BibitemShut {NoStop}%
\bibitem [{\citenamefont {Peña}\ \emph {et~al.}(2023)\citenamefont {Peña}, \citenamefont {Dey}, \citenamefont {Chowdhury}, \citenamefont {Azizimanesh}, \citenamefont {Hou}, \citenamefont {Sewaket}, \citenamefont {Watson}, \citenamefont {Askari},\ and\ \citenamefont {Wu}}]{pena_moire_2023}%
  \BibitemOpen
  \bibfield  {author} {\bibinfo {author} {\bibfnamefont {T.}~\bibnamefont {Peña}}, \bibinfo {author} {\bibfnamefont {A.}~\bibnamefont {Dey}}, \bibinfo {author} {\bibfnamefont {S.~A.}\ \bibnamefont {Chowdhury}}, \bibinfo {author} {\bibfnamefont {A.}~\bibnamefont {Azizimanesh}}, \bibinfo {author} {\bibfnamefont {W.}~\bibnamefont {Hou}}, \bibinfo {author} {\bibfnamefont {A.}~\bibnamefont {Sewaket}}, \bibinfo {author} {\bibfnamefont {C.}~\bibnamefont {Watson}}, \bibinfo {author} {\bibfnamefont {H.}~\bibnamefont {Askari}},\ and\ \bibinfo {author} {\bibfnamefont {S.~M.}\ \bibnamefont {Wu}},\ }\bibfield  {title} {\bibinfo {title} {Moiré engineering in {2D} heterostructures with process-induced strain},\ }\href {https://doi.org/10.1063/5.0142406} {\bibfield  {journal} {\bibinfo  {journal} {Applied Physics Letters}\ }\textbf {\bibinfo {volume} {122}},\ \bibinfo {pages} {143101} (\bibinfo {year} {2023})}\BibitemShut {NoStop}%
\bibitem [{\citenamefont {Sequeira}\ \emph {et~al.}(2024)\citenamefont {Sequeira}, \citenamefont {Barabas}, \citenamefont {Barajas-Aguilar}, \citenamefont {Bacani}, \citenamefont {Nakatsuji}, \citenamefont {Koshino}, \citenamefont {Taniguichi}, \citenamefont {Watanabe},\ and\ \citenamefont {Sanchez-Yamagishi}}]{Sequeira2024}%
  \BibitemOpen
  \bibfield  {author} {\bibinfo {author} {\bibfnamefont {I.}~\bibnamefont {Sequeira}}, \bibinfo {author} {\bibfnamefont {A.~Z.}\ \bibnamefont {Barabas}}, \bibinfo {author} {\bibfnamefont {A.~H.}\ \bibnamefont {Barajas-Aguilar}}, \bibinfo {author} {\bibfnamefont {M.~G.}\ \bibnamefont {Bacani}}, \bibinfo {author} {\bibfnamefont {N.}~\bibnamefont {Nakatsuji}}, \bibinfo {author} {\bibfnamefont {M.}~\bibnamefont {Koshino}}, \bibinfo {author} {\bibfnamefont {T.}~\bibnamefont {Taniguichi}}, \bibinfo {author} {\bibfnamefont {K.}~\bibnamefont {Watanabe}},\ and\ \bibinfo {author} {\bibfnamefont {J.~D.}\ \bibnamefont {Sanchez-Yamagishi}},\ }\bibfield  {title} {\bibinfo {title} {Manipulating {Moirés} by {Controlling} {Heterostrain} in van der {Waals} {Devices}},\ }\href {https://doi.org/10.1021/acs.nanolett.4c04201} {\bibfield  {journal} {\bibinfo  {journal} {Nano Letters}\ }\textbf {\bibinfo {volume} {24}},\ \bibinfo {pages} {15662} (\bibinfo {year} {2024})}\BibitemShut {NoStop}%
\bibitem [{\citenamefont {Huang}\ and\ \citenamefont {Liew}(2025)}]{Huang2025}%
  \BibitemOpen
  \bibfield  {author} {\bibinfo {author} {\bibfnamefont {Z.-C.}\ \bibnamefont {Huang}}\ and\ \bibinfo {author} {\bibfnamefont {K.~M.}\ \bibnamefont {Liew}},\ }\bibfield  {title} {\bibinfo {title} {Strain {Engineering} towards {Enriched} {Surface} {Patterns} in {Graphene} {Twistronics}},\ }\href {https://doi.org/10.1021/acsami.5c00840} {\bibfield  {journal} {\bibinfo  {journal} {ACS Applied Materials \& Interfaces}\ }\textbf {\bibinfo {volume} {17}},\ \bibinfo {pages} {17622} (\bibinfo {year} {2025})}\BibitemShut {NoStop}%
\bibitem [{\citenamefont {Tilak}\ \emph {et~al.}(2022)\citenamefont {Tilak}, \citenamefont {Li}, \citenamefont {Taniguchi}, \citenamefont {Watanabe},\ and\ \citenamefont {Andrei}}]{tilak_moire_2022-1}%
  \BibitemOpen
  \bibfield  {author} {\bibinfo {author} {\bibfnamefont {N.}~\bibnamefont {Tilak}}, \bibinfo {author} {\bibfnamefont {G.}~\bibnamefont {Li}}, \bibinfo {author} {\bibfnamefont {T.}~\bibnamefont {Taniguchi}}, \bibinfo {author} {\bibfnamefont {K.}~\bibnamefont {Watanabe}},\ and\ \bibinfo {author} {\bibfnamefont {E.~Y.}\ \bibnamefont {Andrei}},\ }\bibfield  {title} {\bibinfo {title} {Moiré {Potential}, {Lattice} {Relaxation}, and {Layer} {Polarization} in {Marginally} {Twisted} {MoS}\_2 {Bilayers}},\ }\href {https://doi.org/10.1021/acs.nanolett.2c03676} {\bibfield  {journal} {\bibinfo  {journal} {Nano Letters}\ }\textbf {\bibinfo {volume} {23}},\ \bibinfo {pages} {73} (\bibinfo {year} {2022})}\BibitemShut {NoStop}%
\bibitem [{\citenamefont {Boi}\ \emph {et~al.}(2025)\citenamefont {Boi}, \citenamefont {Odunmbaku}, \citenamefont {Taallah},\ and\ \citenamefont {Wang}}]{Boi2025}%
  \BibitemOpen
  \bibfield  {author} {\bibinfo {author} {\bibfnamefont {F.~S.}\ \bibnamefont {Boi}}, \bibinfo {author} {\bibfnamefont {O.}~\bibnamefont {Odunmbaku}}, \bibinfo {author} {\bibfnamefont {A.}~\bibnamefont {Taallah}},\ and\ \bibinfo {author} {\bibfnamefont {S.}~\bibnamefont {Wang}},\ }\bibfield  {title} {\bibinfo {title} {Quasi-{1D}, rectangular-like and hexagonal moiré superlattices in exfoliated highly oriented pyrolytic graphite},\ }\href {https://doi.org/10.1016/j.diamond.2024.111843} {\bibfield  {journal} {\bibinfo  {journal} {Diamond and Related Materials}\ }\textbf {\bibinfo {volume} {151}},\ \bibinfo {pages} {111843} (\bibinfo {year} {2025})}\BibitemShut {NoStop}%
\bibitem [{\citenamefont {Ou}\ \emph {et~al.}(2025)\citenamefont {Ou}, \citenamefont {Oi}, \citenamefont {Usami}, \citenamefont {Endo}, \citenamefont {Shinokita}, \citenamefont {Kitaura}, \citenamefont {Matsuda}, \citenamefont {Miyata}, \citenamefont {Pu},\ and\ \citenamefont {Takenobu}}]{Ou2025}%
  \BibitemOpen
  \bibfield  {author} {\bibinfo {author} {\bibfnamefont {H.}~\bibnamefont {Ou}}, \bibinfo {author} {\bibfnamefont {K.}~\bibnamefont {Oi}}, \bibinfo {author} {\bibfnamefont {R.}~\bibnamefont {Usami}}, \bibinfo {author} {\bibfnamefont {T.}~\bibnamefont {Endo}}, \bibinfo {author} {\bibfnamefont {K.}~\bibnamefont {Shinokita}}, \bibinfo {author} {\bibfnamefont {R.}~\bibnamefont {Kitaura}}, \bibinfo {author} {\bibfnamefont {K.}~\bibnamefont {Matsuda}}, \bibinfo {author} {\bibfnamefont {Y.}~\bibnamefont {Miyata}}, \bibinfo {author} {\bibfnamefont {J.}~\bibnamefont {Pu}},\ and\ \bibinfo {author} {\bibfnamefont {T.}~\bibnamefont {Takenobu}},\ }\bibfield  {title} {\bibinfo {title} {Continuous {Strain} {Modulation} of {Moiré} {Superlattice} {Symmetry} {From} {Triangle} to {Rectangle}},\ }\href {https://doi.org/10.1002/smll.202407316} {\bibfield  {journal} {\bibinfo  {journal} {Small}\ }\textbf {\bibinfo {volume} {21}},\ \bibinfo {pages} {2407316} (\bibinfo {year} {2025})}\BibitemShut {NoStop}%
\bibitem [{\citenamefont {Varchon}\ \emph {et~al.}(2008)\citenamefont {Varchon}, \citenamefont {Mallet}, \citenamefont {Magaud},\ and\ \citenamefont {Veuillen}}]{Varchon_GrapheneGrowth_PRB2008}%
  \BibitemOpen
  \bibfield  {author} {\bibinfo {author} {\bibfnamefont {F.}~\bibnamefont {Varchon}}, \bibinfo {author} {\bibfnamefont {P.}~\bibnamefont {Mallet}}, \bibinfo {author} {\bibfnamefont {L.}~\bibnamefont {Magaud}},\ and\ \bibinfo {author} {\bibfnamefont {J.-Y.}\ \bibnamefont {Veuillen}},\ }\bibfield  {title} {\bibinfo {title} {Rotational disorder in few-layer graphene films on $6h\text{\ensuremath{-}}\mathrm{Si}\mathrm{C}(000\text{\ensuremath{-}}1)$: A scanning tunneling microscopy study},\ }\href {https://doi.org/10.1103/PhysRevB.77.165415} {\bibfield  {journal} {\bibinfo  {journal} {Phys. Rev. B}\ }\textbf {\bibinfo {volume} {77}},\ \bibinfo {pages} {165415} (\bibinfo {year} {2008})}\BibitemShut {NoStop}%
\bibitem [{SM()}]{SM}%
  \BibitemOpen
  \href@noop {} {}\bibinfo {note} {See supplementary information which includes Refs.\cite{Varchon_GrapheneGrowth_PRB2008, Brihuega_VHs_PRL2012, Faugeras2008, Norimatsu2014, Yazdi2016, Mishra2016, Beechem2014, Sprinkle_PRL2009, Rosenberger2020, Halbertal2022, jong_imaging_2022, Horcas2007, Andrei2020, kogl_moire_2023, Escudero2024, sinner_strain_2022, lopes_dos_santos_graphene_2007, bistritzer_moire_2011, moon_optical_2014, Koshino2015, bi_designing_2019, suzuura_phonons_2002, vozmediano_gauge_2010, choi_effects_2010, long_atomistic_2022, naumis_electronic_2017, Koshino2018, guinea_electrostatic_2018, cea_electrostatic_2022, Cea2019, Pantaleon2021, Nam2017, guinea_continuum_2019, Tersoff1985, Chen1988}}\BibitemShut {NoStop}%
\bibitem [{\citenamefont {Liu}\ \emph {et~al.}(2011)\citenamefont {Liu}, \citenamefont {Pan}, \citenamefont {Fu}, \citenamefont {Zhang}, \citenamefont {Dai},\ and\ \citenamefont {Liu}}]{Liu2011}%
  \BibitemOpen
  \bibfield  {author} {\bibinfo {author} {\bibfnamefont {N.}~\bibnamefont {Liu}}, \bibinfo {author} {\bibfnamefont {Z.}~\bibnamefont {Pan}}, \bibinfo {author} {\bibfnamefont {L.}~\bibnamefont {Fu}}, \bibinfo {author} {\bibfnamefont {C.}~\bibnamefont {Zhang}}, \bibinfo {author} {\bibfnamefont {B.}~\bibnamefont {Dai}},\ and\ \bibinfo {author} {\bibfnamefont {Z.}~\bibnamefont {Liu}},\ }\bibfield  {title} {\bibinfo {title} {The origin of wrinkles on transferred graphene},\ }\href {https://doi.org/10.1007/s12274-011-0156-3} {\bibfield  {journal} {\bibinfo  {journal} {Nano Research}\ }\textbf {\bibinfo {volume} {4}},\ \bibinfo {pages} {996} (\bibinfo {year} {2011})}\BibitemShut {NoStop}%
\bibitem [{\citenamefont {Wang}\ \emph {et~al.}(2013)\citenamefont {Wang}, \citenamefont {Liu}, \citenamefont {Lan},\ and\ \citenamefont {Tan}}]{Wang2013}%
  \BibitemOpen
  \bibfield  {author} {\bibinfo {author} {\bibfnamefont {C.}~\bibnamefont {Wang}}, \bibinfo {author} {\bibfnamefont {Y.}~\bibnamefont {Liu}}, \bibinfo {author} {\bibfnamefont {L.}~\bibnamefont {Lan}},\ and\ \bibinfo {author} {\bibfnamefont {H.}~\bibnamefont {Tan}},\ }\bibfield  {title} {\bibinfo {title} {Graphene wrinkling: formation, evolution and collapse},\ }\href {https://doi.org/10.1039/C3NR00462G} {\bibfield  {journal} {\bibinfo  {journal} {Nanoscale}\ }\textbf {\bibinfo {volume} {5}},\ \bibinfo {pages} {4454} (\bibinfo {year} {2013})}\BibitemShut {NoStop}%
\bibitem [{\citenamefont {Meng}\ \emph {et~al.}(2013)\citenamefont {Meng}, \citenamefont {Su}, \citenamefont {Geng}, \citenamefont {Yu}, \citenamefont {Liu}, \citenamefont {Dou}, \citenamefont {Nie},\ and\ \citenamefont {He}}]{Meng2013}%
  \BibitemOpen
  \bibfield  {author} {\bibinfo {author} {\bibfnamefont {L.}~\bibnamefont {Meng}}, \bibinfo {author} {\bibfnamefont {Y.}~\bibnamefont {Su}}, \bibinfo {author} {\bibfnamefont {D.}~\bibnamefont {Geng}}, \bibinfo {author} {\bibfnamefont {G.}~\bibnamefont {Yu}}, \bibinfo {author} {\bibfnamefont {Y.}~\bibnamefont {Liu}}, \bibinfo {author} {\bibfnamefont {R.-F.}\ \bibnamefont {Dou}}, \bibinfo {author} {\bibfnamefont {J.-C.}\ \bibnamefont {Nie}},\ and\ \bibinfo {author} {\bibfnamefont {L.}~\bibnamefont {He}},\ }\bibfield  {title} {\bibinfo {title} {Hierarchy of graphene wrinkles induced by thermal strain engineering},\ }\href {https://doi.org/10.1063/1.4857115} {\bibfield  {journal} {\bibinfo  {journal} {Applied Physics Letters}\ }\textbf {\bibinfo {volume} {103}},\ \bibinfo {pages} {251610} (\bibinfo {year} {2013})}\BibitemShut {NoStop}%
\bibitem [{\citenamefont {Deng}\ and\ \citenamefont {Berry}(2016)}]{Deng2016}%
  \BibitemOpen
  \bibfield  {author} {\bibinfo {author} {\bibfnamefont {S.}~\bibnamefont {Deng}}\ and\ \bibinfo {author} {\bibfnamefont {V.}~\bibnamefont {Berry}},\ }\bibfield  {title} {\bibinfo {title} {Wrinkled, rippled and crumpled graphene: an overview of formation mechanism, electronic properties, and applications},\ }\href {https://doi.org/10.1016/j.mattod.2015.10.002} {\bibfield  {journal} {\bibinfo  {journal} {Materials Today}\ }\textbf {\bibinfo {volume} {19}},\ \bibinfo {pages} {197} (\bibinfo {year} {2016})}\BibitemShut {NoStop}%
\bibitem [{\citenamefont {Beechem}\ \emph {et~al.}(2014)\citenamefont {Beechem}, \citenamefont {Ohta}, \citenamefont {Diaconescu},\ and\ \citenamefont {Robinson}}]{Beechem2014}%
  \BibitemOpen
  \bibfield  {author} {\bibinfo {author} {\bibfnamefont {T.~E.}\ \bibnamefont {Beechem}}, \bibinfo {author} {\bibfnamefont {T.}~\bibnamefont {Ohta}}, \bibinfo {author} {\bibfnamefont {B.}~\bibnamefont {Diaconescu}},\ and\ \bibinfo {author} {\bibfnamefont {J.~T.}\ \bibnamefont {Robinson}},\ }\bibfield  {title} {\bibinfo {title} {Rotational {Disorder} in {Twisted} {Bilayer} {Graphene}},\ }\href {https://doi.org/10.1021/nn405999z} {\bibfield  {journal} {\bibinfo  {journal} {ACS Nano}\ }\textbf {\bibinfo {volume} {8}},\ \bibinfo {pages} {1655} (\bibinfo {year} {2014})}\BibitemShut {NoStop}%
\bibitem [{\citenamefont {Carr}\ \emph {et~al.}(2017)\citenamefont {Carr}, \citenamefont {Massatt}, \citenamefont {Fang}, \citenamefont {Cazeaux}, \citenamefont {Luskin},\ and\ \citenamefont {Kaxiras}}]{Carr2017}%
  \BibitemOpen
  \bibfield  {author} {\bibinfo {author} {\bibfnamefont {S.}~\bibnamefont {Carr}}, \bibinfo {author} {\bibfnamefont {D.}~\bibnamefont {Massatt}}, \bibinfo {author} {\bibfnamefont {S.}~\bibnamefont {Fang}}, \bibinfo {author} {\bibfnamefont {P.}~\bibnamefont {Cazeaux}}, \bibinfo {author} {\bibfnamefont {M.}~\bibnamefont {Luskin}},\ and\ \bibinfo {author} {\bibfnamefont {E.}~\bibnamefont {Kaxiras}},\ }\bibfield  {title} {\bibinfo {title} {Twistronics: {Manipulating} the electronic properties of two-dimensional layered structures through their twist angle},\ }\href {https://doi.org/10.1103/PhysRevB.95.075420} {\bibfield  {journal} {\bibinfo  {journal} {Physical Review B}\ }\textbf {\bibinfo {volume} {95}},\ \bibinfo {pages} {075420} (\bibinfo {year} {2017})}\BibitemShut {NoStop}%
\bibitem [{\citenamefont {Yang}\ \emph {et~al.}(2020)\citenamefont {Yang}, \citenamefont {Li}, \citenamefont {Yin}, \citenamefont {Xu}, \citenamefont {Mullan}, \citenamefont {Taniguchi}, \citenamefont {Watanabe}, \citenamefont {Geim}, \citenamefont {Novoselov},\ and\ \citenamefont {Mishchenko}}]{Yang2020}%
  \BibitemOpen
  \bibfield  {author} {\bibinfo {author} {\bibfnamefont {Y.}~\bibnamefont {Yang}}, \bibinfo {author} {\bibfnamefont {J.}~\bibnamefont {Li}}, \bibinfo {author} {\bibfnamefont {J.}~\bibnamefont {Yin}}, \bibinfo {author} {\bibfnamefont {S.}~\bibnamefont {Xu}}, \bibinfo {author} {\bibfnamefont {C.}~\bibnamefont {Mullan}}, \bibinfo {author} {\bibfnamefont {T.}~\bibnamefont {Taniguchi}}, \bibinfo {author} {\bibfnamefont {K.}~\bibnamefont {Watanabe}}, \bibinfo {author} {\bibfnamefont {A.~K.}\ \bibnamefont {Geim}}, \bibinfo {author} {\bibfnamefont {K.~S.}\ \bibnamefont {Novoselov}},\ and\ \bibinfo {author} {\bibfnamefont {A.}~\bibnamefont {Mishchenko}},\ }\bibfield  {title} {\bibinfo {title} {In situ manipulation of van der {Waals} heterostructures for twistronics},\ }\href {https://doi.org/10.1126/sciadv.abd3655} {\bibfield  {journal} {\bibinfo  {journal} {Science Advances}\ }\textbf {\bibinfo {volume} {6}},\ \bibinfo {pages} {eabd3655} (\bibinfo {year} {2020})}\BibitemShut {NoStop}%
\bibitem [{\citenamefont {Hennighausen}\ and\ \citenamefont {Kar}(2021)}]{Hennighausen2021}%
  \BibitemOpen
  \bibfield  {author} {\bibinfo {author} {\bibfnamefont {Z.}~\bibnamefont {Hennighausen}}\ and\ \bibinfo {author} {\bibfnamefont {S.}~\bibnamefont {Kar}},\ }\bibfield  {title} {\bibinfo {title} {Twistronics: a turning point in {2D} quantum materials},\ }\href {https://doi.org/10.1088/2516-1075/abd957} {\bibfield  {journal} {\bibinfo  {journal} {Electronic Structure}\ }\textbf {\bibinfo {volume} {3}},\ \bibinfo {pages} {014004} (\bibinfo {year} {2021})}\BibitemShut {NoStop}%
\bibitem [{\citenamefont {Guinea}(2012)}]{guinea_strain_2012}%
  \BibitemOpen
  \bibfield  {author} {\bibinfo {author} {\bibfnamefont {F.}~\bibnamefont {Guinea}},\ }\bibfield  {title} {\bibinfo {title} {Strain engineering in graphene},\ }\href {https://doi.org/10.1016/j.ssc.2012.04.019} {\bibfield  {journal} {\bibinfo  {journal} {Solid State Communications}\ }\textbf {\bibinfo {volume} {152}},\ \bibinfo {pages} {1437} (\bibinfo {year} {2012})}\BibitemShut {NoStop}%
\bibitem [{\citenamefont {Amorim}\ \emph {et~al.}(2016)\citenamefont {Amorim}, \citenamefont {Cortijo}, \citenamefont {Juan}, \citenamefont {Grushin}, \citenamefont {Guinea}, \citenamefont {Gutiérrez-Rubio}, \citenamefont {Ochoa}, \citenamefont {Parente}, \citenamefont {Roldán}, \citenamefont {San-Jose}, \citenamefont {Schiefele}, \citenamefont {Sturla},\ and\ \citenamefont {Vozmediano}}]{amorim_novel_2016}%
  \BibitemOpen
  \bibfield  {author} {\bibinfo {author} {\bibfnamefont {B.}~\bibnamefont {Amorim}}, \bibinfo {author} {\bibfnamefont {A.}~\bibnamefont {Cortijo}}, \bibinfo {author} {\bibfnamefont {F.~d.}\ \bibnamefont {Juan}}, \bibinfo {author} {\bibfnamefont {A.~G.}\ \bibnamefont {Grushin}}, \bibinfo {author} {\bibfnamefont {F.}~\bibnamefont {Guinea}}, \bibinfo {author} {\bibfnamefont {A.}~\bibnamefont {Gutiérrez-Rubio}}, \bibinfo {author} {\bibfnamefont {H.}~\bibnamefont {Ochoa}}, \bibinfo {author} {\bibfnamefont {V.}~\bibnamefont {Parente}}, \bibinfo {author} {\bibfnamefont {R.}~\bibnamefont {Roldán}}, \bibinfo {author} {\bibfnamefont {P.}~\bibnamefont {San-Jose}}, \bibinfo {author} {\bibfnamefont {J.}~\bibnamefont {Schiefele}}, \bibinfo {author} {\bibfnamefont {M.}~\bibnamefont {Sturla}},\ and\ \bibinfo {author} {\bibfnamefont {M.~A.~H.}\ \bibnamefont {Vozmediano}},\ }\bibfield  {title} {\bibinfo {title} {Novel effects of strains in graphene and other two dimensional materials},\ }\href
  {https://doi.org/10.1016/j.physrep.2015.12.006} {\bibfield  {journal} {\bibinfo  {journal} {Physics Reports}\ }\textbf {\bibinfo {volume} {617}},\ \bibinfo {pages} {1} (\bibinfo {year} {2016})}\BibitemShut {NoStop}%
\bibitem [{\citenamefont {Naumis}\ \emph {et~al.}(2017)\citenamefont {Naumis}, \citenamefont {Barraza-Lopez}, \citenamefont {Oliva-Leyva},\ and\ \citenamefont {Terrones}}]{naumis_electronic_2017}%
  \BibitemOpen
  \bibfield  {author} {\bibinfo {author} {\bibfnamefont {G.~G.}\ \bibnamefont {Naumis}}, \bibinfo {author} {\bibfnamefont {S.}~\bibnamefont {Barraza-Lopez}}, \bibinfo {author} {\bibfnamefont {M.}~\bibnamefont {Oliva-Leyva}},\ and\ \bibinfo {author} {\bibfnamefont {H.}~\bibnamefont {Terrones}},\ }\bibfield  {title} {\bibinfo {title} {Electronic and optical properties of strained graphene and other strained {2D} materials: a review},\ }\href {https://doi.org/10.1088/1361-6633/aa74ef} {\bibfield  {journal} {\bibinfo  {journal} {Reports on Progress in Physics}\ }\textbf {\bibinfo {volume} {80}},\ \bibinfo {pages} {096501} (\bibinfo {year} {2017})}\BibitemShut {NoStop}%
\bibitem [{\citenamefont {Benschop}\ \emph {et~al.}(2021)\citenamefont {Benschop}, \citenamefont {de~Jong}, \citenamefont {Stepanov}, \citenamefont {Lu}, \citenamefont {Stalman}, \citenamefont {van~der Molen}, \citenamefont {Efetov},\ and\ \citenamefont {Allan}}]{Benschop2021}%
  \BibitemOpen
  \bibfield  {author} {\bibinfo {author} {\bibfnamefont {T.}~\bibnamefont {Benschop}}, \bibinfo {author} {\bibfnamefont {T.~A.}\ \bibnamefont {de~Jong}}, \bibinfo {author} {\bibfnamefont {P.}~\bibnamefont {Stepanov}}, \bibinfo {author} {\bibfnamefont {X.}~\bibnamefont {Lu}}, \bibinfo {author} {\bibfnamefont {V.}~\bibnamefont {Stalman}}, \bibinfo {author} {\bibfnamefont {S.~J.}\ \bibnamefont {van~der Molen}}, \bibinfo {author} {\bibfnamefont {D.~K.}\ \bibnamefont {Efetov}},\ and\ \bibinfo {author} {\bibfnamefont {M.~P.}\ \bibnamefont {Allan}},\ }\bibfield  {title} {\bibinfo {title} {Measuring local moiré lattice heterogeneity of twisted bilayer graphene},\ }\href {https://doi.org/10.1103/PhysRevResearch.3.013153} {\bibfield  {journal} {\bibinfo  {journal} {Physical Review Research}\ }\textbf {\bibinfo {volume} {3}},\ \bibinfo {pages} {013153} (\bibinfo {year} {2021})}\BibitemShut {NoStop}%
\bibitem [{\citenamefont {Nakatsuji}\ and\ \citenamefont {Koshino}(2022)}]{Nakatsuji2022}%
  \BibitemOpen
  \bibfield  {author} {\bibinfo {author} {\bibfnamefont {N.}~\bibnamefont {Nakatsuji}}\ and\ \bibinfo {author} {\bibfnamefont {M.}~\bibnamefont {Koshino}},\ }\bibfield  {title} {\bibinfo {title} {Moiré disorder effect in twisted bilayer graphene},\ }\href {https://doi.org/10.1103/PhysRevB.105.245408} {\bibfield  {journal} {\bibinfo  {journal} {Physical Review B}\ }\textbf {\bibinfo {volume} {105}},\ \bibinfo {pages} {245408} (\bibinfo {year} {2022})}\BibitemShut {NoStop}%
\bibitem [{\citenamefont {Yu}\ \emph {et~al.}(2024)\citenamefont {Yu}, \citenamefont {Jia}, \citenamefont {Li}, \citenamefont {Zhan}, \citenamefont {Wang}, \citenamefont {Xiao}, \citenamefont {Ju}, \citenamefont {Zhang}, \citenamefont {Hu}, \citenamefont {Guo}, \citenamefont {Lian}, \citenamefont {Tang}, \citenamefont {Pantaleon}, \citenamefont {Zhou}, \citenamefont {Guinea}, \citenamefont {Xue},\ and\ \citenamefont {Li}}]{yu2024twist}%
  \BibitemOpen
  \bibfield  {author} {\bibinfo {author} {\bibfnamefont {J.}~\bibnamefont {Yu}}, \bibinfo {author} {\bibfnamefont {G.}~\bibnamefont {Jia}}, \bibinfo {author} {\bibfnamefont {Q.}~\bibnamefont {Li}}, \bibinfo {author} {\bibfnamefont {Z.}~\bibnamefont {Zhan}}, \bibinfo {author} {\bibfnamefont {Y.}~\bibnamefont {Wang}}, \bibinfo {author} {\bibfnamefont {K.}~\bibnamefont {Xiao}}, \bibinfo {author} {\bibfnamefont {Y.}~\bibnamefont {Ju}}, \bibinfo {author} {\bibfnamefont {H.}~\bibnamefont {Zhang}}, \bibinfo {author} {\bibfnamefont {Z.}~\bibnamefont {Hu}}, \bibinfo {author} {\bibfnamefont {Y.}~\bibnamefont {Guo}}, \bibinfo {author} {\bibfnamefont {B.}~\bibnamefont {Lian}}, \bibinfo {author} {\bibfnamefont {P.}~\bibnamefont {Tang}}, \bibinfo {author} {\bibfnamefont {P.~A.}\ \bibnamefont {Pantaleon}}, \bibinfo {author} {\bibfnamefont {S.}~\bibnamefont {Zhou}}, \bibinfo {author} {\bibfnamefont {F.}~\bibnamefont {Guinea}}, \bibinfo {author} {\bibfnamefont {Q.-K.}\ \bibnamefont {Xue}},\ and\ \bibinfo {author}
  {\bibfnamefont {W.}~\bibnamefont {Li}},\ }\bibfield  {title} {\bibinfo {title} {Strain and twist angle driven electronic structure evolution in twisted bilayer graphene},\ }\href {https://arxiv.org/abs/2406.20040} {\bibfield  {journal} {\bibinfo  {journal} {arXiv}\ } (\bibinfo {year} {2024})},\ \Eprint {https://arxiv.org/abs/2406.20040} {arXiv:2406.20040 [cond-mat.mes-hall]} \BibitemShut {NoStop}%
\bibitem [{\citenamefont {Tersoff}\ and\ \citenamefont {Hamann}(1985)}]{Tersoff1985}%
  \BibitemOpen
  \bibfield  {author} {\bibinfo {author} {\bibfnamefont {J.}~\bibnamefont {Tersoff}}\ and\ \bibinfo {author} {\bibfnamefont {D.~R.}\ \bibnamefont {Hamann}},\ }\bibfield  {title} {\bibinfo {title} {Theory of the scanning tunneling microscope},\ }\href {https://doi.org/10.1103/PhysRevB.31.805} {\bibfield  {journal} {\bibinfo  {journal} {Physical Review B}\ }\textbf {\bibinfo {volume} {31}},\ \bibinfo {pages} {805} (\bibinfo {year} {1985})}\BibitemShut {NoStop}%
\bibitem [{\citenamefont {Chen}(1988)}]{Chen1988}%
  \BibitemOpen
  \bibfield  {author} {\bibinfo {author} {\bibfnamefont {C.~J.}\ \bibnamefont {Chen}},\ }\bibfield  {title} {\bibinfo {title} {Theory of scanning tunneling spectroscopy},\ }\href {https://doi.org/10.1116/1.575444} {\bibfield  {journal} {\bibinfo  {journal} {Journal of Vacuum Science \& Technology A}\ }\textbf {\bibinfo {volume} {6}},\ \bibinfo {pages} {319} (\bibinfo {year} {1988})}\BibitemShut {NoStop}%
\bibitem [{\citenamefont {Choi}\ \emph {et~al.}(2019)\citenamefont {Choi}, \citenamefont {Kemmer}, \citenamefont {Peng}, \citenamefont {Thomson}, \citenamefont {Arora}, \citenamefont {Polski}, \citenamefont {Zhang}, \citenamefont {Ren}, \citenamefont {Alicea}, \citenamefont {Refael}, \citenamefont {von Oppen}, \citenamefont {Watanabe}, \citenamefont {Taniguchi},\ and\ \citenamefont {Nadj-Perge}}]{choi_electronic_2019}%
  \BibitemOpen
  \bibfield  {author} {\bibinfo {author} {\bibfnamefont {Y.}~\bibnamefont {Choi}}, \bibinfo {author} {\bibfnamefont {J.}~\bibnamefont {Kemmer}}, \bibinfo {author} {\bibfnamefont {Y.}~\bibnamefont {Peng}}, \bibinfo {author} {\bibfnamefont {A.}~\bibnamefont {Thomson}}, \bibinfo {author} {\bibfnamefont {H.}~\bibnamefont {Arora}}, \bibinfo {author} {\bibfnamefont {R.}~\bibnamefont {Polski}}, \bibinfo {author} {\bibfnamefont {Y.}~\bibnamefont {Zhang}}, \bibinfo {author} {\bibfnamefont {H.}~\bibnamefont {Ren}}, \bibinfo {author} {\bibfnamefont {J.}~\bibnamefont {Alicea}}, \bibinfo {author} {\bibfnamefont {G.}~\bibnamefont {Refael}}, \bibinfo {author} {\bibfnamefont {F.}~\bibnamefont {von Oppen}}, \bibinfo {author} {\bibfnamefont {K.}~\bibnamefont {Watanabe}}, \bibinfo {author} {\bibfnamefont {T.}~\bibnamefont {Taniguchi}},\ and\ \bibinfo {author} {\bibfnamefont {S.}~\bibnamefont {Nadj-Perge}},\ }\bibfield  {title} {\bibinfo {title} {Electronic correlations in twisted bilayer graphene near the magic angle},\ }\href
  {https://doi.org/10.1038/s41567-019-0606-5} {\bibfield  {journal} {\bibinfo  {journal} {Nature Physics}\ }\textbf {\bibinfo {volume} {15}},\ \bibinfo {pages} {1174} (\bibinfo {year} {2019})}\BibitemShut {NoStop}%
\bibitem [{\citenamefont {Suzuura}\ and\ \citenamefont {Ando}(2002)}]{suzuura_phonons_2002}%
  \BibitemOpen
  \bibfield  {author} {\bibinfo {author} {\bibfnamefont {H.}~\bibnamefont {Suzuura}}\ and\ \bibinfo {author} {\bibfnamefont {T.}~\bibnamefont {Ando}},\ }\bibfield  {title} {\bibinfo {title} {Phonons and electron-phonon scattering in carbon nanotubes},\ }\href {https://doi.org/10.1103/PhysRevB.65.235412} {\bibfield  {journal} {\bibinfo  {journal} {Phys. Rev. B}\ }\textbf {\bibinfo {volume} {65}},\ \bibinfo {pages} {235412} (\bibinfo {year} {2002})}\BibitemShut {NoStop}%
\bibitem [{\citenamefont {Choi}\ \emph {et~al.}(2010)\citenamefont {Choi}, \citenamefont {Jhi},\ and\ \citenamefont {Son}}]{choi_effects_2010}%
  \BibitemOpen
  \bibfield  {author} {\bibinfo {author} {\bibfnamefont {S.-M.}\ \bibnamefont {Choi}}, \bibinfo {author} {\bibfnamefont {S.-H.}\ \bibnamefont {Jhi}},\ and\ \bibinfo {author} {\bibfnamefont {Y.-W.}\ \bibnamefont {Son}},\ }\bibfield  {title} {\bibinfo {title} {Effects of strain on electronic properties of graphene},\ }\href {https://doi.org/10.1103/PhysRevB.81.081407} {\bibfield  {journal} {\bibinfo  {journal} {Phys. Rev. B}\ }\textbf {\bibinfo {volume} {81}},\ \bibinfo {pages} {081407} (\bibinfo {year} {2010})}\BibitemShut {NoStop}%
\bibitem [{\citenamefont {Vozmediano}\ \emph {et~al.}(2010)\citenamefont {Vozmediano}, \citenamefont {Katsnelson},\ and\ \citenamefont {Guinea}}]{vozmediano_gauge_2010}%
  \BibitemOpen
  \bibfield  {author} {\bibinfo {author} {\bibfnamefont {M.~A.~H.}\ \bibnamefont {Vozmediano}}, \bibinfo {author} {\bibfnamefont {M.~I.}\ \bibnamefont {Katsnelson}},\ and\ \bibinfo {author} {\bibfnamefont {F.}~\bibnamefont {Guinea}},\ }\bibfield  {title} {\bibinfo {title} {Gauge fields in graphene},\ }\href {https://doi.org/10.1016/j.physrep.2010.07.003} {\bibfield  {journal} {\bibinfo  {journal} {Physics Reports}\ }\textbf {\bibinfo {volume} {496}},\ \bibinfo {pages} {109} (\bibinfo {year} {2010})}\BibitemShut {NoStop}%
\bibitem [{\citenamefont {Guinea}\ and\ \citenamefont {Walet}(2018)}]{guinea_electrostatic_2018}%
  \BibitemOpen
  \bibfield  {author} {\bibinfo {author} {\bibfnamefont {F.}~\bibnamefont {Guinea}}\ and\ \bibinfo {author} {\bibfnamefont {N.~R.}\ \bibnamefont {Walet}},\ }\bibfield  {title} {\bibinfo {title} {Electrostatic effects, band distortions, and superconductivity in twisted graphene bilayers},\ }\href {https://doi.org/10.1073/pnas.1810947115} {\bibfield  {journal} {\bibinfo  {journal} {Proceedings of the National Academy of Sciences}\ }\textbf {\bibinfo {volume} {115}},\ \bibinfo {pages} {13174} (\bibinfo {year} {2018})}\BibitemShut {NoStop}%
\bibitem [{\citenamefont {Cea}\ \emph {et~al.}(2022)\citenamefont {Cea}, \citenamefont {Pantaleón}, \citenamefont {Walet},\ and\ \citenamefont {Guinea}}]{cea_electrostatic_2022}%
  \BibitemOpen
  \bibfield  {author} {\bibinfo {author} {\bibfnamefont {T.}~\bibnamefont {Cea}}, \bibinfo {author} {\bibfnamefont {P.~A.}\ \bibnamefont {Pantaleón}}, \bibinfo {author} {\bibfnamefont {N.~R.}\ \bibnamefont {Walet}},\ and\ \bibinfo {author} {\bibfnamefont {F.}~\bibnamefont {Guinea}},\ }\bibfield  {title} {\bibinfo {title} {Electrostatic interactions in twisted bilayer graphene},\ }\href {https://doi.org/10.1016/j.nanoms.2021.10.001} {\bibfield  {journal} {\bibinfo  {journal} {Nano Materials Science}\ }\textbf {\bibinfo {volume} {4}},\ \bibinfo {pages} {27} (\bibinfo {year} {2022})}\BibitemShut {NoStop}%
\bibitem [{\citenamefont {Cea}\ \emph {et~al.}(2019)\citenamefont {Cea}, \citenamefont {Walet},\ and\ \citenamefont {Guinea}}]{Cea2019}%
  \BibitemOpen
  \bibfield  {author} {\bibinfo {author} {\bibfnamefont {T.}~\bibnamefont {Cea}}, \bibinfo {author} {\bibfnamefont {N.~R.}\ \bibnamefont {Walet}},\ and\ \bibinfo {author} {\bibfnamefont {F.}~\bibnamefont {Guinea}},\ }\bibfield  {title} {\bibinfo {title} {Electronic band structure and pinning of {Fermi} energy to {Van} {Hove} singularities in twisted bilayer graphene: {A} self-consistent approach},\ }\href {https://doi.org/10.1103/PhysRevB.100.205113} {\bibfield  {journal} {\bibinfo  {journal} {Physical Review B}\ }\textbf {\bibinfo {volume} {100}},\ \bibinfo {pages} {205113} (\bibinfo {year} {2019})}\BibitemShut {NoStop}%
\bibitem [{\citenamefont {Pantaleón}\ \emph {et~al.}(2021)\citenamefont {Pantaleón}, \citenamefont {Cea}, \citenamefont {Brown}, \citenamefont {Walet},\ and\ \citenamefont {Guinea}}]{Pantaleon2021}%
  \BibitemOpen
  \bibfield  {author} {\bibinfo {author} {\bibfnamefont {P.~A.}\ \bibnamefont {Pantaleón}}, \bibinfo {author} {\bibfnamefont {T.}~\bibnamefont {Cea}}, \bibinfo {author} {\bibfnamefont {R.}~\bibnamefont {Brown}}, \bibinfo {author} {\bibfnamefont {N.~R.}\ \bibnamefont {Walet}},\ and\ \bibinfo {author} {\bibfnamefont {F.}~\bibnamefont {Guinea}},\ }\bibfield  {title} {\bibinfo {title} {Narrow bands, electrostatic interactions and band topology in graphene stacks},\ }\href {https://doi.org/10.1088/2053-1583/ac1b6d} {\bibfield  {journal} {\bibinfo  {journal} {2D Materials}\ }\textbf {\bibinfo {volume} {8}},\ \bibinfo {pages} {044006} (\bibinfo {year} {2021})}\BibitemShut {NoStop}%
\bibitem [{\citenamefont {Koshino}\ \emph {et~al.}(2018)\citenamefont {Koshino}, \citenamefont {Yuan}, \citenamefont {Koretsune}, \citenamefont {Ochi}, \citenamefont {Kuroki},\ and\ \citenamefont {Fu}}]{Koshino2018}%
  \BibitemOpen
  \bibfield  {author} {\bibinfo {author} {\bibfnamefont {M.}~\bibnamefont {Koshino}}, \bibinfo {author} {\bibfnamefont {N.~F.}\ \bibnamefont {Yuan}}, \bibinfo {author} {\bibfnamefont {T.}~\bibnamefont {Koretsune}}, \bibinfo {author} {\bibfnamefont {M.}~\bibnamefont {Ochi}}, \bibinfo {author} {\bibfnamefont {K.}~\bibnamefont {Kuroki}},\ and\ \bibinfo {author} {\bibfnamefont {L.}~\bibnamefont {Fu}},\ }\bibfield  {title} {\bibinfo {title} {Maximally {Localized} {Wannier} {Orbitals} and the {Extended} {Hubbard} {Model} for {Twisted} {Bilayer} {Graphene}},\ }\href {https://doi.org/10.1103/PhysRevX.8.031087} {\bibfield  {journal} {\bibinfo  {journal} {Physical Review X}\ }\textbf {\bibinfo {volume} {8}},\ \bibinfo {pages} {031087} (\bibinfo {year} {2018})}\BibitemShut {NoStop}%
\bibitem [{\citenamefont {Long}\ \emph {et~al.}(2024)\citenamefont {Long}, \citenamefont {Jimeno-Pozo}, \citenamefont {Sainz-Cruz}, \citenamefont {Pantaleón},\ and\ \citenamefont {Guinea}}]{Long2024Evolution}%
  \BibitemOpen
  \bibfield  {author} {\bibinfo {author} {\bibfnamefont {M.}~\bibnamefont {Long}}, \bibinfo {author} {\bibfnamefont {A.}~\bibnamefont {Jimeno-Pozo}}, \bibinfo {author} {\bibfnamefont {H.}~\bibnamefont {Sainz-Cruz}}, \bibinfo {author} {\bibfnamefont {P.~A.}\ \bibnamefont {Pantaleón}},\ and\ \bibinfo {author} {\bibfnamefont {F.}~\bibnamefont {Guinea}},\ }\bibfield  {title} {\bibinfo {title} {Evolution of superconductivity in twisted graphene multilayers},\ }\bibfield  {journal} {\bibinfo  {journal} {Proceedings of the National Academy of Sciences}\ }\textbf {\bibinfo {volume} {121}},\ \href {https://doi.org/10.1073/pnas.2405259121} {10.1073/pnas.2405259121} (\bibinfo {year} {2024})\BibitemShut {NoStop}%
\bibitem [{\citenamefont {Faugeras}\ \emph {et~al.}(2008)\citenamefont {Faugeras}, \citenamefont {Nerrière}, \citenamefont {Potemski}, \citenamefont {Mahmood}, \citenamefont {Dujardin}, \citenamefont {Berger},\ and\ \citenamefont {de~Heer}}]{Faugeras2008}%
  \BibitemOpen
  \bibfield  {author} {\bibinfo {author} {\bibfnamefont {C.}~\bibnamefont {Faugeras}}, \bibinfo {author} {\bibfnamefont {A.}~\bibnamefont {Nerrière}}, \bibinfo {author} {\bibfnamefont {M.}~\bibnamefont {Potemski}}, \bibinfo {author} {\bibfnamefont {A.}~\bibnamefont {Mahmood}}, \bibinfo {author} {\bibfnamefont {E.}~\bibnamefont {Dujardin}}, \bibinfo {author} {\bibfnamefont {C.}~\bibnamefont {Berger}},\ and\ \bibinfo {author} {\bibfnamefont {W.~A.}\ \bibnamefont {de~Heer}},\ }\bibfield  {title} {\bibinfo {title} {Few-layer graphene on {SiC}, pyrolitic graphite, and graphene: {A} {Raman} scattering study},\ }\href {https://doi.org/10.1063/1.2828975} {\bibfield  {journal} {\bibinfo  {journal} {Applied Physics Letters}\ }\textbf {\bibinfo {volume} {92}},\ \bibinfo {pages} {011914} (\bibinfo {year} {2008})}\BibitemShut {NoStop}%
\bibitem [{\citenamefont {Norimatsu}\ and\ \citenamefont {Kusunoki}(2014)}]{Norimatsu2014}%
  \BibitemOpen
  \bibfield  {author} {\bibinfo {author} {\bibfnamefont {W.}~\bibnamefont {Norimatsu}}\ and\ \bibinfo {author} {\bibfnamefont {M.}~\bibnamefont {Kusunoki}},\ }\bibfield  {title} {\bibinfo {title} {Epitaxial graphene on {SiC}\{0001\}: advances and perspectives},\ }\href {https://doi.org/10.1039/C3CP54523G} {\bibfield  {journal} {\bibinfo  {journal} {Physical Chemistry Chemical Physics}\ }\textbf {\bibinfo {volume} {16}},\ \bibinfo {pages} {3501} (\bibinfo {year} {2014})}\BibitemShut {NoStop}%
\bibitem [{\citenamefont {Yazdi}\ \emph {et~al.}(2016)\citenamefont {Yazdi}, \citenamefont {Iakimov},\ and\ \citenamefont {Yakimova}}]{Yazdi2016}%
  \BibitemOpen
  \bibfield  {author} {\bibinfo {author} {\bibfnamefont {G.~R.}\ \bibnamefont {Yazdi}}, \bibinfo {author} {\bibfnamefont {T.}~\bibnamefont {Iakimov}},\ and\ \bibinfo {author} {\bibfnamefont {R.}~\bibnamefont {Yakimova}},\ }\bibfield  {title} {\bibinfo {title} {Epitaxial {Graphene} on {SiC}: {A} {Review} of {Growth} and {Characterization}},\ }\href {https://doi.org/10.3390/cryst6050053} {\bibfield  {journal} {\bibinfo  {journal} {Crystals}\ }\textbf {\bibinfo {volume} {6}},\ \bibinfo {pages} {53} (\bibinfo {year} {2016})}\BibitemShut {NoStop}%
\bibitem [{\citenamefont {Mishra}\ \emph {et~al.}(2016)\citenamefont {Mishra}, \citenamefont {Boeckl}, \citenamefont {Motta},\ and\ \citenamefont {Iacopi}}]{Mishra2016}%
  \BibitemOpen
  \bibfield  {author} {\bibinfo {author} {\bibfnamefont {N.}~\bibnamefont {Mishra}}, \bibinfo {author} {\bibfnamefont {J.}~\bibnamefont {Boeckl}}, \bibinfo {author} {\bibfnamefont {N.}~\bibnamefont {Motta}},\ and\ \bibinfo {author} {\bibfnamefont {F.}~\bibnamefont {Iacopi}},\ }\bibfield  {title} {\bibinfo {title} {Graphene growth on silicon carbide: {A} review},\ }\href {https://doi.org/10.1002/pssa.201600091} {\bibfield  {journal} {\bibinfo  {journal} {physica status solidi (a)}\ }\textbf {\bibinfo {volume} {213}},\ \bibinfo {pages} {2277} (\bibinfo {year} {2016})}\BibitemShut {NoStop}%
\bibitem [{\citenamefont {Sprinkle}\ \emph {et~al.}(2009)\citenamefont {Sprinkle}, \citenamefont {Siegel}, \citenamefont {Hu}, \citenamefont {Hicks}, \citenamefont {Tejeda}, \citenamefont {Taleb-Ibrahimi}, \citenamefont {Le~F\`evre}, \citenamefont {Bertran}, \citenamefont {Vizzini}, \citenamefont {Enriquez}, \citenamefont {Chiang}, \citenamefont {Soukiassian}, \citenamefont {Berger}, \citenamefont {de~Heer}, \citenamefont {Lanzara},\ and\ \citenamefont {Conrad}}]{Sprinkle_PRL2009}%
  \BibitemOpen
  \bibfield  {author} {\bibinfo {author} {\bibfnamefont {M.}~\bibnamefont {Sprinkle}}, \bibinfo {author} {\bibfnamefont {D.}~\bibnamefont {Siegel}}, \bibinfo {author} {\bibfnamefont {Y.}~\bibnamefont {Hu}}, \bibinfo {author} {\bibfnamefont {J.}~\bibnamefont {Hicks}}, \bibinfo {author} {\bibfnamefont {A.}~\bibnamefont {Tejeda}}, \bibinfo {author} {\bibfnamefont {A.}~\bibnamefont {Taleb-Ibrahimi}}, \bibinfo {author} {\bibfnamefont {P.}~\bibnamefont {Le~F\`evre}}, \bibinfo {author} {\bibfnamefont {F.}~\bibnamefont {Bertran}}, \bibinfo {author} {\bibfnamefont {S.}~\bibnamefont {Vizzini}}, \bibinfo {author} {\bibfnamefont {H.}~\bibnamefont {Enriquez}}, \bibinfo {author} {\bibfnamefont {S.}~\bibnamefont {Chiang}}, \bibinfo {author} {\bibfnamefont {P.}~\bibnamefont {Soukiassian}}, \bibinfo {author} {\bibfnamefont {C.}~\bibnamefont {Berger}}, \bibinfo {author} {\bibfnamefont {W.~A.}\ \bibnamefont {de~Heer}}, \bibinfo {author} {\bibfnamefont {A.}~\bibnamefont {Lanzara}},\ and\ \bibinfo {author} {\bibfnamefont {E.~H.}\
  \bibnamefont {Conrad}},\ }\bibfield  {title} {\bibinfo {title} {First direct observation of a nearly ideal graphene band structure},\ }\href {https://doi.org/10.1103/PhysRevLett.103.226803} {\bibfield  {journal} {\bibinfo  {journal} {Phys. Rev. Lett.}\ }\textbf {\bibinfo {volume} {103}},\ \bibinfo {pages} {226803} (\bibinfo {year} {2009})}\BibitemShut {NoStop}%
\bibitem [{\citenamefont {Rosenberger}\ \emph {et~al.}(2020)\citenamefont {Rosenberger}, \citenamefont {Chuang}, \citenamefont {Phillips}, \citenamefont {Oleshko}, \citenamefont {McCreary}, \citenamefont {Sivaram}, \citenamefont {Hellberg},\ and\ \citenamefont {Jonker}}]{Rosenberger2020}%
  \BibitemOpen
  \bibfield  {author} {\bibinfo {author} {\bibfnamefont {M.~R.}\ \bibnamefont {Rosenberger}}, \bibinfo {author} {\bibfnamefont {H.-J.}\ \bibnamefont {Chuang}}, \bibinfo {author} {\bibfnamefont {M.}~\bibnamefont {Phillips}}, \bibinfo {author} {\bibfnamefont {V.~P.}\ \bibnamefont {Oleshko}}, \bibinfo {author} {\bibfnamefont {K.~M.}\ \bibnamefont {McCreary}}, \bibinfo {author} {\bibfnamefont {S.~V.}\ \bibnamefont {Sivaram}}, \bibinfo {author} {\bibfnamefont {C.~S.}\ \bibnamefont {Hellberg}},\ and\ \bibinfo {author} {\bibfnamefont {B.~T.}\ \bibnamefont {Jonker}},\ }\bibfield  {title} {\bibinfo {title} {Twist {Angle}-{Dependent} {Atomic} {Reconstruction} and {Moiré} {Patterns} in {Transition} {Metal} {Dichalcogenide} {Heterostructures}},\ }\href {https://doi.org/10.1021/acsnano.0c00088} {\bibfield  {journal} {\bibinfo  {journal} {ACS Nano}\ }\textbf {\bibinfo {volume} {14}},\ \bibinfo {pages} {4550} (\bibinfo {year} {2020})}\BibitemShut {NoStop}%
\bibitem [{\citenamefont {Halbertal}\ \emph {et~al.}(2022)\citenamefont {Halbertal}, \citenamefont {Shabani}, \citenamefont {Passupathy},\ and\ \citenamefont {Basov}}]{Halbertal2022}%
  \BibitemOpen
  \bibfield  {author} {\bibinfo {author} {\bibfnamefont {D.}~\bibnamefont {Halbertal}}, \bibinfo {author} {\bibfnamefont {S.}~\bibnamefont {Shabani}}, \bibinfo {author} {\bibfnamefont {A.~N.}\ \bibnamefont {Passupathy}},\ and\ \bibinfo {author} {\bibfnamefont {D.~N.}\ \bibnamefont {Basov}},\ }\bibfield  {title} {\bibinfo {title} {Extracting the {Strain} {Matrix} and {Twist} {Angle} from the {Moiré} {Superlattice} in van der {Waals} {Heterostructures}},\ }\href {https://doi.org/10.1021/acsnano.1c09789} {\bibfield  {journal} {\bibinfo  {journal} {ACS Nano}\ }\textbf {\bibinfo {volume} {16}},\ \bibinfo {pages} {1471} (\bibinfo {year} {2022})}\BibitemShut {NoStop}%
\bibitem [{\citenamefont {Jong}\ \emph {et~al.}(2022)\citenamefont {Jong}, \citenamefont {Benschop}, \citenamefont {Chen}, \citenamefont {Krasovskii}, \citenamefont {Dood}, \citenamefont {Tromp}, \citenamefont {Allan},\ and\ \citenamefont {Molen}}]{jong_imaging_2022}%
  \BibitemOpen
  \bibfield  {author} {\bibinfo {author} {\bibfnamefont {T.~A.~d.}\ \bibnamefont {Jong}}, \bibinfo {author} {\bibfnamefont {T.}~\bibnamefont {Benschop}}, \bibinfo {author} {\bibfnamefont {X.}~\bibnamefont {Chen}}, \bibinfo {author} {\bibfnamefont {E.~E.}\ \bibnamefont {Krasovskii}}, \bibinfo {author} {\bibfnamefont {M.~J. A.~d.}\ \bibnamefont {Dood}}, \bibinfo {author} {\bibfnamefont {R.~M.}\ \bibnamefont {Tromp}}, \bibinfo {author} {\bibfnamefont {M.~P.}\ \bibnamefont {Allan}},\ and\ \bibinfo {author} {\bibfnamefont {S.~J. v.~d.}\ \bibnamefont {Molen}},\ }\bibfield  {title} {\bibinfo {title} {Imaging moiré deformation and dynamics in twisted bilayer graphene},\ }\bibfield  {journal} {\bibinfo  {journal} {Nature Communications}\ }\textbf {\bibinfo {volume} {13}},\ \href {https://doi.org/10.1038/s41467-021-27646-1} {10.1038/s41467-021-27646-1} (\bibinfo {year} {2022})\BibitemShut {NoStop}%
\bibitem [{\citenamefont {Horcas}\ \emph {et~al.}(2007)\citenamefont {Horcas}, \citenamefont {Fernández}, \citenamefont {Gómez-Rodríguez}, \citenamefont {Colchero}, \citenamefont {Gómez-Herrero},\ and\ \citenamefont {Baro}}]{Horcas2007}%
  \BibitemOpen
  \bibfield  {author} {\bibinfo {author} {\bibfnamefont {I.}~\bibnamefont {Horcas}}, \bibinfo {author} {\bibfnamefont {R.}~\bibnamefont {Fernández}}, \bibinfo {author} {\bibfnamefont {J.~M.}\ \bibnamefont {Gómez-Rodríguez}}, \bibinfo {author} {\bibfnamefont {J.}~\bibnamefont {Colchero}}, \bibinfo {author} {\bibfnamefont {J.}~\bibnamefont {Gómez-Herrero}},\ and\ \bibinfo {author} {\bibfnamefont {A.~M.}\ \bibnamefont {Baro}},\ }\bibfield  {title} {\bibinfo {title} {{WSXM}: A software for scanning probe microscopy and a tool for nanotechnology},\ }\href {https://doi.org/10.1063/1.2432410} {\bibfield  {journal} {\bibinfo  {journal} {Rev. Sci. Instrum.}\ }\textbf {\bibinfo {volume} {78}},\ \bibinfo {pages} {013705} (\bibinfo {year} {2007})}\BibitemShut {NoStop}%
\bibitem [{\citenamefont {Long}\ \emph {et~al.}(2022)\citenamefont {Long}, \citenamefont {Pantaleón}, \citenamefont {Zhan}, \citenamefont {Guinea}, \citenamefont {Silva-Guillén},\ and\ \citenamefont {Yuan}}]{long_atomistic_2022}%
  \BibitemOpen
  \bibfield  {author} {\bibinfo {author} {\bibfnamefont {M.}~\bibnamefont {Long}}, \bibinfo {author} {\bibfnamefont {P.~A.}\ \bibnamefont {Pantaleón}}, \bibinfo {author} {\bibfnamefont {Z.}~\bibnamefont {Zhan}}, \bibinfo {author} {\bibfnamefont {F.}~\bibnamefont {Guinea}}, \bibinfo {author} {\bibfnamefont {J.~A.}\ \bibnamefont {Silva-Guillén}},\ and\ \bibinfo {author} {\bibfnamefont {S.}~\bibnamefont {Yuan}},\ }\bibfield  {title} {\bibinfo {title} {An atomistic approach for the structural and electronic properties of twisted bilayer graphene-boron nitride heterostructures},\ }\bibfield  {journal} {\bibinfo  {journal} {npj Computational Materials}\ }\textbf {\bibinfo {volume} {8}},\ \href {https://doi.org/10.1038/s41524-022-00763-1} {10.1038/s41524-022-00763-1} (\bibinfo {year} {2022})\BibitemShut {NoStop}%
\bibitem [{\citenamefont {Nam}\ and\ \citenamefont {Koshino}(2017)}]{Nam2017}%
  \BibitemOpen
  \bibfield  {author} {\bibinfo {author} {\bibfnamefont {N.~N.~T.}\ \bibnamefont {Nam}}\ and\ \bibinfo {author} {\bibfnamefont {M.}~\bibnamefont {Koshino}},\ }\bibfield  {title} {\bibinfo {title} {Lattice relaxation and energy band modulation in twisted bilayer graphene},\ }\href {https://doi.org/10.1103/PhysRevB.96.075311} {\bibfield  {journal} {\bibinfo  {journal} {Physical Review B}\ }\textbf {\bibinfo {volume} {96}},\ \bibinfo {pages} {075311} (\bibinfo {year} {2017})}\BibitemShut {NoStop}%
\bibitem [{\citenamefont {Guinea}\ and\ \citenamefont {Walet}(2019)}]{guinea_continuum_2019}%
  \BibitemOpen
  \bibfield  {author} {\bibinfo {author} {\bibfnamefont {F.}~\bibnamefont {Guinea}}\ and\ \bibinfo {author} {\bibfnamefont {N.~R.}\ \bibnamefont {Walet}},\ }\bibfield  {title} {\bibinfo {title} {Continuum models for twisted bilayer graphene: {Effect} of lattice deformation and hopping parameters},\ }\href {https://doi.org/10.1103/PhysRevB.99.205134} {\bibfield  {journal} {\bibinfo  {journal} {Phys. Rev. B}\ }\textbf {\bibinfo {volume} {99}},\ \bibinfo {pages} {205134} (\bibinfo {year} {2019})}\BibitemShut {NoStop}%
\end{thebibliography}
\end{document}